\documentclass[11pt,a4paper]{article}
\usepackage[T1]{fontenc}
\usepackage[utf8]{inputenc}
\usepackage{lmodern}
\usepackage[margin=24mm]{geometry}
\usepackage{amsmath,amssymb,bm}
\usepackage{graphicx}
\usepackage{caption,subcaption}
\usepackage{algorithm,algpseudocode}
\usepackage{booktabs}
\usepackage{placeins}
\usepackage{xcolor,listings}
\usepackage[authoryear,longnamesfirst]{natbib}

\usepackage{authblk}
\usepackage{microtype}
\usepackage[hidelinks]{hyperref}
\hypersetup{
  pdftitle={A regularized lattice Boltzmann model for the solid-liquid phase change},
  pdfauthor={Qiang Fu, Yu Han, Xuhui Li, Xiaowen Shan}
}
\lstdefinestyle{cuda}{
	language=C++,
	basicstyle=\small\ttfamily,
	keywordstyle=\color{blue}\bfseries,
	commentstyle=\color{gray}\itshape,
	stringstyle=\color{teal},
	showstringspaces=false,
	breaklines=true,
	morekeywords={
		cudaEventRecord, cudaStreamWaitEvent, cudaStream_t,
		cudaEvent_t, __global__, __device__, __host__
	}
}

\DeclareMathOperator{\erf}{erf}
\DeclareMathOperator{\erfc}{erfc}

\newcommand{\ba}[1]{\bm{a}^{(#1)}}

\newcommand{\bu}{\bm{u}}

\newcommand{\bxi}{\bm{\xi}}

\newcommand{\bH}[1]{{\cal H}^{(#1)}}

\title{A regularized lattice Boltzmann model for the solid-liquid phase change}
\author[1]{Qiang Fu}
\author[1]{Yu Han}
\author[2]{Xuhui Li\thanks{Corresponding author. Email: \href{mailto:lixuhui@hrbeu.edu.cn}{lixuhui@hrbeu.edu.cn}. ORCID: \href{https://orcid.org/0000-0001-5421-3579}{0000-0001-5421-3579}.}}
\author[3]{Xiaowen Shan}
\affil[1]{Northeast Agricultural University, Harbin 150001, Heilongjiang, China}
\affil[2]{Harbin Engineering University, Harbin 150001, Heilongjiang, China}
\affil[3]{Beijing Normal-Hong Kong Baptist University, Zhuhai 519000, Guangdong, China}
\date{}
\begin{document}
\maketitle
\begin{abstract}
In this study, a regularized lattice Boltzmann method is proposed for solid-liquid phase-change simulations. The governing total-enthalpy equation is solved within a regularized lattice Boltzmann (RLB) framework, which enables accurate tracking of the solid-liquid interface while effectively suppressing numerical diffusion. In addition, the multi-relaxation-time RLB (MRT-RLB) model with standard lattices is developed to solve the flow field. To enhance computational efficiency, an efficient multi-GPU (graphics processing unit) implementation based on CUDA C (Compute Unified Device Architecture) is developed, which is particularly suited for computationally intensive three-dimensional phase-change simulations. The proposed model is validated against a series of one-, two-, and three-dimensional benchmark problems. The results demonstrate that the present solver provides a robust, accurate, and efficient framework for simulating solid-liquid phase-change heat transfer.
\end{abstract}

\section{Introduction}\label{Introduction}
Solid-liquid phase change, characterized by minimal volume variation, a nearly isothermal transition, and substantial latent heat exchange, is extensively utilized in both daily life and engineering applications, ranging from freeze-thaw cycles of permafrost in cold regions to thermal energy storage in green buildings. However, the dynamically evolving and highly nonlinear phase interface, coupled with the intricate, multi-physical nature of fluid flow and heat transfer, presents formidable challenges to investigating these phenomena. For such complex phase-change problems, experimental approaches are often constrained by the difficulty of replicating specific operating conditions and capturing localized transport characteristics, thereby rendering numerical simulation an indispensable tool. Compared with conventional computational fluid dynamics (CFD) methods, the lattice Boltzmann method (LBM)~\citep{Chen1998,Cyrus2010}, by virtue of its inherent mesoscopic nature, exhibits distinct advantages in handling complex geometries, multi-physics coupling, and parallel computation. Consequently, LBM has emerged as a highly promising alternative for simulating complex solid-liquid phase-change processes. Nevertheless, as a developing mesoscopic numerical framework, LBM still faces several critical challenges when applied to solid-liquid phase-change scenarios.

In 1998, de Fabritiis et al.\citep{deFabritiis1998} pioneered the application of the lattice Boltzmann method to solid-liquid phase-change problems. In their model, the solid and liquid phases were treated as two distinct components governed by separate lattice Boltzmann equations, with the phase transition realized by incorporating a reaction source term into the evolution equations. Since this seminal work, extensive research has been dedicated to applying LBM to solid-liquid phase-change phenomena. Generally, existing lattice Boltzmann approaches for modeling these processes can be broadly classified into three categories: (1) phase-field-based methods\citep{Miller2001}, (2) enthalpy-based methods~\citep{Jiaung2001}, and (3) interface-tracking methods~\citep{Huang2014}. In phase-field-based methods, an auxiliary order parameter is introduced to capture the phase interface, which is represented by the continuous spatial variation of this parameter~\citep{Miller2001}. In enthalpy-based LBMs~\citep{Jiaung2001}, the temperature field is reconstructed from the enthalpy, which also serves to determine the local liquid fraction at the interface. Meanwhile, interface-tracking methods, as developed by Huang et al.\citep{Huang2014} and Li et al.\citep{Li2015}, offer an alternative paradigm by explicitly locating and updating the phase boundary.

Owing to its simplicity and robustness, the enthalpy-based lattice Boltzmann approach has been widely adopted for the simulation of solid-liquid phase-change problems. In 2001, Jiaung et al.\citep{Jiaung2001} proposed the first enthalpy-based lattice Boltzmann model for solving the convection-diffusion equation in solid-liquid phase-change problems. In this pioneering enthalpy-based model, the latent heat source term is treated using an implicit iterative scheme, whereas the convection term is not rigorously considered. Following this framework, a series of subsequent studies were conducted. These include the introduction of a relaxation factor to enhance the convergence of the implicit iteration by Chatterjee and Chakraborty\citep{Chatterjee2008}, the further incorporation of viscous dissipation by Chatterjee and  Chakraborty~\citep{Chatterjee2009,Chatterjee2010}, and the compulsory single-iteration treatment proposed by Huber et al.\citep{Huber2008}. Subsequently, Eshraghi and Felicelli\citep{Eshraghi2012} proposed another implicit lattice Boltzmann model, in which the latent heat source term is resolved by solving a system of linear equations at each lattice node instead of using an iterative procedure. In general, the aforementioned enthalpy-based lattice Boltzmann approaches rely on various implicit iterative techniques to treat the latent heat source term. Huang et al.~\citep{Huang2013,Huang2015} developed a total-enthalpy-based lattice Boltzmann approach, where the convection term was neglected in the treatment of the source term. In this model, the total enthalpy, including both sensible and latent enthalpy, is taken as the target macroscopic variable, and an elaborately designed discrete equilibrium function is introduced to recover the corresponding macroscopic partial differential equation.

To mitigate numerical diffusion at the phase interface, various advanced collision models have been developed.
In the work of Huang et al.~\citep{Huang2015}, the so-called magic parameters~\citep{Ginzburg2003} were incorporated into the multiple-relaxation-time (MRT) collision model to suppress numerical diffusion.
Li et al.~\citep{Li2017} and Liu et al.~\citep{Liu2019} employed a non-orthogonal MRT model to solve the enthalpy equation based on the D3Q7 lattice.
Lu et al.~\citep{Lu2019} adopted the two-relaxation-time (TRT) collision model, coupled with a magic parameter~\citep{Ginzburg2010}, to suppress unphysical numerical diffusion.
Liu et al.~\citep{Liu2022} utilized a cascaded MRT model to solve the enthalpy equation, thereby suppressing numerical diffusion and enhancing numerical stability.
To simulat the droplet impact and freezing on solid surfaces, Xu et al.~\citep{Xu2025} applied an enhanced cascaded lattice Boltzmann method to calculate the flow field and a multiple-relaxation-time scheme to solve the liquid-solid phase change model.
Jing et al.~\citep{Jing2026} proposed a multiphase model that integrated the phase-field method, the enthalpy-based method, and the lattice Boltzmann method to simulate the droplet solidification on cold substrates.
Overall, these collision models and integrated strategy have alleviated the unphysical diffusion at the solid-liquid interface and promoted the model close to the real physical problems.

Since solid-liquid phase-change processes are inherently slow-evolving physical phenomena, their simulation using the lattice Boltzmann method (LBM) is generally computationally demanding.
To address this, Liu et al.~\citep{Liu2025} developed a multi-GPU code for simulating high-Rayleigh-number, convection-dominated melting, based on a TRT lattice Boltzmann evolution equation with a source term to ensure the strict recovery of the macroscopic total-enthalpy equation.
In recent years, regularized lattice Boltzmann collision models have been extensively and successfully employed in high-Reynolds-number flow simulations, owing to their outstanding ability to enhance the robustness and accuracy of violent flow simulations.
Among the existing regularized collision models, recursive regularized model for weakly compressible flow~\citep{Malaspinas2015} and compressible flow ~\citep{Mattila2017,Coreixas2017}, hybrid recursive regularized model for weakly compressible flow~\citep{Jacob2018}, and temperature-scaled regularized multiple-relaxation-time models for compressible flow~\citep{Li2019,Shan2021,Li2021} represent the primary categories.

In the present work, a regularized lattice Boltzmann model is designed to solve the enthalpy equation, while the flow field is simultaneously resolved using the weakly compressible multi-relaxation-time regularized lattice Boltzmann (MRT-RLB) model on standard lattices(D3Q27). This implementation builds upon our previous high-order lattice model~\citep{Li2022, Li2024} to enhance the numerical accuracy and stability of the flow simulation. The numerical results demonstrate that the proposed model not only provides robust numerical stability for flow-field simulations but also effectively suppresses numerical diffusion at the solid--liquid interface. In general, this work represents the first application of advanced regularized collision model to both the flow field and the enthalpy-based governing equation for solid-liquid phase-change.

The remainder of this paper is organized as follows. Section 2 presents the theoretical model, numerical implementation details, and the GPU acceleration strategy. Section 3 validates the proposed model through several benchmark problems, encompassing one-, two-, and three-dimensional cases. Finally, Section 4 summarizes the primary conclusions of this work.

\section{Theoretical Model}

In the present study, the following assumptions are made for the solid-liquid phase change physics.
Compressibility effects are assumed to be negligible, and the Boussinesq approximation is adopted to describe the buoyancy force arising from temperature variations;
 Volume expansion resulting from density variations during the phase-change process is neglected;
The thermophysical properties of the solid and liquid phases are assumed to be constant within each phase. This assumption is valid for phase-change processes involving moderate temperature differences. The effects of viscous dissipation and pressure work are neglected.

For the evolution of temperature field, the macroscopic partial differential equation of total enthalpy is employed~\cite{Huang2015}:
\begin{equation}
\label{eq:total_enthalpy0}
\frac{\partial \rho En}{\partial t} + \nabla \cdot (\rho En \mathbf{u}) = \nabla \cdot (\lambda \nabla T)
\end{equation}
in which $T$ is the temperature and $En = h + \Delta En$ is the total enthalpy, including the sensible enthalpy $h = c_p T$ and the latent enthalpy $\Delta En = f_l L_a$, with $f_l$ the liquid fraction and $L_a$ the latent heat. $\lambda$ is the thermal conductivity.  For the phase change process of the pure phase-change materials, the flux of the latent enthalpy can be usually negelected since the mushy zone acts like solid. Thus the governing equation of total enthalpy can be simplified as
\begin{equation}
\label{eq:total_enthalpy1}
\frac{\partial \rho En}{\partial t} + \nabla \cdot (\rho h \mathbf{u}) = \nabla \cdot (\lambda \nabla T)
\end{equation}

The solid-liquid interface can be characterized by the latent enthalpy or, equivalently, the liquid fraction. In the solid phase, both the liquid fraction and the latent enthalpy are zero, whereas in the liquid phase, the liquid fraction is unity and the latent enthalpy equals the latent heat. Within the mushy zone (or interface region), the liquid fraction ranges continuously between zero and unity. The temperature is subsequently evaluated from the total enthalpy, which is governed by an independent distribution function:
\begin{equation}
T =
\begin{cases}
\dfrac{En}{c_{ps}}, & En \le En_s \\[8pt]
T_s + \dfrac{En - En_s}{En_l - En_s}\left(T_l - T_s\right), & En_s < En < En_l \\[8pt]
T_l + \dfrac{En - En_l}{c_{pl}}, & En \ge En_l
\end{cases}
\end{equation}
with $En_s = c_{ps} T_s$ and $En_l = c_{pl} T_l + L_a$ the solidus and liquidus enthalpy, $T_s$ and $T_l$ are the corresponding temperatures. The liquid fraction can also be determined by the enthalpy as following:
\begin{equation}
f_l =
\begin{cases}
0, & En \le En_s \\
\dfrac{En - En_s}{En_l - En_s}, & En_s < En < En_l \\
1, & En \ge En_l
\end{cases}
\end{equation}

The thermophysical properties across the solid-liquid interface can be defined by the linearly interpolation form such as the thermal conductivity:
\begin{equation}
\lambda = (1 - f_l)\lambda_s + f_l \lambda_l
\end{equation}
and similar definition can be applied to other thermophysical parameters not mentioned.

The governing equations for the flow field are as following:
\begin{equation}
\frac{\partial \rho}{\partial t} + \nabla \cdot (\rho \mathbf{u}) = 0
\end{equation}
\begin{equation}
\frac{\partial (\rho \mathbf{u})}{\partial t} + \nabla \cdot (\rho \mathbf{u}\mathbf{u}) = -\nabla p + \nabla \cdot \left[ \mu (\nabla \mathbf{u} + \nabla \mathbf{u}^T) \right] + \mathbf{F}
\end{equation}
in which $\rho, \mathbf{u}, p, \mathbf{F}, \mu$ are the density, velocity, pressure, body force and dynamic viscosity, respectively. The body force is defined with the Boussinesq approximation as:
\begin{equation}
\mathbf{F} = - \rho_{\text{ref}} \beta \mathbf{g} (T - T_{\text{ref}})
\end{equation}
in which $\beta$ is the thermal expansion coefficient, $\mathbf{g}$ is the gravitational acceleration, and $\rho_{\text{ref}}$ and $T_{\text{ref}}$ are the reference density and temperature.

To characterize the flow and temperature fields, several dimensionless parameters are introduced, including the Prandtl number, Rayleigh number, Stefan number, and Fourier number:
\begin{align}
&\text{Pr} = \frac{\nu_l}{\kappa_l}, \quad \text{Ra} = \frac{\beta_l g (T_h - T_l) L_c^3}{\nu_l \kappa_l}, \nonumber \\
&\text{Ste} = \frac{c_{pl}(T_h - T_m)}{L_a}, \quad \text{Fo} = \frac{\kappa_l t}{L_c^2}
\end{align}
in which $\nu_l$ is the kinematic viscosity, $\kappa_l$ the thermal diffusivity, $L_c$ the characteristic length and $T_m$ the melt temperature, $T_h, T_l$ are the imposed high and low temperature on the boundaries.

\subsection{Regularized LB model to solve total enthalpy}
The total enthalpy distribution function can be defined as follows~\cite{Liu2019}:
\begin{equation}
\label{eq:geq}
g_i^{\text{eq}}(\mathbf{x}, t) =
\begin{cases}
En - (1 - w_0) c_{p,ref} T , & i = 0 \\[8pt]
w_i c_{p,ref} T +  w_i c_p T \dfrac{\mathbf{e}_i \cdot \mathbf{u}}{c_{s,T}^2} , & i \neq 0
\end{cases}
\end{equation}

The discrete velocity set, corresponding weights, and lattice sound speed for the two-dimensional D2Q5 model are given as follows:
\begin{equation}
\mathbf{e}_x = [0, 1, 0, -1, 0], \quad \mathbf{e}_y = [0, 0, 1, 0, -1]
\end{equation}
\begin{equation}
w_i = \left[\frac{1}{2}, \frac{1}{8}, \frac{1}{8}, \frac{1}{8}, \frac{1}{8}\right], \quad c_{s,T}^2 = \frac{1}{4}
\end{equation}

The discrete velocity set, corresponding weights, and lattice sound speed for the three-dimensional D3Q7 model are given as follows:
\begin{gather}
\label{eq:lattice_d3q7}
\mathbf{e}_x = [0, 1, -1, 0, 0, 0, 0], \nonumber \\
\mathbf{e}_y = [0, 0, 0, 1, -1, 0, 0], \nonumber \\
\mathbf{e}_z = [0, 0, 0, 0, 0, 1, -1]
\end{gather}
\begin{equation}
w_i = \left[\frac{1}{4}, \frac{1}{8}, \frac{1}{8}, \frac{1}{8}, \frac{1}{8}, \frac{1}{8}, \frac{1}{8}\right], \quad c_{s,T}^2 = \frac{1}{4}
\end{equation}

The zero-order, first-order, and second-order moments of the equilibrium distribution function of the total enthalpy based on the discrete velocity set and weights mentioned above can be easily obtained via some algebraic operations:
\begin{subequations}
\label{eq:eqmoments}
	\begin{align}
	\sum_{i} g_i^{\text{eq}} &= \text{En} \\
	\sum_{i} g_i^{\text{eq}} \mathbf{e}_i &= c_p T \mathbf{u} \\
	\sum_{i} g_i^{\text{eq}} \mathbf{e}_i \mathbf{e}_i &=  c_{p,\text{ref}} T c_{s,T}^2 \mathbf{I}
	\end{align}
\end{subequations}

The regularized lattice Boltzmann evolution equation can be written as follows:
\begin{eqnarray}
&&g_i(\mathbf{x} + \mathbf{e}_i \delta_t, t + \delta_t)
= g_i(\mathbf{x}, t) - \frac{1}{\tau_g}\left[g_i(\mathbf{x}, t)-g_i^{\text{eq}}(\mathbf{x} , t)\right]
\nonumber \\
&&= g_i^{\text{eq}}(\mathbf{x},t) +\left(1 - \frac{1}{\tau_g}\right) g_i^{\text{neq,reg}}(\mathbf{x}, t)
\end{eqnarray}

In the above lattice Boltzmann equation, the equilibrium distribution function in Eq. \eqref{eq:geq} is employed, and the corresponding non-equilibrium distribution function can be written as follows:
\begin{equation}
\label{eq:gneq_reg}
g_i^{\text{neq,reg}}(\mathbf{x}, t) = w_i \sum_{n=0}^1 \frac{1}{c_{s,T}^{2n}n!} \mathbf{a}^{(n)}_{\text{neq}} : \mathcal{H}^{(n)}(\mathbf{e}_i)
\end{equation}
with
\begin{equation}
\mathbf{a}^{(n)}_{\text{neq}} = \sum_{i} \left[ g_i(\mathbf{x}, t) - g_i^{\text{eq}}(\mathbf{x}, t) \right] \mathcal{H}^{(n)}(\mathbf{e}_i)
\end{equation}

The zero-order and first-order non-equilibrium Hermite coefficients can be explicitly written as:
\begin{subequations}
    \begin{align}
    \mathbf{a}_{\text{neq}}^{(0)} &= \sum_{i} \left[ g_i - g_i^{\text{eq}} \right] = 0 \\
    \mathbf{a}_{\text{neq}}^{(1)} &= \sum_{i} \left[ g_i - g_i^{\text{eq}} \right] \mathbf{e}_i
    \end{align}
\end{subequations}

Thus, Eq.\eqref{eq:gneq_reg} can be explicitly written as:
\begin{equation}
g_i^{\text{neq,reg}}(\mathbf{x}, t) = w_i \frac{1}{c_{s,T}^2} \mathbf{a}_{\text{neq}}^{(1)} : \mathbf{e}_i
\end{equation}

To eliminate the deviation term arising in the recovery of the diffusion term in the total-enthalpy partial differential equation, a source term is introduced, and the regularized lattice Boltzmann evolution equation can then be written as follows:

\begin{align}
\label{eq:reg_lbe_sourceterm}
&g_i(\mathbf{x} + \mathbf{e}_i \delta_t, t + \delta_t)   \nonumber \\
&= g_i(\mathbf{x}, t) -\frac{1}{\tau_g} \left[ g_i(\mathbf{x}, t) - g_i^{\text{eq}}(\mathbf{x}, t) \right] + \delta_t S_i \nonumber \\
&= g_i^{\text{eq}}(\mathbf{x}, t) + (1-\frac{1}{\tau_g})  g_i^{\text{neq,reg}}(\mathbf{x}, t) +  \delta_t S_i  \nonumber \\
&= g_i^{\text{eq}}(\mathbf{x}, t) + (1-\frac{1}{\tau_g} ) \frac{w_i}{c_{s,T}^2}\left[\mathbf{a}_{neq}^{(1)}:\mathbf{e}_i\right] \nonumber \\
&+ \delta_t \left(1 - \frac{1}{2\tau_g}\right) \frac{w_i}{c_{s,T}^2}  \mathbf{e}_i \cdot \frac{\partial(c_pT\mathbf{u})}{\partial t}  \\
&= g_i(\mathbf{x}, t) -\frac{1}{\tau_g} \frac{w_i}{c_{s,T}^2}\left[\mathbf{a}_{neq}^{(1)}:\mathbf{e}_i\right] \nonumber \\
&+ \delta_t \left(1 - \frac{1}{2\tau_g}\right) \frac{w_i}{c_{s,T}^2}  \mathbf{e}_i \cdot \frac{\partial(c_pT\mathbf{u})}{\partial t}
\end{align}

To facilitate the Taylor expansion, the regularized lattice Boltzmann evolution equation is recast into the form presented in the final line of the preceding equation, which does affect the following derivations. Subsequently, a Chapman--Enskog expansion is carried out to demonstrate that the regularized lattice Boltzmann equation Eq.\eqref{eq:reg_lbe_sourceterm} correctly recovers the macroscopic partial differential equation for the total enthalpy.

The zero-order and first-order moments of the source term can be easily evaluated via some algebraic operation as:
\begin{subequations}
    \begin{align}
    \sum_{i} S_i &= 0 \\
    \sum_{i} S_i \mathbf{e}_i &= \left(1 - \frac{1}{2\tau_g}\right) \frac{\partial(c_p T \mathbf{u})}{\partial t}
    \end{align}
\end{subequations}

In the numerical implementation, the temporal derivative can be discretised through the explicit Euler scheme. In our simulations, this source term has been igored directly.

In the multiscale expansion, the temporal and spatial derivatives, distribution functions, and source term are expressed in the following scale-separated forms:
\begin{eqnarray}
\label{eq:multiscale}
\frac{\partial}{\partial t} = \epsilon \frac{\partial}{\partial t_1} + \epsilon^2 \frac{\partial}{\partial t_2}, \quad \nabla = \epsilon \nabla_1, \nonumber \\
g_i = g_i^{(0)} + \epsilon g_i^{(1)} + \epsilon^2 g_i^{(2)}, \quad S_i = \epsilon S_i^{(1)}
\end{eqnarray}

Apply the Taylor expansion to Eq.\eqref{eq:reg_lbe_sourceterm}, the following equation can be obtained:
\begin{eqnarray}
\delta_t \left( \frac{\partial}{\partial t} + \mathbf{e}_i \cdot \nabla \right) g_i + \frac{\delta_t^2}{2} \left( \frac{\partial}{\partial t} + \mathbf{e}_i \cdot \nabla \right)^2 g_i
+ \mathcal{O}(\delta_t^3) \nonumber \\
= -\frac{1}{\tau_g} \frac{w_i}{c_{s,T}^2}\left[\mathbf{a}_{neq}^{(1)}:\mathbf{e}_i\right] + \delta_t S_i
\end{eqnarray}

Substituting the multiscale expansions given in Eq.\eqref{eq:multiscale} into the above equation and collecting terms of different orders yields the following equations at the corresponding scales:
\begin{subequations}
\begin{align}
\label{eq:CE_0th}
\mathcal{O}(\epsilon^0): &\quad g_i^{(0)} = g_i^{\text{eq}} \\
\label{eq:CE_1st}
\mathcal{O}(\epsilon^1): &\quad \left( \frac{\partial}{\partial t_1} + \mathbf{e}_i \cdot \nabla_1 \right) g_i^{(0)} = -\frac{1}{\delta_t \tau_g} \frac{w_i}{c_{s,T}^2}\left[\mathbf{a}_{neq}^{(1)}:\mathbf{e}_i\right] + \text{S}_i^{(1)} \\
\label{eq:CE_2rd}
\mathcal{O}(\epsilon^2): &\quad \frac{\partial g_i^{(0)}}{\partial t_2} + \left( \frac{\partial}{\partial t_1} + \mathbf{e}_i \cdot \nabla_1 \right) \left( 1 - \frac{1}{2\tau_g} \right) \frac{w_i}{c_{s,T}^2}\left[\mathbf{a}_{neq}^{(1)}:\mathbf{e}_i\right] \nonumber \\
&= -\frac{1}{\delta_t \tau_g} g_i^{(2)} - \frac{\delta t}{2}\left( \frac{\partial}{\partial t_1} + \mathbf{e}_i \cdot \nabla_1 \right) \text{S}_i^{(1)}
\end{align}
\end{subequations}

The zero-order moment of Eq.\eqref{eq:CE_1st} can be obtained attributed to Eq.\eqref{eq:eqmoments}:
\begin{equation}
\label{eq:pde_t1}
\frac{\partial En}{\partial t_1} + \nabla_1 \cdot (c_p T \mathbf{u}) = 0
\end{equation}

Similarly, the zero-order moment of Eq.\eqref{eq:CE_2rd} can be obtained attributed to Eq.\eqref{eq:eqmoments}:
\begin{equation}
\label{eq:CE_2rd_moment}
\frac{\partial En}{\partial t_2} + \nabla_1 \cdot \left[ \left(1 - \frac{1}{2\tau_g}\right) \mathbf{a}_{\text{neq}}^{(1)} \right] = - \frac{\delta t}{2} \left(1 - \frac{1}{2\tau_g}\right) \nabla_1 \cdot \left(\partial_{t_1} \left(c_pT\mathbf{u}\right)\right)
\end{equation}

According to Eq.\eqref{eq:CE_1st}, the first-order moment of the non-equilibrium distribution function in the above equation can be rewritten in the following form:

\begin{multline}
\mathbf{a}_{\text{neq}}^{(1)} = -\delta_t \tau_g \left[\partial_{t_1} \sum_i \mathbf{e}_ig_i^{eq} + \nabla_1 \cdot \sum_i \mathbf{e}_i \mathbf{e}_i g_i^{eq} - \sum_i \mathbf{e}_i \text{S}_i^{(1)}\right] \\
= -\delta_t \tau_g \left[ \frac{\partial}{\partial t_1} (c_p T \mathbf{u}) + \nabla_1 \cdot \left(c_{p,\text{ref}} T c_{s,T}^2 \mathbf{I}\right) \right. \\
\left. \quad - \left(1 - \frac{1}{2\tau_g}\right) \frac{\partial \left(c_p T \mathbf{u}\right)}{\partial_{t_1}} \right] \\
= -\delta_t \tau_g  \nabla_1 \cdot \left( c_{p,\text{ref}} T c_{s,T}^2 \mathbf{I} \right)
- \frac{\delta t}{2} \frac{\partial \left(c_p T \mathbf{u}\right)}{\partial_{t_1}}
\end{multline}

Substituting the above equation into Eq.~\eqref{eq:CE_2rd_moment}, the compact form of the equation at the second-order scale can be expressed as follows through straightforward algebraic manipulations:
\begin{equation}
\label{eq:pde_t2}
\frac{\partial En}{\partial t_2} = \nabla_1 \cdot \left[ \delta_t \left(\tau_g - \frac{1}{2}\right) c_{p,\text{ref}} c_{s,T}^2 \nabla_1 T \right]
\end{equation}

By combing Eq.\eqref{eq:pde_t1} and Eq.\eqref{eq:pde_t2} at different scales, the desired macroscopic partial differential equation governing total enthalpy can be recovered without introducing any deviation term
\begin{equation}
\frac{\partial En}{\partial t} + \nabla \cdot (c_p T \mathbf{u}) = \nabla \cdot \left( \lambda \nabla T \right)
\end{equation}
with
\begin{equation}
\frac{\lambda}{\rho c_{p,\text{ref}}} =  \delta t c_{s,T}^2 \left(\tau_g - \frac{1}{2} \right).
\end{equation}

\subsection{Regularized LB model to solve the flow field}

For weakly compressible flows, the multiple-relaxation-time regularized lattice Boltzmann model (MRT-RLB) with the standard lattice $\text{D3Q27}$, based on our previous high-order lattice regularized collision model with force term~\citep{Li2024}, is employed here to guarantee the robustness and accuracy of flow-field simulations. Unlike fully compressible thermal flows, the standard lattice version is designed to deal with isothermal flows, where the Boussinesq approximation is adopted to incorporate thermal effects with a body force term. Consequently, conducting the collision process in the Hermite central moment space is sufficient.

The standard lattice $\text{D3Q27}$ in the present work is written as

\begin{align}
\label{eq:lattice_d3q27}
\mathbf{e}_x &= [0, 1,-1, 0, 0, 0, 0, 1, -1, 1, -1, 1, -1, \nonumber \\
&\quad\; 1, -1, 0, 0, 0, 0, 1, -1, 1, -1, 1, -1, 1, -1], \nonumber \\
\mathbf{e}_y &= [0, 0, 0, 1, -1, 0, 0, 1, 1, -1, -1, 0, 0, \nonumber \\
&\quad\; 0, 0, 1, -1, 1, -1, 1, 1, -1, -1, 1, 1, -1, -1], \nonumber \\
\mathbf{e}_z &= [0, 0, 0, 0, 0, 1, -1, 0, 0, 0, 0, 1, 1, \nonumber \\
&\quad\; -1, -1, 1, 1, -1, -1, 1, 1, 1, 1, -1, -1, -1, -1].
\end{align}

The multiple-relaxation-time recursive collision operators from~\citep{Li2024} with $\theta = 1$ can be written as follows:

\begin{subequations}
	\label{eq:RLB-MRT}
	\begin{align}
	\ba{1}_{\Omega} &= \frac{\bm{F}}{2} \\
	\ba{2}_{\Omega} &= \left(1 - \frac{1}{\tau_2}\right)\left(\ba{2}_1 - 2\bu\ba{1}_1 \right) + \bm{F}\bu  \\
	&= \ba{2}_1
	-\left(\frac{1}{\tau_{21}}\ba{2,1}_1
	+ \frac{1}{\tau_{22}}\ba{2,2}_1\right) \nonumber \\
	&\quad + \ba{2}_{\bm{F}}
	-\left(\frac{1}{2\tau_{21}}\ba{2,1}_{\bm{F}}
	+ \frac{1}{2\tau_{22}}\ba{2,2}_{\bm{F}}\right) \\
	\ba{3}_{\Omega}
	&= \left(1 - \frac{1}{\tau_3}\right)\left(\bm{a}^{(3)}_1
	- 3\bm{u} \bm{a}^{(2)}_1
	+ 3\bm{u}\bm{u}\bm{a}^{(1)}_1 \right) \nonumber \\
	&\quad + 3\bu \ba{2}_{\Omega}
	- 3\bu\bu \ba{1}_{\Omega} \\
	&= \left(1 - \frac{1}{\tau_3}\right) \ba{3}_1
	+ \left( \frac{1}{\tau_3} 3 \bu\ba{2}_1
	-\frac{1}{\tau_{21}} 3 \bu\ba{2,1}_1 \right. \nonumber \\
	&\left. \quad -\frac{1}{\tau_{22}} 3 \bu\ba{2,2}_1 \right)
	+ \left(1 - \frac{1}{2\tau_3}\right) \ba{3}_{\bm{F}} \nonumber \\
	&\quad + \left( \frac{1}{2\tau_3} 3 \bu\ba{2}_{\bm{F}}
	-\frac{1}{2\tau_{21}} 3 \bu\ba{2,1}_{\bm{F}}
	-\frac{1}{2\tau_{22}} 3 \bu\ba{2,2}_{\bm{F}} \right) \\
	\ba{4}_{\Omega}
	&= \left(1 - \frac{1}{\tau_4}\right) \left( \bm{a}^{(4)}_1
	- 4\bm{u} \bm{a}^{(3)}_1
	+ 6\bm{u}\bm{u}\bm{a}^{(2)}_1
	- 4 \bm{u}\bm{u}\bm{u} \bm{a}^{(1)}_1 \right) \nonumber \\
	&\quad + 4\bu \ba{3}_{\Omega}
	- 6\bu\bu \ba{2}_{\Omega}
	+ 4 \bu\bu\bu \ba{1}_{\Omega} \\
	&= \left(1 - \frac{1}{\tau_4}\right) \ba{4}_1
	+ \left(\frac{1}{\tau_4} - \frac{1}{\tau_3}\right)4\bu \ba{3}_1 \nonumber \\
	&\quad -\left(\frac{1}{\tau_4} - \frac{2}{\tau_3}\right)6\bu\bu\ba{2}_1
	- 6\bu\bu\left(
	\frac{1}{\tau_{21}} \ba{2,1}_1
	+ \frac{1}{\tau_{22}}\ba{2,2}_1
	\right) \nonumber\\
	&\quad + \left(1 - \frac{1}{2\tau_4}\right) \ba{4}_{\bm{F}}
	+ \left(\frac{1}{2\tau_4} - \frac{1}{2\tau_3}\right)4\bu \ba{3}_{\bm{F}} \nonumber\\
	&\quad -\left(\frac{1}{2\tau_4} - \frac{2}{2\tau_3}\right)6\bu\bu\ba{2}_{\bm{F}} \nonumber\\
	&\quad - 6\bu\bu\left(
	\frac{1}{2\tau_{21}} \ba{2,1}_{\bm{F}}
	+ \frac{1}{2\tau_{22}}\ba{2,2}_{\bm{F}}
	\right)
	\end{align}
\end{subequations}
in which $\ba{2,1}_1$ and $\ba{2,2}_1$ represent the traceless and trace components of the second-order nonequilibrium moment, respectively, whereas $\ba{2,1}_{\bm{F}}$ and $\ba{2,2}_{\bm{F}}$ denote the corresponding traceless and trace components of the second-order force term, and $\bm{F}$ is the body force with the dimension of density multiplied by acceleration.
Due to the insufficient velocity-space resolution of standard lattices (such as D2Q9 and D3Q27), the present MRT-RLB model adopts components identical to those in the recursive regularized collision model~\cite{Malaspinas2015}.

In the MRT-RLB collision model, the raw-moment Hermite coefficients of the force term are as follows \cite{Li2022, Li2024}
\begin{subequations}
	\begin{align}
	\label{eq:af}
	\ba{0}_{\bm{F}} &= 0, \quad \ba{1}_{\bm{F}} = \bm{F}, \quad \ba{2}_{\bm{F}} = 2 \bm{F}\bu, \nonumber \\
	&\quad \ba{3}_{\bm{F}} = 3\bm{F}\bu\bu, \quad \ba{4}_{\bm{F}} = 4\bm{F}\bu\bu\bu
	\end{align}
\end{subequations}
in which the symmetric tensor $2\bm{F}\bu$ denotes $ F_{\alpha}u_{\beta} + F_{\beta}u_{\alpha}$ and others are similar.

The raw-moment Hermite coefficients of the nonequilibrium distribution function are
\begin{equation}
\label{eq:nonequilibrium-moments}
\ba{0}_1 = 0, \ba{1}_1 = -\frac{\bm{F}}{2}, \ba{n}_1 = \sum_i\left[f_i - f_i^{(0)}\right]  \bH{n}_{i}, n \geq 2.
\end{equation}

The equilibrium distribution function can be calcualted by the sixth-order recursive form in \cite{Malaspinas2015}
\begin{equation}
\begin{aligned}
f_i^{(0)}
&=
w_i \rho
\Bigg[
1
+
\frac{\bm{\xi}_i\cdot\bm{u}}{c_s^2}
+
\frac{1}{2c_s^4}
\bH{2}_i:\bm{u}\bm{u}
\\
&
+
\frac{1}{2c_s^6}
\Big(
\bH{3}_{i,xxy}u_x^2u_y
+
\bH{3}_{i,xxz}u_x^2u_z
+
\bH{3}_{i,xyy}u_xu_y^2
\\
&+
\bH{3}_{i,xzz}u_xu_z^2
+
\bH{3}_{i,yzz}u_yu_z^2
+
\bH{3}_{i,yyz}u_y^2u_z
+
2\bH{3}_{i,xyz}u_xu_yu_z
\Big)
\\
&
+
\frac{1}{4c_s^8}
\Big(
\bH{4}_{i,xxyy}u_x^2u_y^2
+
\bH{4}_{i,xxzz}u_x^2u_z^2
+
\bH{4}_{i,yyzz}u_y^2u_z^2
\\
&
+
2\big(
\bH{4}_{i,xyzz}u_xu_yu_z^2
+
\bH{4}_{i,xyyz}u_xu_y^2u_z
+
\bH{4}_{i,xxyz}u_x^2u_yu_z
\big)
\Big)
\\
&
+
\frac{1}{4c_s^{10}}
\Big(
\bH{5}_{i,xxyzz}u_x^2u_yu_z^2
+
\bH{5}_{i,xxyyz}u_x^2u_y^2u_z
+
\bH{5}_{i,xyyzz}u_xu_y^2u_z^2
\Big)
\\
&
+
\frac{1}{8c_s^{12}}
\bH{6}_{i,xxyyzz}u_x^2u_y^2u_z^2
\Bigg].
\end{aligned}
\label{eq:equilibrium_pdf}
\end{equation}

The non-equilibrium distribution function with relaxations can be reconstructed with the collision coefficients as
\begin{equation}
\begin{aligned}
f_i^{(1)}
&=
w_i
\Bigg[
\frac{1}{c_s^2}\bxi_{i} \cdot \bm{a}^{(1)}_{{\Omega}}
+ \frac{1}{2c_s^4} \bH{2}_{i}:\bm{a}^{(2)}_{{\Omega}}
\\
&\quad
+
\frac{1}{2c_s^6}
\Big(
\bH{3}_{i,xxy}a^{(3)}_{{\Omega},xxy}
+
\bH{3}_{i,xxz}a^{(3)}_{{\Omega},xxz}
\\
&\qquad\qquad
+
\bH{3}_{i,xyy}a^{(3)}_{{\Omega},xyy}
+
\bH{3}_{i,xzz}a^{(3)}_{{\Omega},xzz}
\\
&\qquad\qquad
+
\bH{3}_{i,yzz}a^{(3)}_{{\Omega},yzz}
+
\bH{3}_{i,yyz}a^{(3)}_{{\Omega},yyz}
+
2\bH{3}_{i,xyz}a^{(3)}_{{\Omega},xyz}
\Big)
\\
&\quad
+
\frac{1}{4c_s^8}
\Big(
\bH{4}_{i,xxyy}a^{(4)}_{{\Omega},xxyy}
+
\bH{4}_{i,xxzz}a^{(4)}_{{\Omega},xxzz}
\\
&\qquad\qquad
+
\bH{4}_{i,yyzz}a^{(4)}_{{\Omega},yyzz}
+
2\Big(
\bH{4}_{i,xyzz}a^{(4)}_{{\Omega},xyzz}
\\
&\qquad\qquad\quad
+
\bH{4}_{i,xyyz}a^{(4)}_{{\Omega},xyyz}
+
\bH{4}_{i,xxyz}a^{(4)}_{{\Omega},xxyz}
\Big)
\Big)
\Bigg]
\end{aligned}
\label{eq:nonequilibrium_pdf}
\end{equation}
in which orders higher than the fourth order are regularized.

The post-collision distribution functions of the present MRT-RLB model can be written as follows
\begin{equation}
f^p_i(\bm{x}, t)= f_i^{(0)}(\bm{x}, t) + f_i^{(1)}(\bm{x}, t),
\label{eq:RLB-force}
\end{equation}
which can be streamed to the neighboring nodes at next time step.

In terms of the boundary conditions in the numerical simulations, Guo's extrapolation bounadry condition \cite{Guo2002} is applied to deal with the no-slip boundary for flow field and imposed temperature on the wall for the temperature field. Other boundary conditions will be described in the corresponding benchmarks. In dealing with the heat flux for the evaluation of Nusselt number, the one-side finite difference method with boundary node and two neighboring nodes is adopted. In terms of the no-slip boundary condition on the solid-liquid phase interface, the volumetric boundary condition in \cite{Huang2016} is applied in the present work. Take the effect of the solid phase into consideration, the density distribution function is redefined as
\begin{equation}
f_i^+ = f_l f_i + (1-f_l) f_i^{eq}\left(\rho,\bm{u}_s\right),
\end{equation}
in which the superscript $+$ denotes the updated distribution function on the interface, and $\bm{u}_s = 0$ is the velocity of the solid phase. The density can be evaluated as
\begin{equation}
\rho = \sum_{i=0} f_i.
\end{equation}

Then the velocity can be updated as
\begin{equation}
 \rho \bm{u} = \sum_{i=0} f_i^+ \bm{e}_i + \frac{\delta t}{2} \rho \bm{F}.
\end{equation}

\subsection{GPU implementation details}
All simulations in this work were conducted on NVIDIA A800 GPU cards. A Structure-of-Arrays (SoA) layout is adopted for the particle distribution functions, ensuring that data corresponding to a given lattice direction are stored contiguously in global memory. Specifically, the global memory index $\mathcal{I}$ of a distribution function is computed as:
\begin{equation}
\mathcal{I} = b \cdot \text{size\_domain} + k \cdot (\text{xDim} \cdot \text{yDim}) + j \cdot \text{xDim} + i,
\end{equation}
where $b$ denotes the discrete lattice direction, and $i, j, k$ represent the spatial indices of the fluid nodes.

A crucial aspect of GPU implementation is the mapping between the computational lattice and GPU threads. In this study, a one-dimensional (1D) linear thread-mapping strategy is adopted for both two-dimensional (2D) and three-dimensional (3D) simulations.

The global linear thread index $\text{threadID}$ is computed via CUDA built-in variables as
\begin{equation}
\text{threadID} = \mathrm{blockDim.x} \cdot \mathrm{blockIdx.x} + \mathrm{threadIdx.x}.
\end{equation}

From $\text{threadID}$, the spatial indices of the lattice nodes are reconstructed as follows:

\noindent \textbf{2D case:}
\begin{align}
j &= \left\lfloor \frac{\text{threadID}}{\text{xDim}} \right\rfloor, \\
i &= \text{threadID} \bmod \text{xDim},
\end{align}
where $\text{xDim}$ is the lattice resolution in the $x$-direction, $\lfloor \cdot \rfloor$ denotes the floor function (integer division), and $\bmod$ represents the modulo operation.

\noindent \textbf{3D case:}
\begin{align}
k &= \left\lfloor \frac{\text{threadID}}{\text{xDim} \cdot \text{yDim}} \right\rfloor, \\
j &= \left\lfloor \frac{\text{threadID} - k \cdot \text{xDim} \cdot \text{yDim}}{\text{xDim}} \right\rfloor, \\
i &= \text{threadID} \bmod \text{xDim},
\end{align}
where $\text{yDim}$ denotes the lattice resolution in the $y$-direction.

The kernel execution configuration, namely, the number of threads per block ($B$) and the number of thread blocks per grid ($G$)—is specified as:
\begin{align}
B &= 128, \\
G &= \left\lceil \frac{\text{size\_domain}}{B} \right\rceil = \left\lfloor \frac{\text{size\_domain} + B - 1}{B} \right\rfloor,
\end{align}
where $\text{size\_domain}$ is the total number of fluid nodes, and $\lceil \cdot \rceil$ denotes the ceiling function. Similar mapping strategies and execution configurations are also employed for boundary condition treatments.

In the present work, the collision and streaming steps are fused into a unified evolution kernel to minimize global memory traffic, as lattice Boltzmann implementations on GPUs are typically memory bandwidth bound. Because the pull scheme (uncoalesced reads and coalesced writes) generally outperforms the push scheme (coalesced reads and uncoalesced writes), it is adopted to gather the particle distribution functions from neighboring lattice nodes during evolution. Between consecutive time steps, the distribution functions for the flow field $f,f^{temp}$ and temperature field $g,g^{temp}$ are updated efficiently via pointer swapping.

Another key consideration for the regularized lattice Boltzmann model is that the discrete velocities and Hermite basis should be directly pre-substituted and explicitly expanded for each lattice direction for the equilibrium distribution, the forcing term, the non-equilibrium Hermite coefficients, and the regularized non-equilibrium distribution. This per-direction analytical expansion substantially minimizes memory lookups and eliminates redundant floating-point operations associated with zero-valued components.

A single-GPU implementation provides sufficient performance for 2D problems, whereas the computationally demanding 3D solid-liquid phase-change simulations require multi-GPU implementation. Algorithm~\ref{alg:mpi_cuda_lbm} outlines the workflow of the present regularized total-enthalpy-based solver, which employs a 1D domain decomposition along the
$z$-axis.

In the multi-GPU implementation, to hide communication overhead, the domain is partitioned into inner($k=2\to zDim-1$), halo($k=1,zDim$) regions and ghost layer($k=0,zDim+1$). The inner and halo regions are coordinated across the compute and communication streams via two fine-grained CUDA event synchronizations:

\begin{lstlisting}[style=cuda]
// 1. Compute -> Comm: Ensure BC updates complete before packing
cudaEventRecord(hc->evReady, hc->compute);
cudaStreamWaitEvent(hc->comm, hc->evReady, 0);
\end{lstlisting}

First, boundary condition (BC) kernels update the boundary
(\(k = 1 \to zDim\)) in the compute stream, which are subsequently read by the packing kernel in the communication stream. To prevent a Read-After-Write (RAW) hazard, \texttt{evReady} ensures packing starts only after the BC updates finish. Crucially, this dependency blocks only the communication stream, allowing the compute-intensive inner domain (\(k = 2, \dots, zDim - 1\)) to execute concurrently with packing and MPI transfers.

\begin{lstlisting}[style=cuda]
// 2. Comm -> Compute: Ensure unpacking completes before next collision
cudaEventRecord(hc->evHalo, hc->comm);
cudaStreamWaitEvent(hc->compute, hc->evHalo, 0);
\end{lstlisting}

Conversely, once the halo exchange pipeline ($\text{pack} \to \text{MPI transfer} \to \text{unpack}$) completes, the unpacking kernel writes the received data into the ghost layers (\(k = 0, n_z + 1\)). A reverse dependency enforced by \texttt{evHalo} ensures that the subsequent evolution step of halo region waits until the ghost layers are fully populated, thereby avoiding reading stale distribution data.

For the purpose of convinience, conventionally the computational speed is denoted by the million nodes updated per second (MNUPS):
\begin{equation}
\text{MNUPS} = \frac{\text{xDim} \cdot \text{yDim} \cdot \text{zDim}}{\text{clocktime} \times 1000000} \cdot \text{timestep}
\end{equation}

The effecitve bandwidth (BW) is determined by the bytes of reading and writting per second as
\begin{equation}
\text{Effecitve BW} = \frac{\text{MNUPS} \cdot \text{Bytes-per-node}}{\text{1000}}
\end{equation}
with
\begin{equation}
\text{Bytes-per-node} = \left(2Q + 5 + 4 + 2QT + 3 + 3\right)\times 8.
\end{equation}
in which $2Q + 5 + 4$ denotes the number of double-precision varaibles reading of $f(Q),\allowbreak usx,\allowbreak usy,\allowbreak usz,\allowbreak T,\allowbreak FL$ and the number of double-precision varaibles writting of $f^{temp}(Q), rho, ux, uy, uz$ for flow field, while $2QT + 3 + 3$ denotes the number of double-precision varaibles reading of $g(QT), ux, uy, uz$ and the number of double-precision varaibles writting of $g^{temp}(QT), T, H, FL$ for teperature field.

\begin{algorithm}[t]
	\caption{Total enthalpy RLB model with Multi-GPU implementation(1D domain partitioning)}
	\label{alg:mpi_cuda_lbm}
	\begin{algorithmic}[1]
		\State \textbf{Allocate} memory on Host and Device
		\State \textbf{Initialize} macroscopic variables and distributions
		\State \textbf{Transfer} data: Host $\to$ Device (\texttt{cudaMemcpy})
		\State \Call{HaloInit}{$hc, \mathrm{rank}, \mathrm{nprocs}, \dots$}

		\For{$t = 1$ \textbf{to} $T_{\max}$}
		\Statex \hspace{1em}\textit{\textbf{--- Stage 1: Flow Field Evolution ($f$) ---}}
		\State Launch boundary kernels along $x$- and $y$-directions
		\State Launch $z$-start boundary kernel
		\State Launch $z$-end boundary kernel

		\State \Call{HaloFlowStart}{$hc, f$} \Comment{Pack \& \texttt{MPI\_Isend/Irecv}}
		\State Launch \Call{CollisionStreaming-Inner}{$f$} \Comment{Overlap}
		\State \Call{HaloFlowFinish}{$hc, f$} \Comment{\texttt{MPI\_Waitall} \& Unpack}
		\State Launch \Call{CollisionStreaming-Halo}{$f$}
		\State Swap distribution pointers: $f \longleftrightarrow f^{\mathrm{temp}}$

		\Statex \hspace{1em}\textit{\textbf{--- Stage 2: Temperature Field Evolution ($g$) ---}}
		\State Launch thermal boundary kernels along $x$- and $y$-directions
		\State Launch thermal $z$-start boundary kernel
		\State Launch thermal $z$-end boundary kernel
		\State \Call{HaloTempStart}{$hc, g$} \Comment{Pack \& \texttt{MPI\_Isend/Irecv}}
		\State Launch \Call{ThermalCollision-Inner}{$g$} \Comment{Overlap}
		\State \Call{HaloTempFinish}{$hc, g$} \Comment{\texttt{MPI\_Waitall} \& Unpack}
		\State Launch \Call{ThermalCollision-Halo}{$g$}
		\State Swap temperature pointers: $g \longleftrightarrow g^{\mathrm{temp}}$
		\EndFor
	\end{algorithmic}
\end{algorithm}

In the present work, the parallel performance of the present GPU code will be presented only for the three-dimensional benchmarks.

\section{Numerical Results}
In this section, a one-dimensional, two two-dimensional and one three-dimensional benchmark cases are considered to validate the present regularzied lattice Boltzmann model for the flow field and temperature field.
\subsection{Two-phase conduction melting in a semi-infinite space}

\begin{figure}[htbp]
	\centering
	\includegraphics[width=0.62\textwidth]{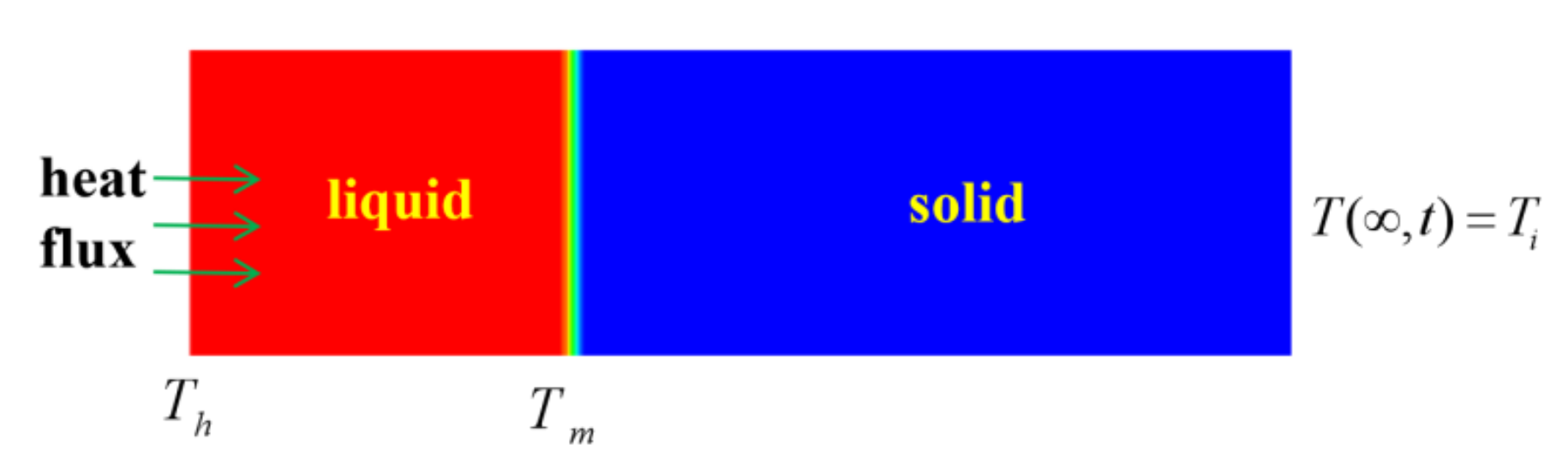}
	\caption{Schematic diagram of two-phase conduction melting in a semi-infinite space.}
	\label{fig:two_phase_melting_schematic}
\end{figure}

In this subsection, the two-phase conduction melting problem in a semi-infinite
domain is simulated using both the RLB model and the LBGK
model, as illustrated in Fig.~\ref{fig:two_phase_melting_schematic}. Initially, the
phase-change material is set to the temperature $T_i$, which is below the melting
temperature $T_m$ ($T_l=T_s=T_m$). The temperature at the left boundary is fixed
at a high temperature $T_h$ to introduce energy into the solid material, while the
right boundary is sufficiently far from the left boundary and remains at the initial
temperature.

For this simple case, an analytical solution is available, as given below ~\cite{Hu1996}:
\begin{equation}
T(x,t)=
\begin{cases}
T_h-\dfrac{(T_h-T_m)}{\erf(\eta)}
\erf\left(\dfrac{x}{2\sqrt{\kappa_l t}}\right),
 0\leq x\leq X(t), \text{liquid}, \\[12pt]
T_i+\dfrac{(T_m-T_i)}
{\erfc\left(\eta\sqrt{\dfrac{\kappa_l}{\kappa_s}}\right)}\erfc\left(\dfrac{x}{2\sqrt{\kappa_s t}}\right),
 x>X(t), \text{solid}.
\end{cases}
\label{eq:temp_ana_solution}
\end{equation}

Here, the error function and its complementary form are defined as
\begin{equation}
\erf(\eta)
=
\frac{2}{\sqrt{\pi}}
\int_{0}^{\eta}\exp\left(-\vartheta^2\right)\,\mathrm{d}\vartheta,
\label{eq:erf}
\end{equation}
and
\begin{equation}
\erfc(\eta)=1-\erf(\eta).
\label{eq:erfc}
\end{equation}

The analytical interface position is given by
\begin{equation}
X(t)=2\eta\sqrt{\kappa_l t}.
\label{eq:interface_position}
\end{equation}

The parameter $\eta$ can be obtained by solving the following transcendental
equation:
\begin{equation}
\frac{\mathrm{Ste}_l}{\sqrt{\pi}}
\frac{\exp(-\eta^2)}{\erf(\eta)}
-
\frac{\mathrm{Ste}_s}{\sqrt{\pi}}
\sqrt{\frac{\kappa_s}{\kappa_l}}\,
\frac{\exp\left(-\eta^2\dfrac{\kappa_l}{\kappa_s}\right)}
{\erfc\left(\eta\sqrt{\dfrac{\kappa_l}{\kappa_s}}\right)}
=
\eta.
\label{eq:eta_transcendental}
\end{equation}

In Eq.~\eqref{eq:eta_transcendental},
\begin{equation}
\mathrm{Ste}_l
=
\frac{c_{pl}(T_h-T_m)}{L_a},
\qquad
\mathrm{Ste}_s
=
\frac{c_{ps}(T_m-T_i)}{L_a},
\label{eq:stefan_numbers}
\end{equation}
are the Stefan numbers of the liquid and solid phases, respectively. The thermal
diffusivities of the liquid and solid phases are defined as
\begin{equation}
\kappa_l=\frac{\lambda_l}{\rho c_{pl}},
\qquad
\kappa_s=\frac{\lambda_s}{\rho c_{ps}}.
\label{eq:thermal_diffusivities}
\end{equation}

In the simulations, the temperature-related parameters are specified as
\begin{equation}
T_h=1,
\qquad
T_m=0,
\qquad
T_i=-1.
\label{eq:temperature_parameters}
\end{equation}

The thermophysical parameters are set as
\begin{equation}
c_{pl}=c_{ps}=1,
\quad
\kappa_s=0.02,
\quad
\mathrm{Ste}_l=1.0,
\label{eq:thermophysical_parameters}
\end{equation}
and the thermal diffusivity in the interfacial region is evaluated by
\begin{equation}
\kappa=f_l\kappa_l+(1-f_l)\kappa_s,
\label{eq:interpolated_diffusivity}
\end{equation}
where $f_l$ denotes the liquid fraction.

The relaxation times related to temperature and flow field can be set as
\begin{equation}
\tau_g
=
\frac{1}{\omega_g}
=
\frac{\kappa}{c_{s,T}^2\Delta t}
+\frac{1}{2},
\qquad
\tau_{\nu}=1.0.
\label{eq:relaxation_time}
\end{equation}

In the simulations, different ratios of thermal diffusivity between the liquid and solid phases are investigated:
\begin{equation}
\frac{\kappa_l}{\kappa_s}=1,\ 2,\ 3,\ 5.
\label{eq:diffusivity_ratios}
\end{equation}

Furthermore, different relaxation times for the temperature evolution are considered
for both the RLB collision model and the LBGK model:
\begin{equation}
\tau_{g,l}=0.58,\ 1.0,\ 2.5,\ 4.5,\ 10.5.
\label{eq:relaxation_rates}
\end{equation}

\begin{figure}[htbp]
	\centering
	\includegraphics[width=0.62\textwidth]{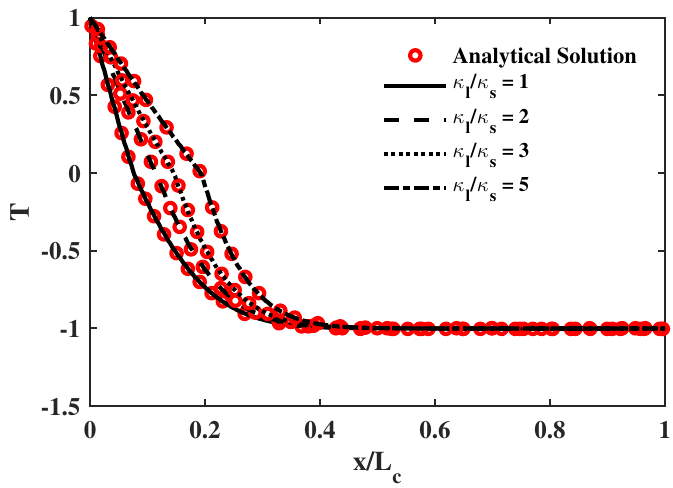}
	\caption{Temperature profiles across the domain for different thermal diffusivity
		ratios($\kappa_l/\kappa_s=1,2,3,5$) with $\kappa_s=0.02$ at $\mathrm{Fo}=0.01$.}
	\label{fig:temperature_profiles}
\end{figure}

The computational domain consists of $448\times 6$ lattice nodes. Periodic boundary
conditions are applied at the upper and lower boundaries. Figure~\ref{fig:temperature_profiles}
compares the temperature profiles predicted by the present RLB model with the
analytical solutions at
\begin{equation}
\mathrm{Fo}=\frac{\kappa_s t}{L_c^2}=0.01,
\qquad
L_c=448,
\label{eq:fourier_number}
\end{equation}
where $L_c$ is measured in lattice units.
It can be found that excellent agreements are achieved between the results of present model and the analytical solutions.

\begin{figure}[htbp]
	\centering
	\begin{subfigure}[b]{0.48\textwidth}
		\centering
		\includegraphics[width=\textwidth]{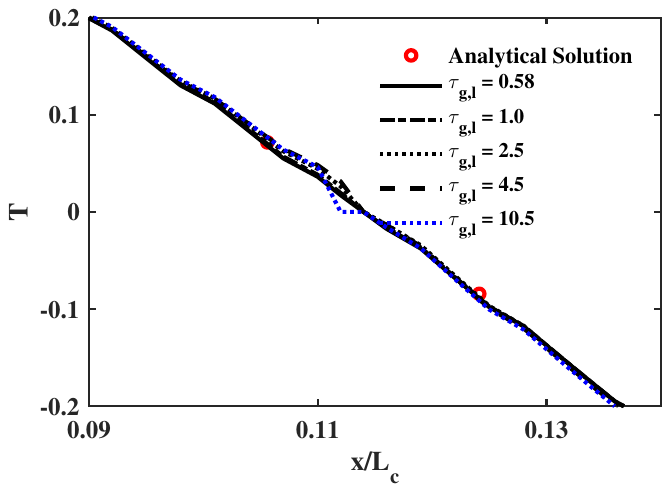}
		\caption{RLB model}
		\label{fig:3-panel-1}
	\end{subfigure}\hfill
\begin{subfigure}[b]{0.48\textwidth}
		\centering
		\includegraphics[width=\textwidth]{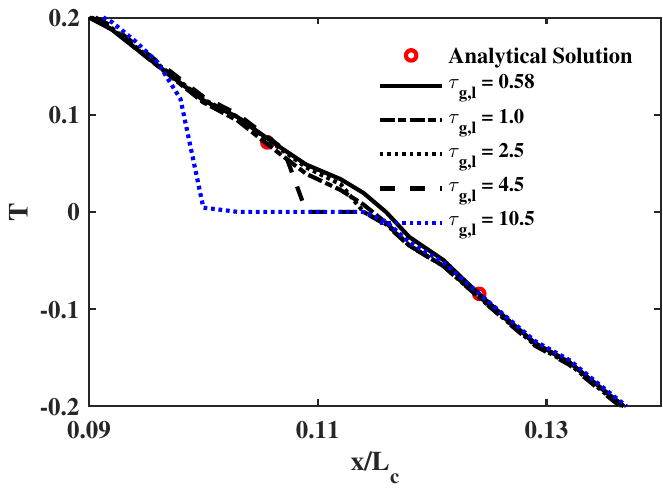}
		\caption{LBGK model}
		\label{fig:3-panel-2}
	\end{subfigure}
	\caption{Comparisons of the temperature profiles between the present RLB
		model (a) and the LBGK model (b) for $\kappa_l/\kappa_s=2$ using different
		relaxation times($\tau_{g,l}=0.58,1.0,2.5,4.5,10.5$) at $\mathrm{Fo}=0.01$.}
	\label{fig:temperature_comparison_alpha2}
\end{figure}

\begin{figure}[htbp]
	\centering
	\begin{subfigure}[b]{0.48\textwidth}
		\centering
	    \includegraphics[width=\textwidth]{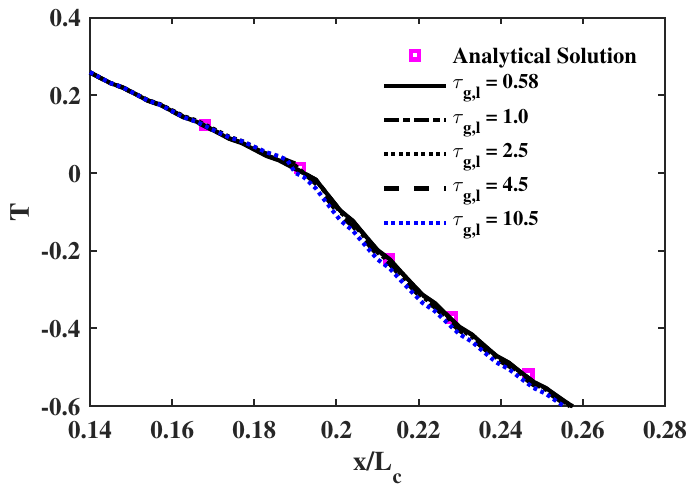}
	    \caption{RLB model}
	    \label{fig:4-panel-1}
    \end{subfigure}\hfill
\begin{subfigure}[b]{0.48\textwidth}
	    \centering
	    \includegraphics[width=\textwidth]{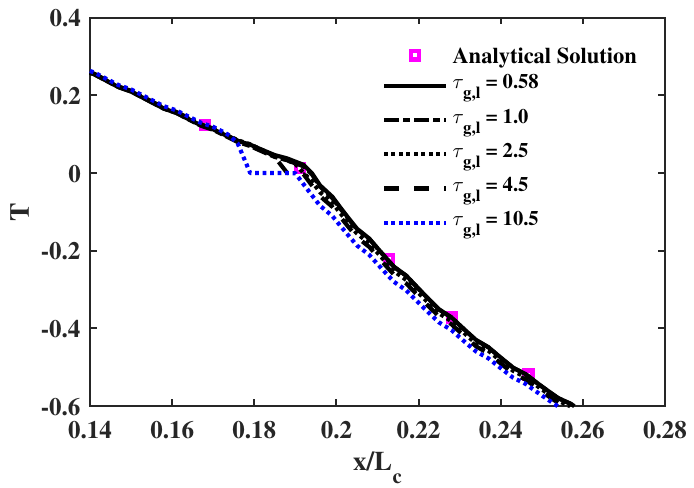}
	    \caption{LBGK model}
	    \label{fig:4-panel-2}
	\end{subfigure}
	\caption{Comparisons of the temperature profiles between the present RLB model
		and the LBGK model for $\kappa_l/\kappa_s=5$ using different
		relaxation times($\tau_{g,l}=0.58,1.0,2.5,4.5,10.5$) at $\mathrm{Fo}=0.01$.}
	\label{fig:temperature_comparison_alpha5}
\end{figure}

Figures~\ref{fig:temperature_comparison_alpha2} and \ref{fig:temperature_comparison_alpha5} show comparisons of the temperature profiles predicted by the present RLB model and the LBGK model using different thermal relaxation times for the thermal diffusivity ratios
$\kappa_l/\kappa_s=2$ and $\kappa_l/\kappa_s=5$, respectively. The temperature
profiles predicted by the present RLB model agree well with the analytical solution,
whereas those obtained by the LBGK model exhibit noticeable deviations near the
phase interface.

\begin{figure}[htbp]
	\centering
	\begin{subfigure}[b]{0.48\textwidth}
		\centering
	    \includegraphics[width=\textwidth]{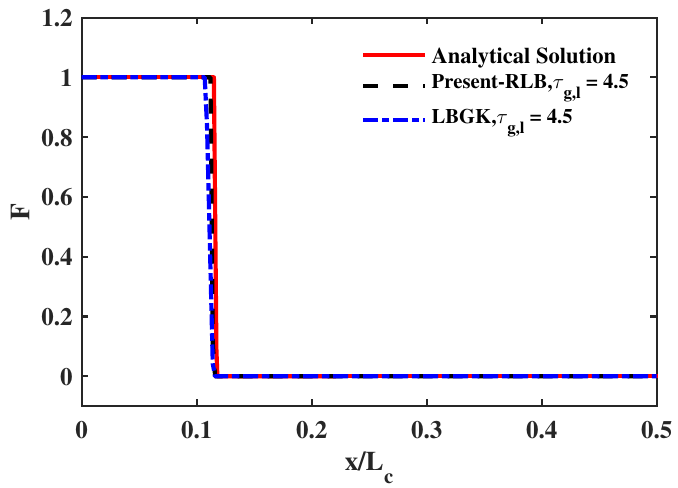}
	    \caption{}
	    \label{fig:5-panel-1}
    \end{subfigure}\hfill
\begin{subfigure}[b]{0.48\textwidth}
	    \centering
	    \includegraphics[width=\textwidth]{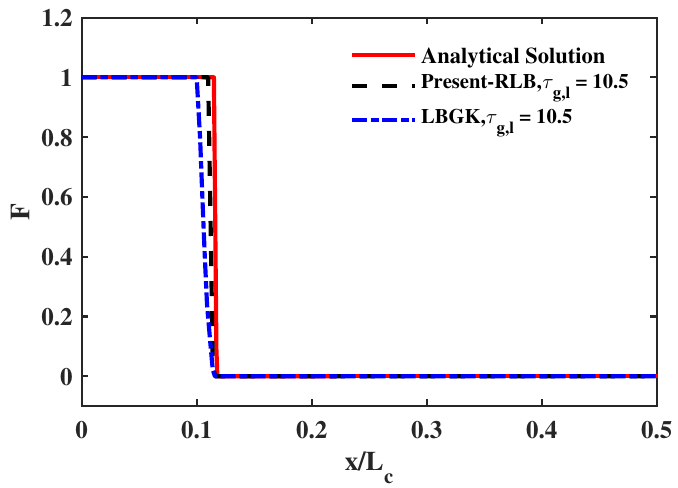}
	    \caption{}
	    \label{fig:5-panel-2}
	\end{subfigure}
	\caption{Comparisons of the liquid fraction between the present RLB model and
		the LBGK model for $\tau_{g,l} = 4.5$  and $\tau_{g,l} = 10.5$ , with
		$\kappa_l/\kappa_s=2$ at $\mathrm{Fo}=0.01$.}
	\label{fig:liquid_fraction}
\end{figure}

In terms of the interface position, Fig.~\ref{fig:liquid_fraction} shows that the
interface thickness predicted by the LBGK model is significantly larger than one
lattice cell, whereas those obtained by the present RLB model remains confined to one lattice cell. This demonstrates that, even with a single relaxation
rate, the desinged RLB model for D2Q5 velocity set for the the temperature evolution can effectively suppress numerical diffusion at the solid--liquid interface, which has the same anti-diffusion effect of the MRT~\citep{Huang2015} and TRT~\citep{Lu2019} models.

\subsection{Solidification in a semi-infinite corner}

\begin{figure}[htbp]
	\centering
	\includegraphics[width=0.62\textwidth]{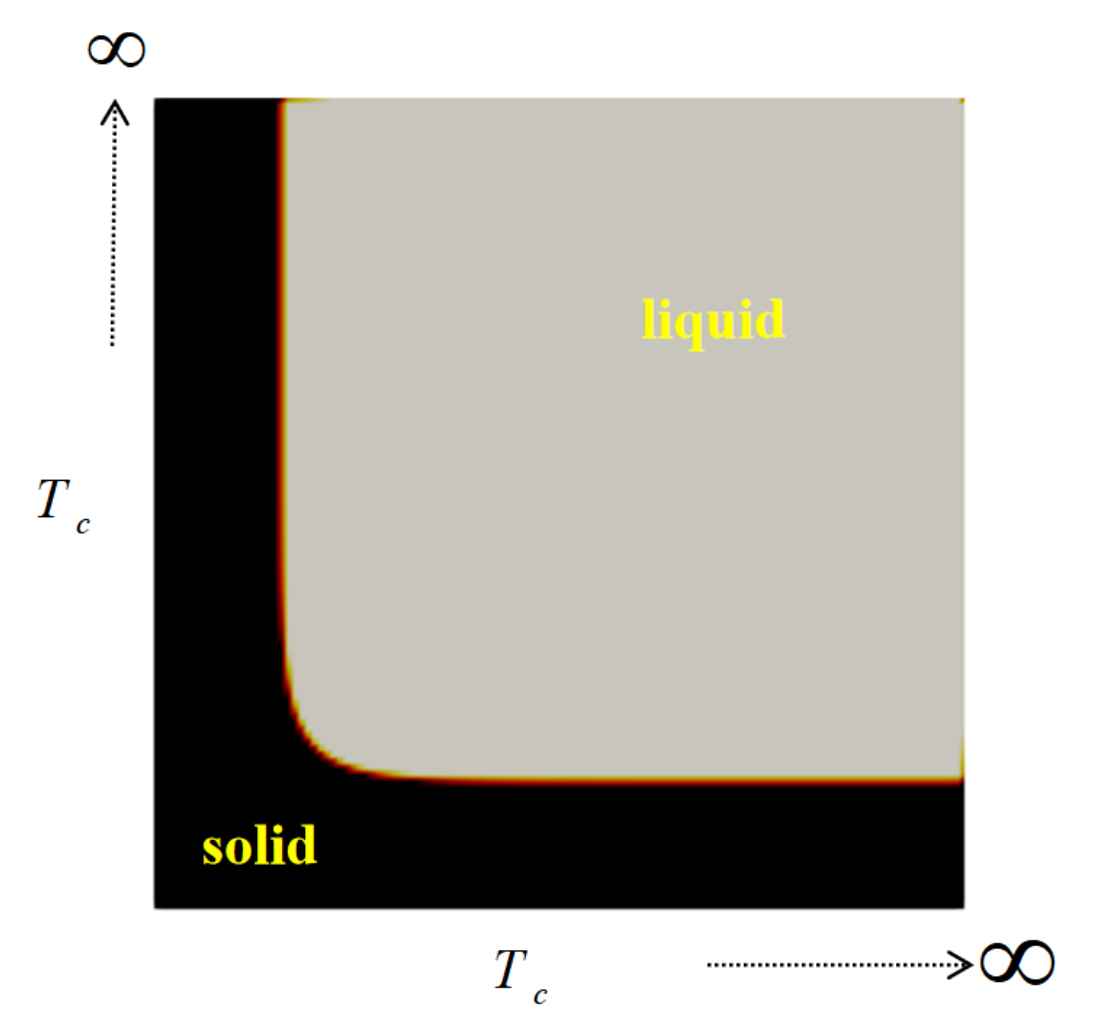}
	\caption{Schematic diagram of solidification in a semi-infinite corner.The left and bottom walls are imposed with cold temperature while the temperature of the top and right sides is matained as the initial state.}
	\label{fig:semi_corner_schematic}
\end{figure}

In this subsection, a two-dimensional benchmark problem involving solidification in a semi-infinite corner is investigated using the present RLB model. Initially, the liquid phase occupies the corner with temperature $T_i$, then the temperatures of the left side and the bottom side are exposed as $T_c$ which is lower than the melting temperature. Gradually, solidification occurs near the left and bottom boundaries, while the temperatures at the top and right boundaries, far from the solidification region, are maintained at their initial values, as shown in Fig.~\ref{fig:semi_corner_schematic}. In the present simulations, the temperature parameters are as following:
\begin{equation}
T_i = 0.3, \quad T_m = 0, \quad T_c = -1.0, \quad T_s = -0.01, \quad T_l = 0.01
\end{equation}

The thermophysical parameters are set as following:
\begin{eqnarray}
c_{pl} = 1, \quad c_{ps} = 1, \quad \kappa_l = 0.2, \quad \kappa_s = 0.2, \nonumber \\
\text{Ste} = \frac{c_{pl}(T_m - T_c)}{L_a} = 4.0
\end{eqnarray}

The relaxation rates across the interface is weighted by the liquid fraction. The simulation domain is set as $L_x = L_y = 128$ and the characteristic length is chosen as $L_c = 0.5 L_x$.

\begin{figure}[htbp]
	\centering
	\begin{subfigure}[b]{0.31\textwidth}
		\centering
		\includegraphics[width=\textwidth]{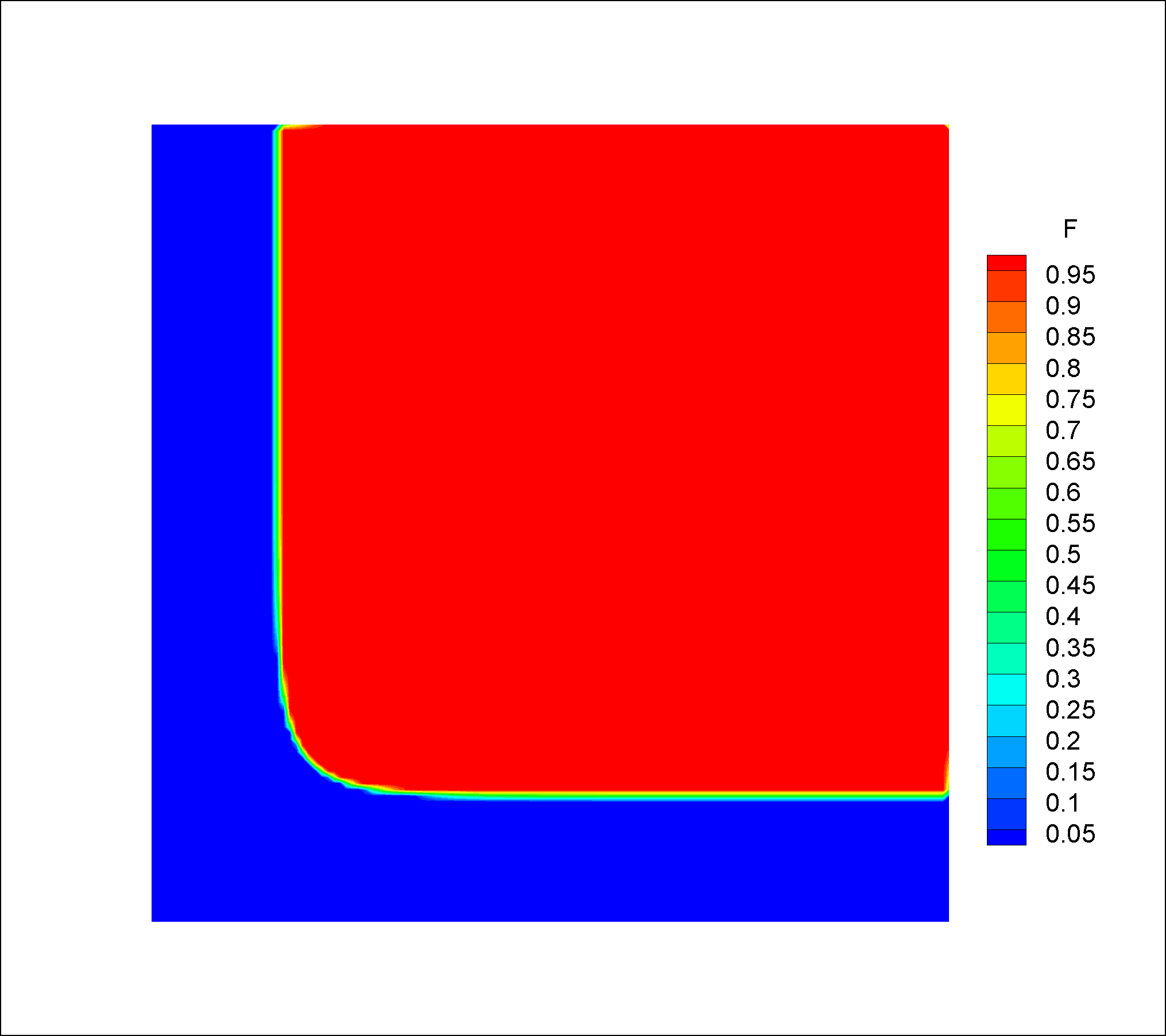}
		\caption{$\text{Fo} = 0.05$}
		\label{fig:7-panel-1}
	\end{subfigure}\hfill
\begin{subfigure}[b]{0.31\textwidth}
		\centering
		\includegraphics[width=\textwidth]{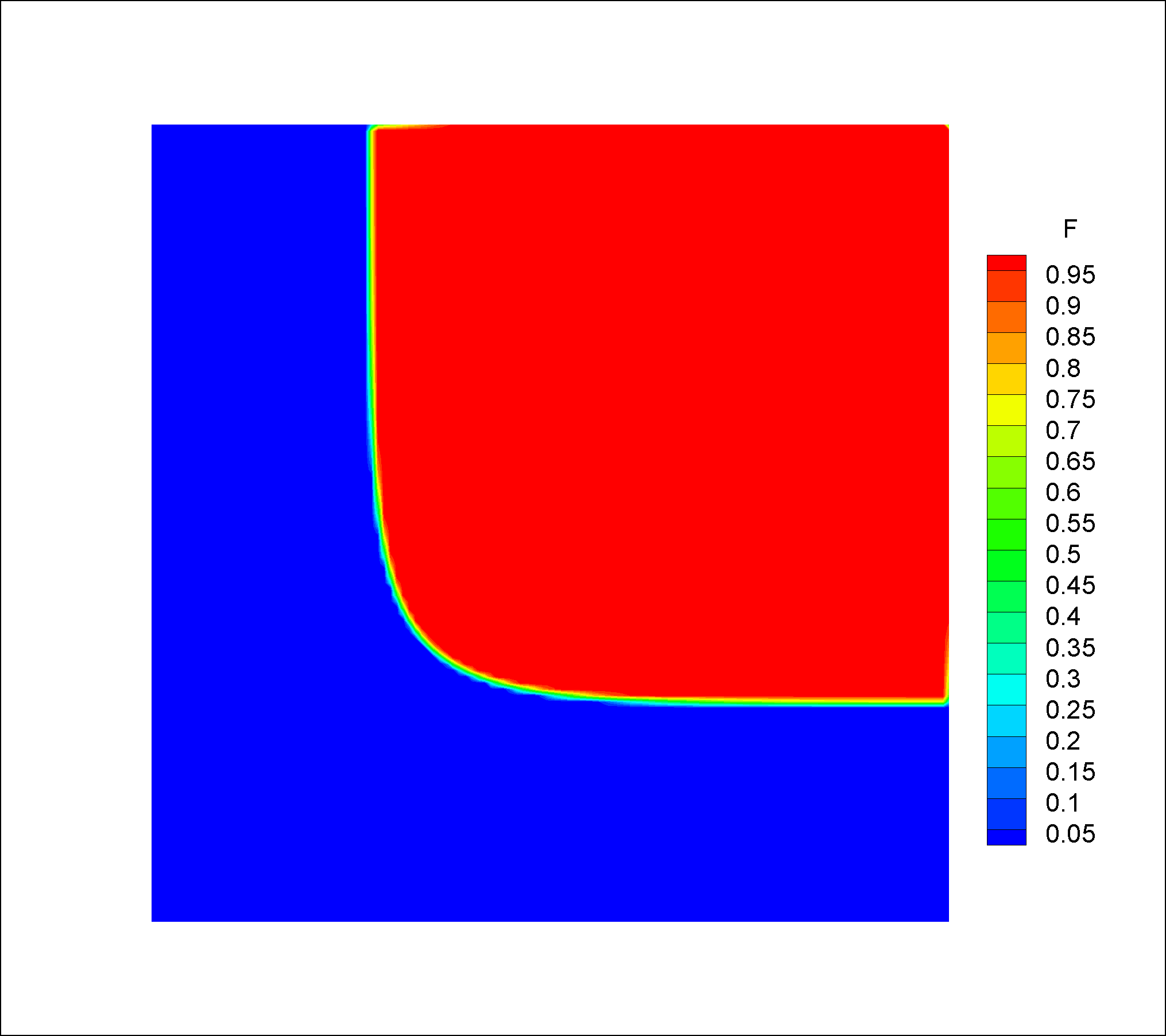}
		\caption{$\text{Fo} = 0.15$}
		\label{fig:7-panel-2}
	\end{subfigure}\hfill
\begin{subfigure}[b]{0.31\textwidth}
		\centering
		\includegraphics[width=\textwidth]{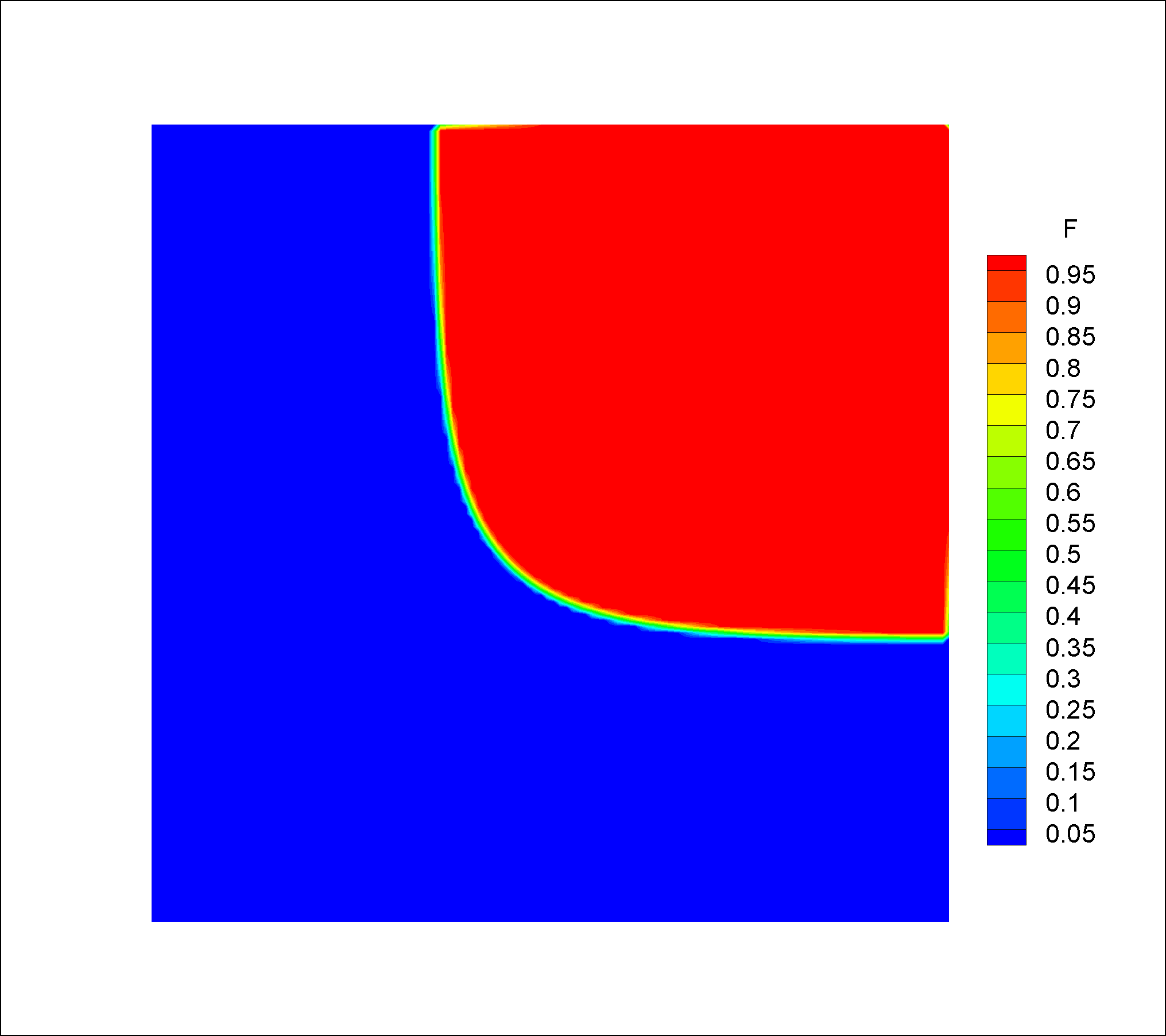}
		\caption{$\text{Fo} = 0.25$}
		\label{fig:7-panel-3}
	\end{subfigure}

	\caption{Fraction of liquid in the field at specific time intervals for different $\text{Fo}$ numbers,$Fo = 0.05,0.15,0.25$.}
	\label{fig:semi_corner_frac}
\end{figure}

\begin{figure}[htbp]
	\centering
	\includegraphics[width=0.62\textwidth]{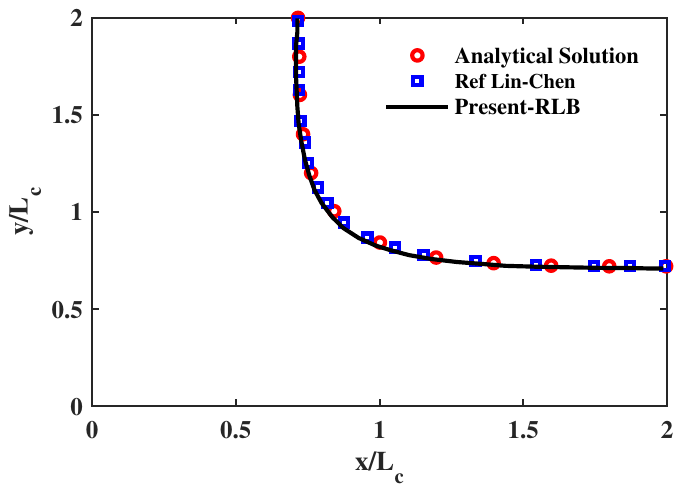}
	\caption{Comparison of the solid-liquid interface at $\text{Fo} = 0.25$ for the present RLB model and the analytical solution~\cite{RathjenandJiji1971} and referred numerical results~\cite{Lin1997}}
	\label{fig:semi_corner_interface}
\end{figure}

\begin{figure}[htbp]
	\centering
	\includegraphics[width=0.62\textwidth]{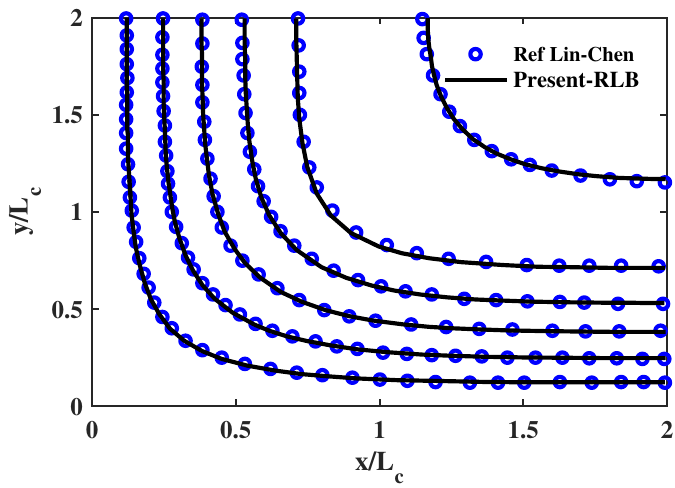}
	\caption{Comparison of the isotherms ($T = -0.8$ to $T = 0.2$ from the bottom to top) at $\text{Fo} = 0.25$ for the present RLB model and referred numerical results~\cite{Lin1997}}
	\label{fig:semi_corner_isotherms}
\end{figure}

Fig.~\ref{fig:semi_corner_frac} qualitatively illustrates the solidification process in the semi-infinite corner at different time instants. As shown in Fig.~\ref{fig:semi_corner_interface}, the solid-liquid interface obtained by the present RLB model at $\text{Fo} = 0.25$ is compared with the analytical solution~\cite{RathjenandJiji1971} and the numerical results reported in Ref.~\cite{Lin1997}, and good agreement is observed. Furthermore, the temperature isotherms in the computational domain at $\text{Fo} = 0.25$ are extracted with an interval of 0.2. It can be seen from Fig.~\ref{fig:semi_corner_isotherms} that the predicted temperature profiles are in good agreement with the reference numerical results.

\subsection{Convection melting in a two dimensional square cavity}

\begin{figure}[htbp]
	\centering
	\includegraphics[width=0.62\textwidth]{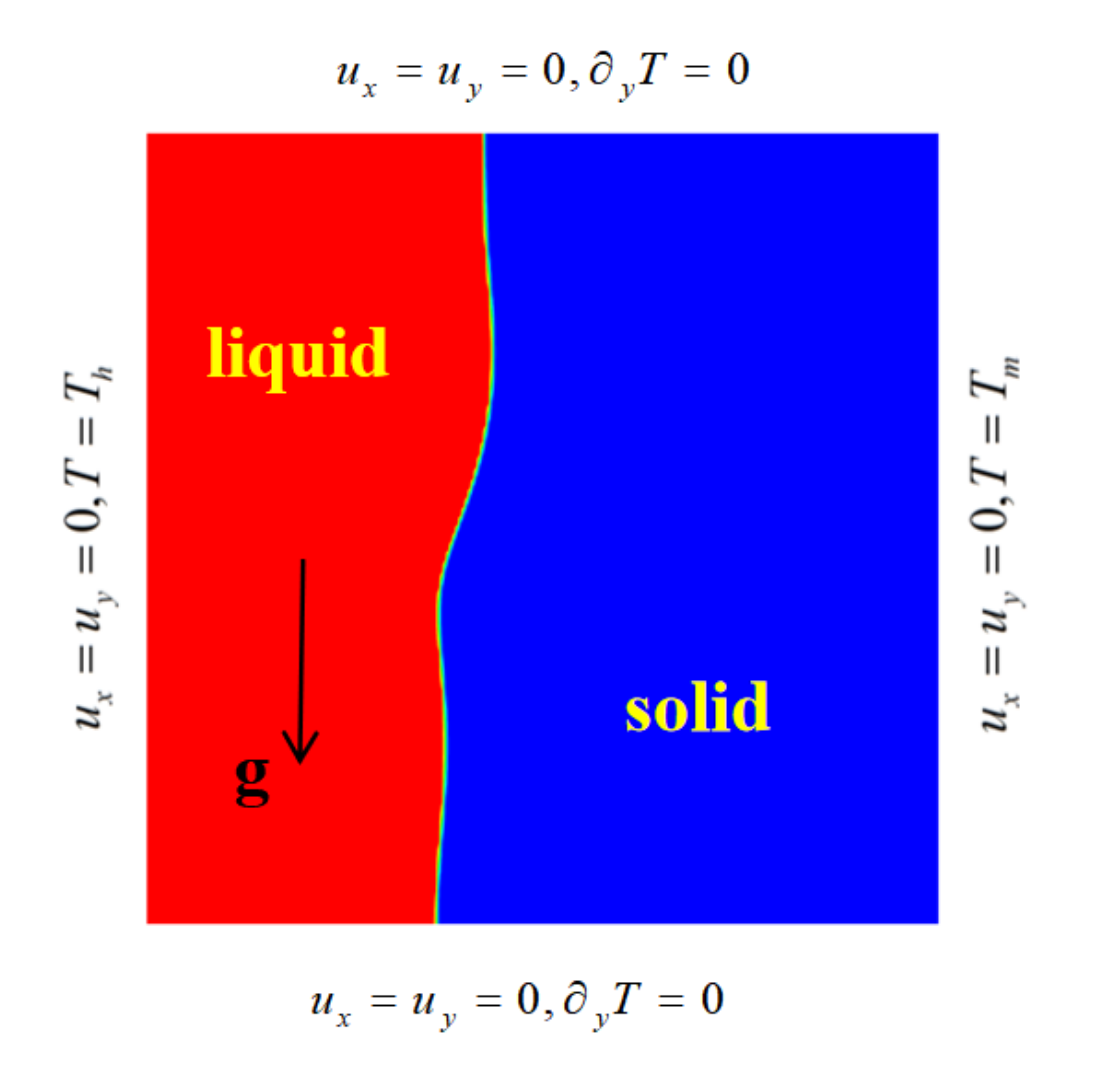}
	\caption{Schematic diagram of the 2D cavity melting. The left wall is the heat wall and no-slip boundary, the right wall is the cold wall and no-slip boundary, the top and bottom walls are of no-slip velocity boundary and zero-normal temperature gradient boundary.}
	\label{fig:cavity_schematic}
\end{figure}

In this subsection, a two-dimensional square cavity melting problem is investigated to account for both conduction and convection effects during the phase-change process. The benchmark configuration is illustrated in Fig.~\ref{fig:cavity_schematic}. Initially, the computational domain is entirely occupied by a solid-state pure phase-change material at the melting temperature $T_m$. The bottom and top walls are subjected to Neumann boundary conditions for both temperature and enthalpy (zero normal derivative), whereas the left and right walls are maintained at a high temperature $T_h$ and the melting temperature $T_m$, respectively. This physical problem is governed by several key dimensionless parameters, namely the Rayleigh, Prandtl, and Stefan numbers.

In the present simulations, the dimensionless temperature parameters are set as follows:
\begin{equation}
T_h = 1, \quad T_m = 0, \quad T_{\text{ref}} = 0, \quad T_s = T_m, \quad T_l = T_m
\end{equation}

The thermophysical properties are specified as:
\begin{equation}
c_{ps} = 1, \quad c_{pl} = 1, \quad \kappa_s = 0.2, \quad \kappa_l = 0.2
\end{equation}

To evaluate the influence of natural convection, two cases with different Rayleigh numbers are investigated:
\begin{align}
\text{Case 1:} \quad &\text{Ra} = 2.5 \times 10^4, \, \text{Pr} = 0.02, \, \text{Ste} = 0.01 \\
\text{Case 2:} \quad &\text{Ra} = 2.5 \times 10^5, \, \text{Pr} = 0.02, \, \text{Ste} = 0.01
\end{align}

In the numerical simulations, the computational domain is discretized using a $128 \times 128, 256 \times 256, 512\times 512$ lattices, with the side length of the cavity serving as the characteristic length.

\begin{figure}[!t]
	\centering
	\begin{subfigure}[b]{0.31\textwidth}
		\centering
		\includegraphics[width=\linewidth]{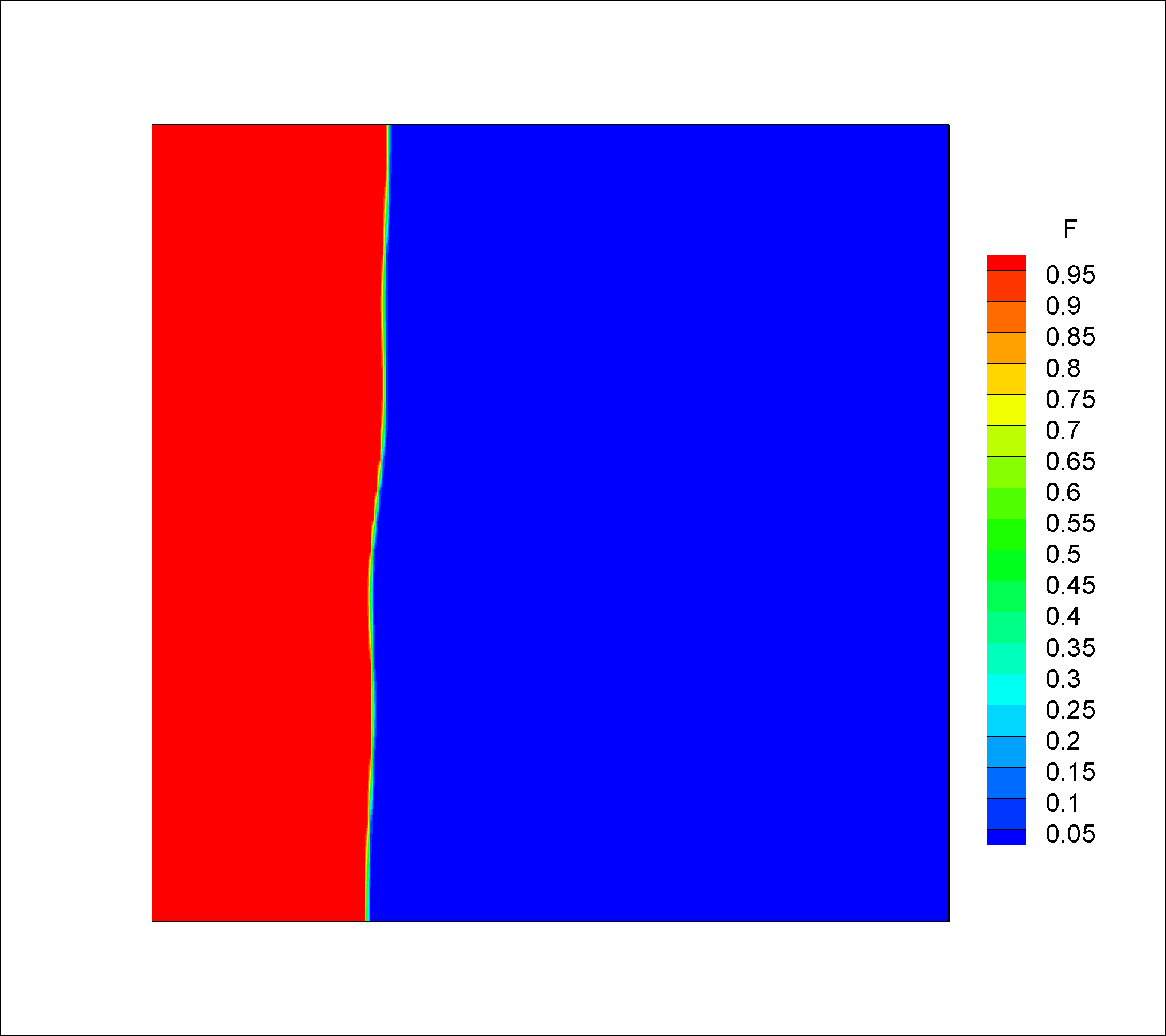}
		\caption{$\text{Fo} = 4.0$ (Liquid fraction)}
		\label{fig:11-panel-1}
	\end{subfigure}\hfill
	\begin{subfigure}[b]{0.31\textwidth}
		\centering
		\includegraphics[width=\linewidth]{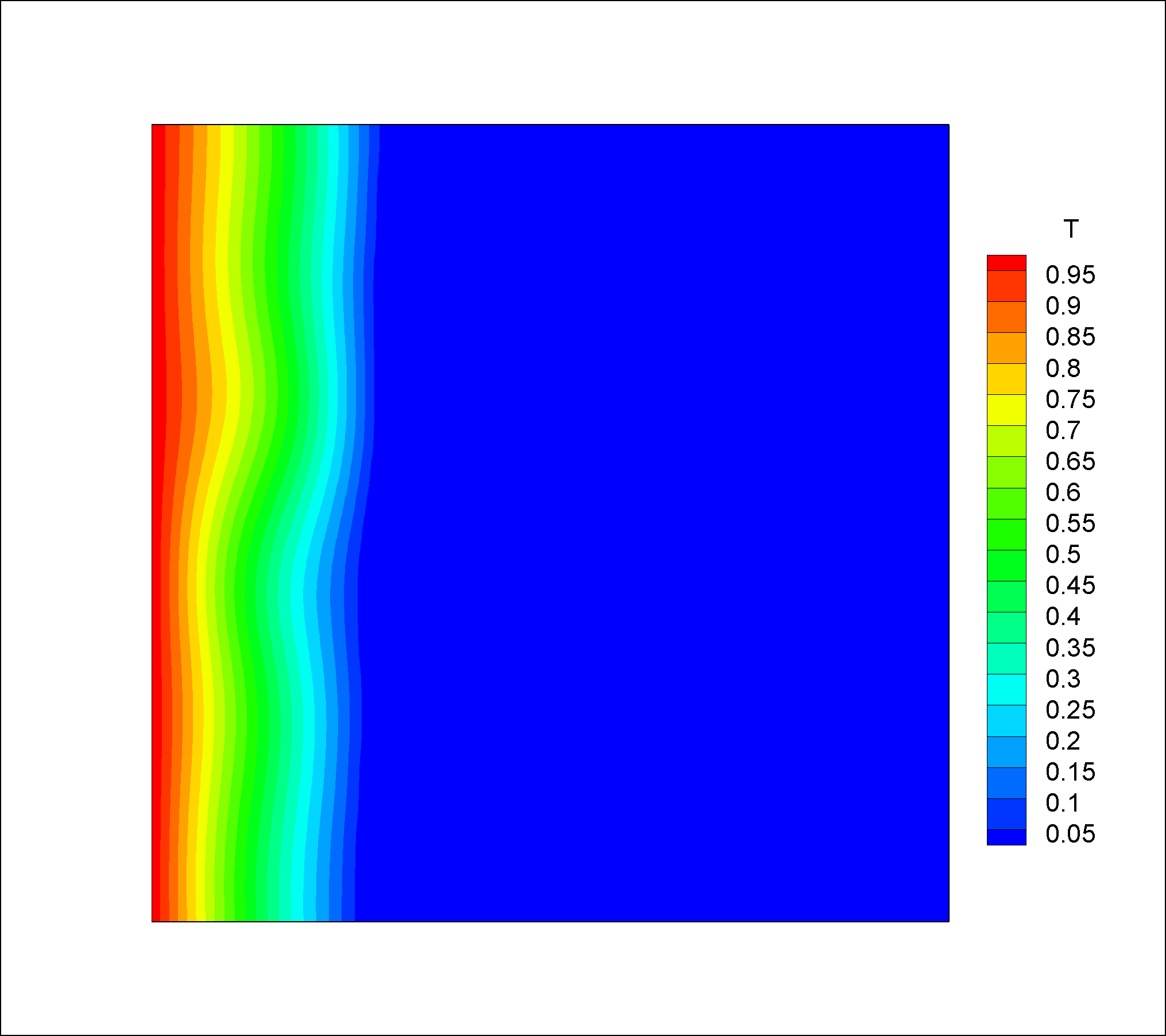}
		\caption{$\text{Fo} = 4.0$ (Temperature)}
		\label{fig:11-panel-2}
	\end{subfigure}\hfill
	\begin{subfigure}[b]{0.31\textwidth}
	    \centering
	    \includegraphics[width=\linewidth]{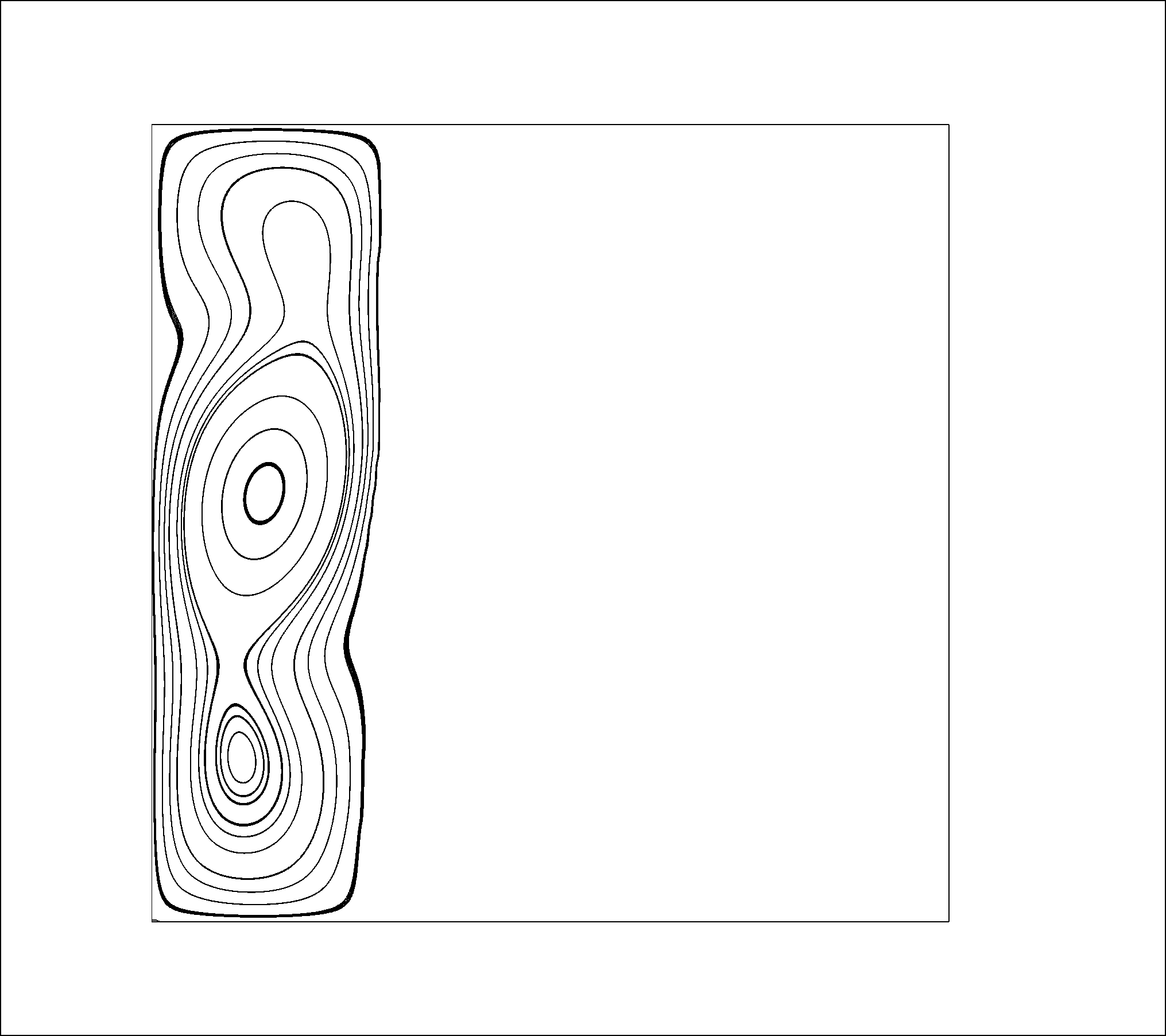}
	    \caption{$\text{Fo} = 4.0$ (Streamline)}
	    \label{fig:11-panel-3}
    \end{subfigure}

	\vspace{0.5em}

	\begin{subfigure}[b]{0.31\textwidth}
		\centering
		\includegraphics[width=\linewidth]{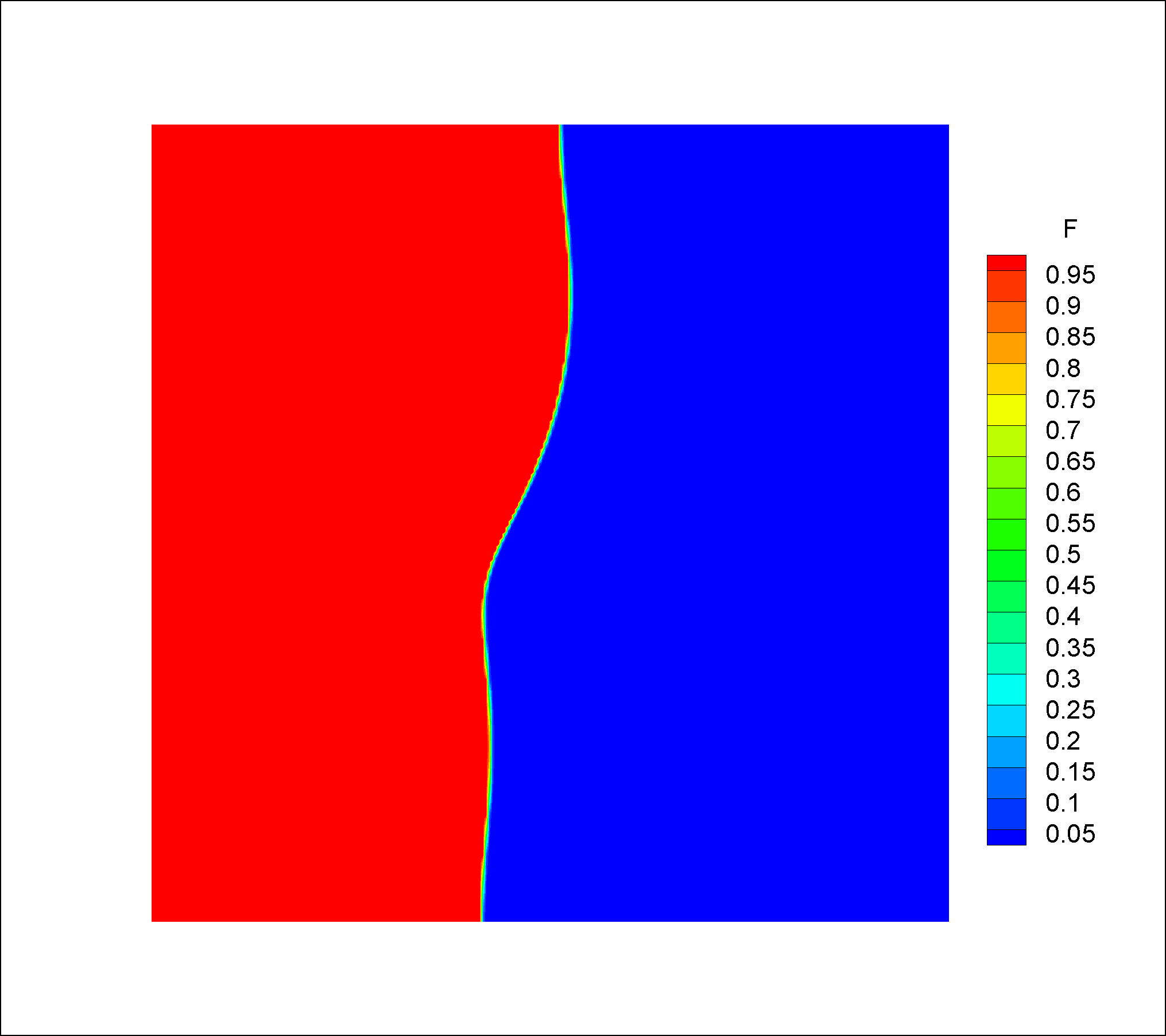}
		\caption{$\text{Fo} = 10.0$ (Liquid fraction)}
		\label{fig:11-panel-4}
	\end{subfigure}\hfill
	\begin{subfigure}[b]{0.31\textwidth}
		\centering
		\includegraphics[width=\linewidth]{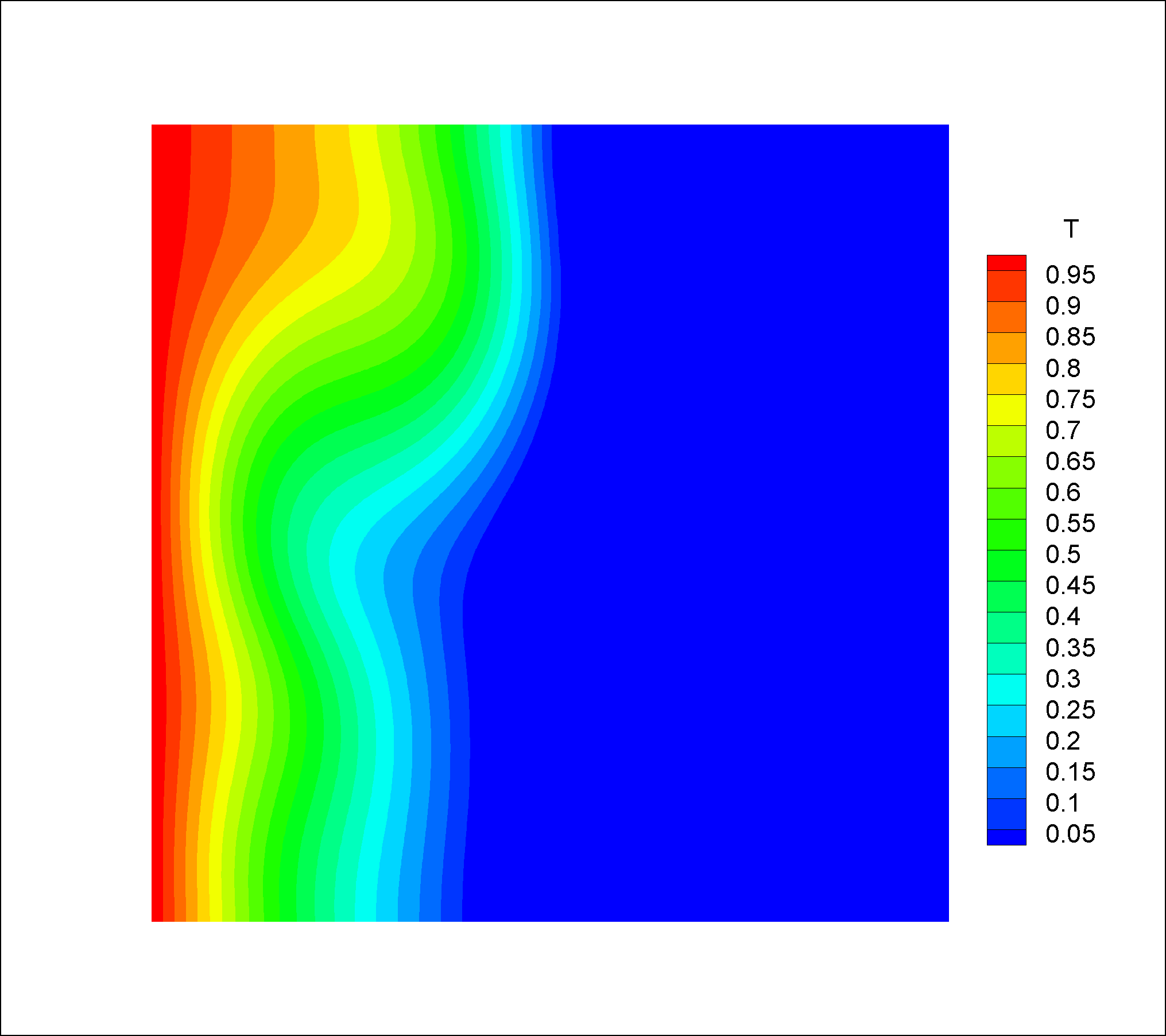}
		\caption{$\text{Fo} = 10.0$ (Temperature)}
		\label{fig:11-panel-5}
	\end{subfigure}\hfill
    \begin{subfigure}[b]{0.31\textwidth}
    	\centering
    	\includegraphics[width=\linewidth]{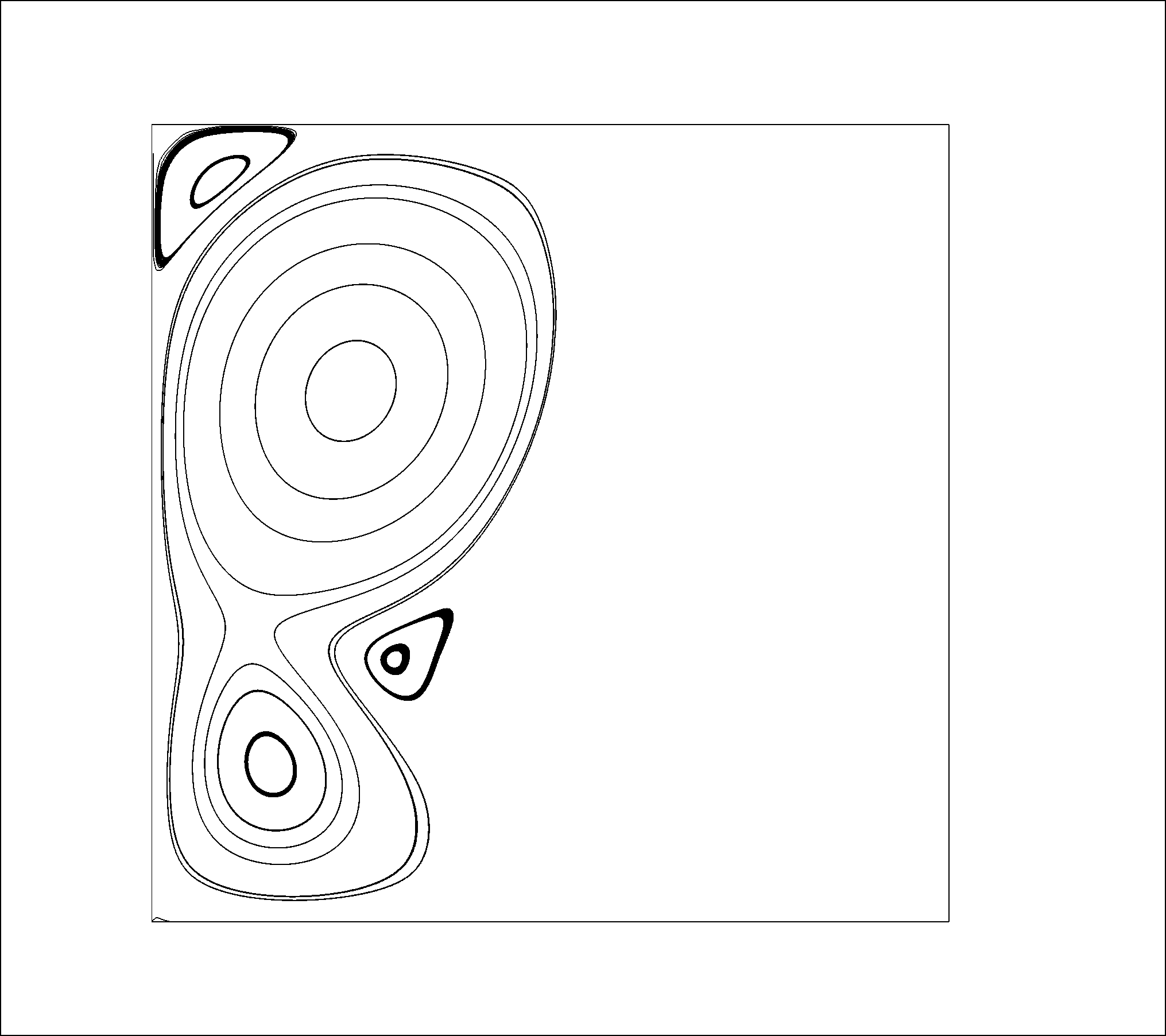}
    	\caption{$\text{Fo} = 10.0$ (Streamline)}
    	\label{fig:11-panel-6}
    \end{subfigure}

	\vspace{0.5em}

	\begin{subfigure}[b]{0.31\textwidth}
		\centering
		\includegraphics[width=\linewidth]{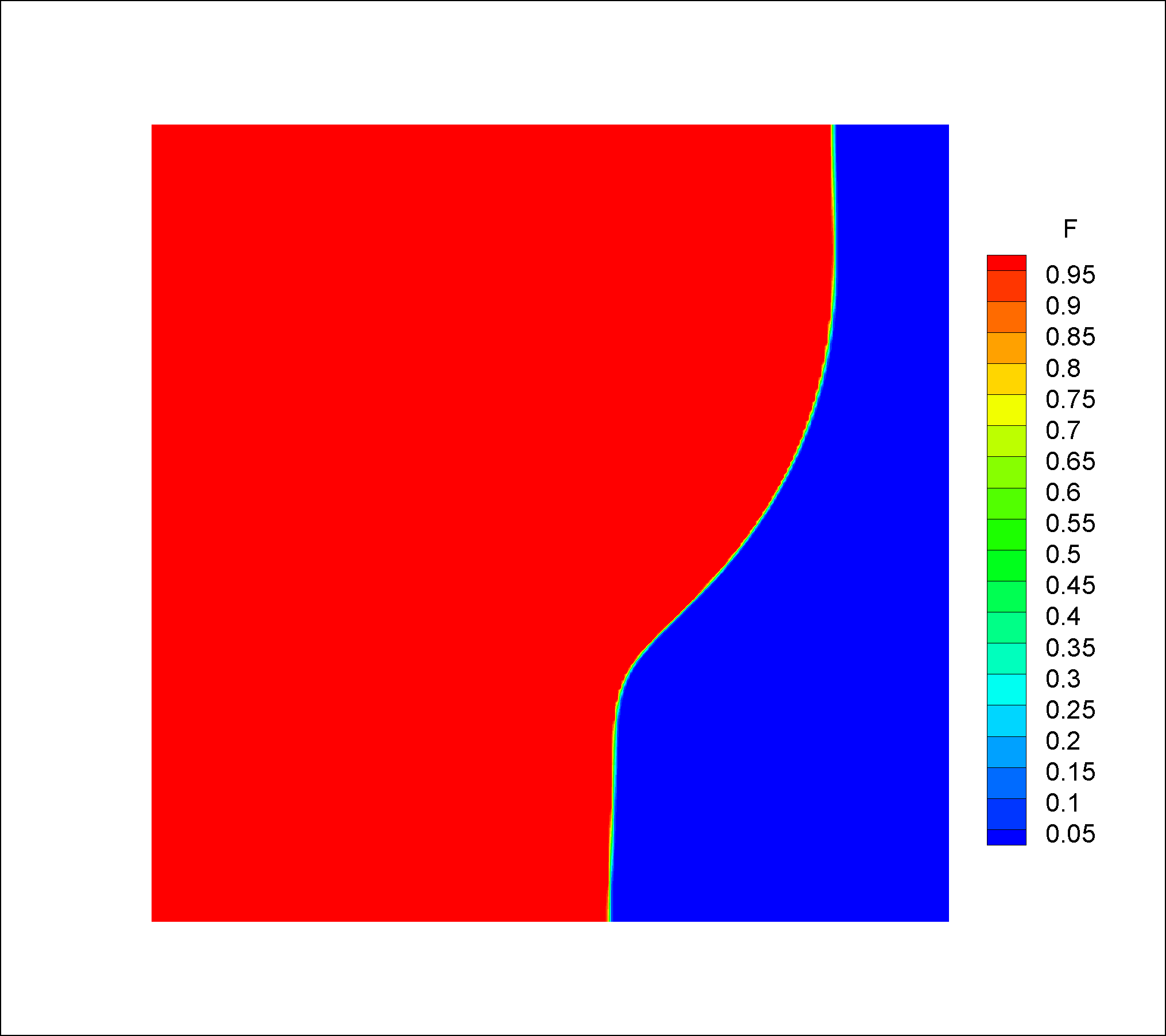}
		\caption{$\text{Fo} = 20.0$ (Liquid fraction)}
		\label{fig:11-panel-7}
	\end{subfigure}\hfill
	\begin{subfigure}[b]{0.31\textwidth}
		\centering
		\includegraphics[width=\linewidth]{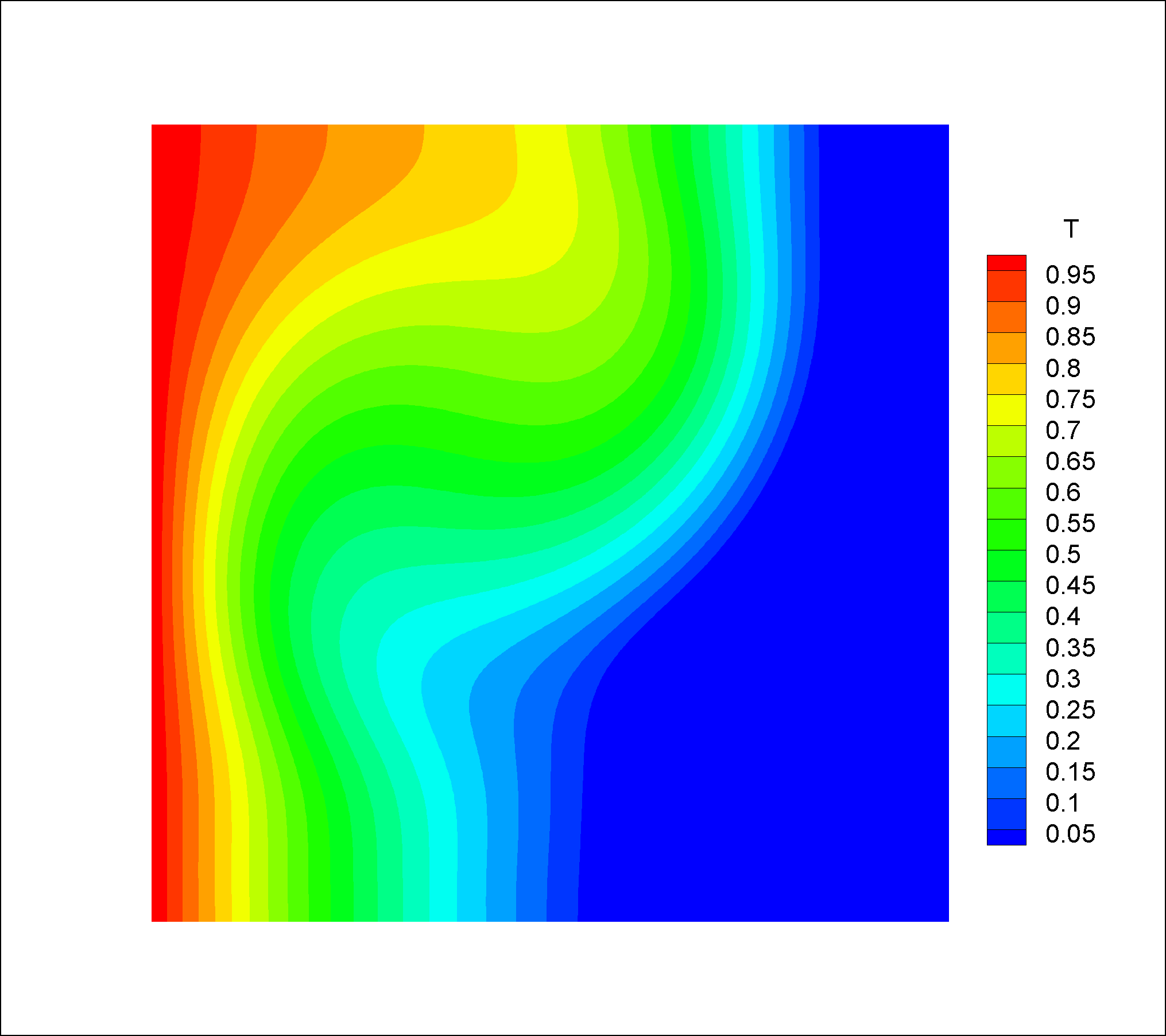}
		\caption{$\text{Fo} = 20.0$ (Temperature)}
		\label{fig:11-panel-8}
	\end{subfigure}\hfill
    	\begin{subfigure}[b]{0.31\textwidth}
    	\centering
    	\includegraphics[width=\linewidth]{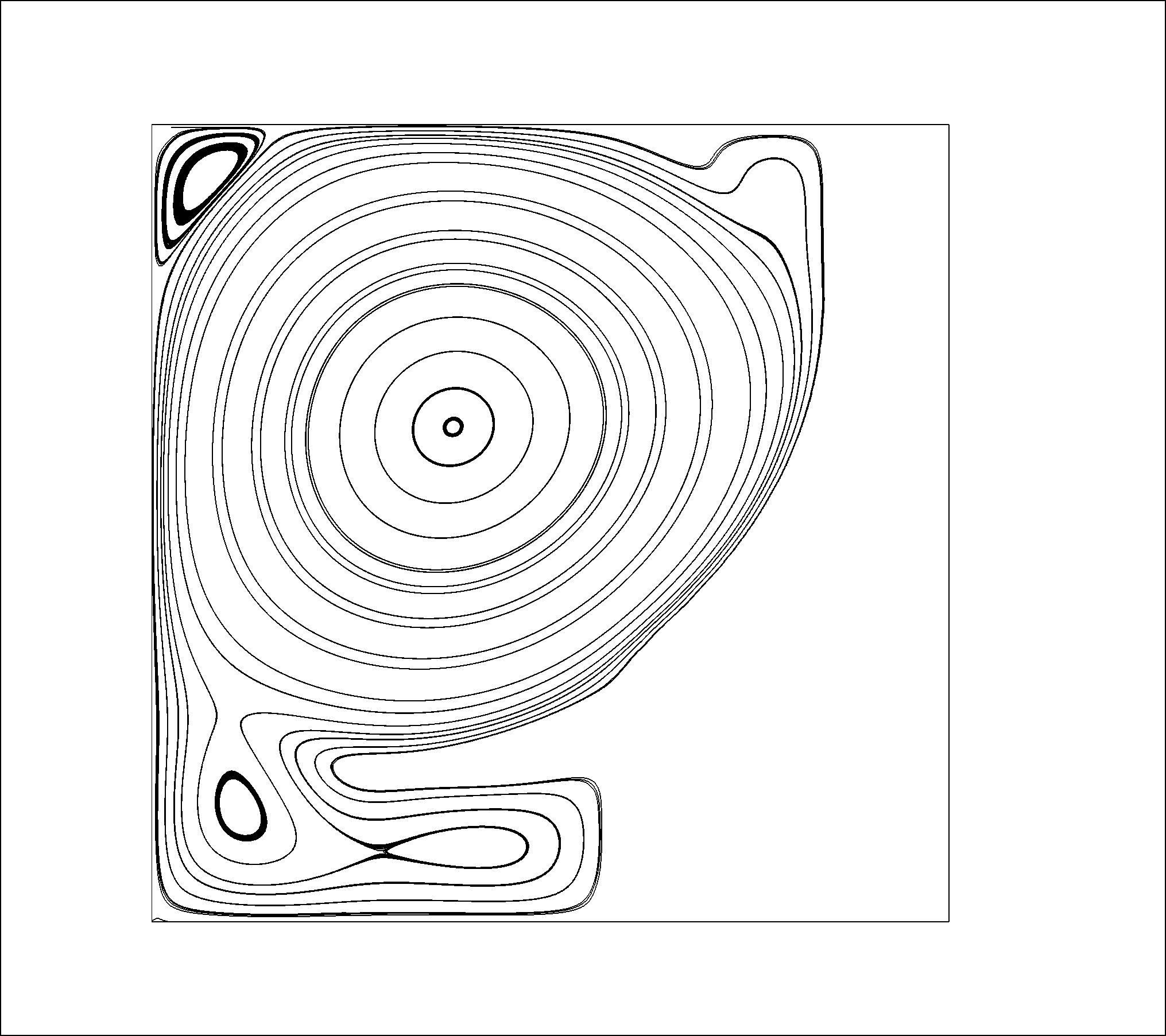}
    	\caption{$\text{Fo} = 20.0$ (Streamline)}
    	\label{fig:11-panel-9}
    \end{subfigure}

	\vspace{0.5em}

	\begin{subfigure}[b]{0.31\textwidth}
		\centering
		\includegraphics[width=\linewidth]{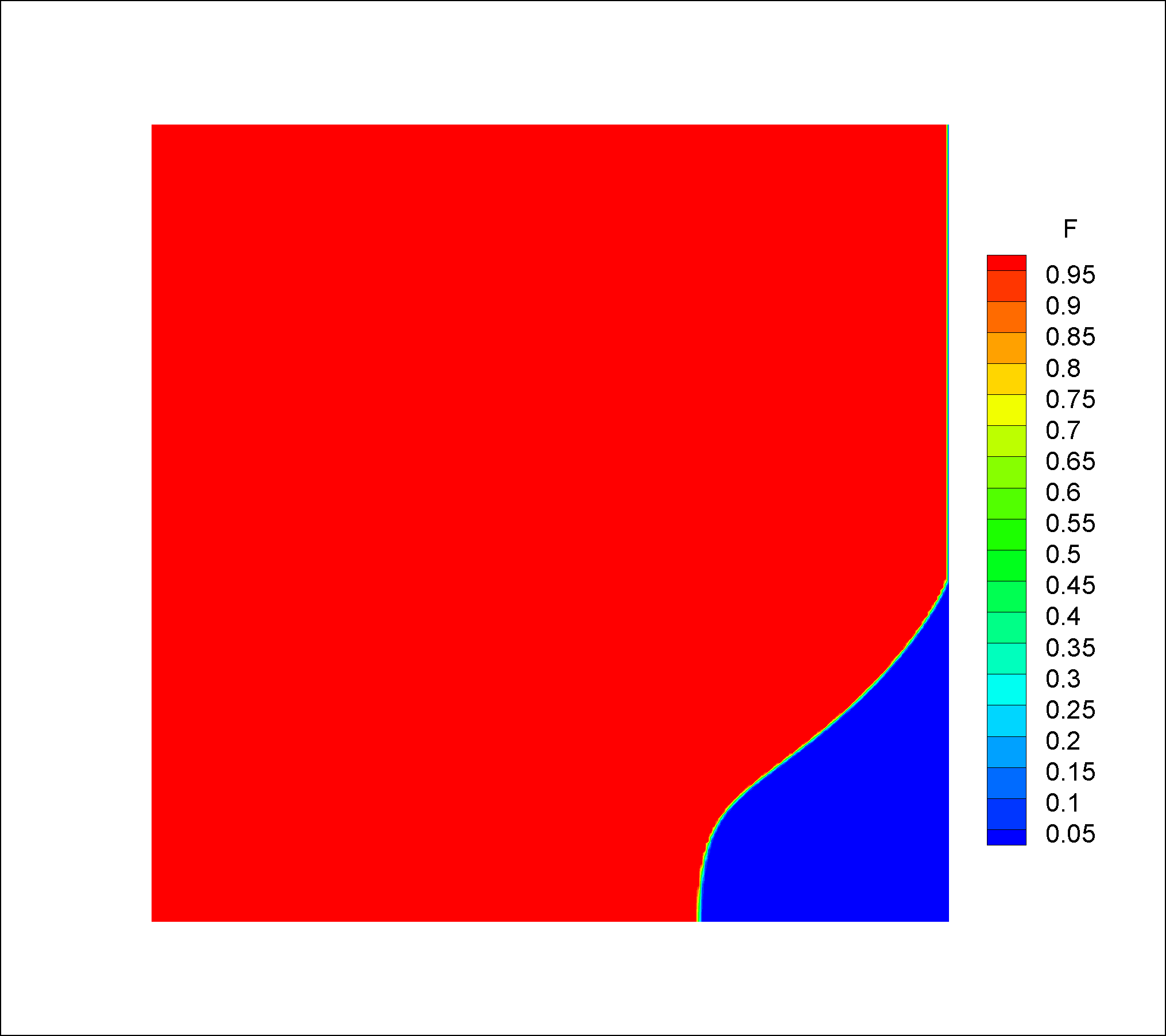}
		\caption{$\text{Fo} = 30.0$ (Liquid fraction)}
		\label{fig:11-panel-10}
	\end{subfigure}\hfill
	\begin{subfigure}[b]{0.31\textwidth}
		\centering
		\includegraphics[width=\linewidth]{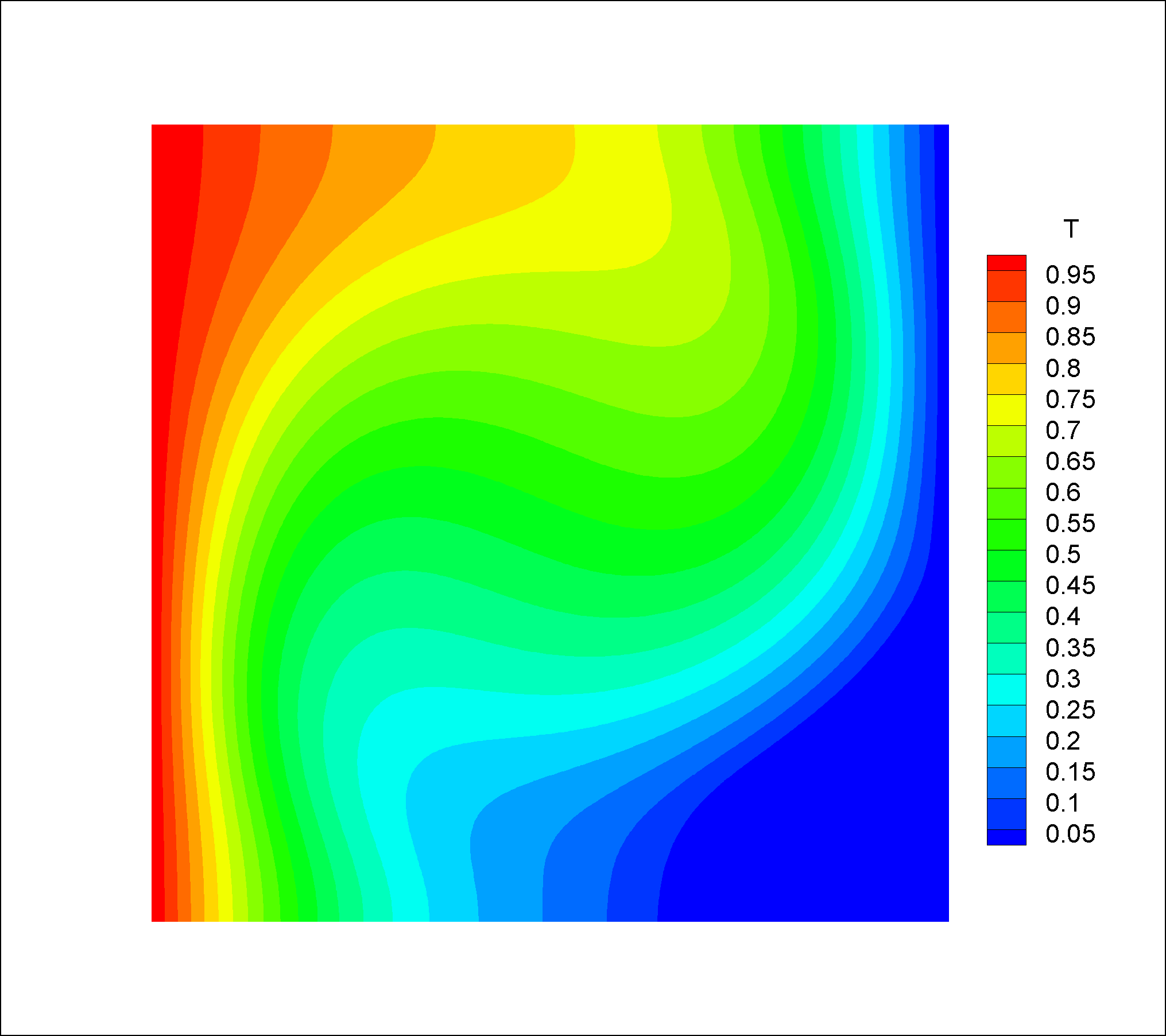}
		\caption{$\text{Fo} = 30.0$ (Temperature)}
		\label{fig:11-panel-11}
	\end{subfigure}\hfill
    	\begin{subfigure}[b]{0.31\textwidth}
    	\centering
    	\includegraphics[width=\linewidth]{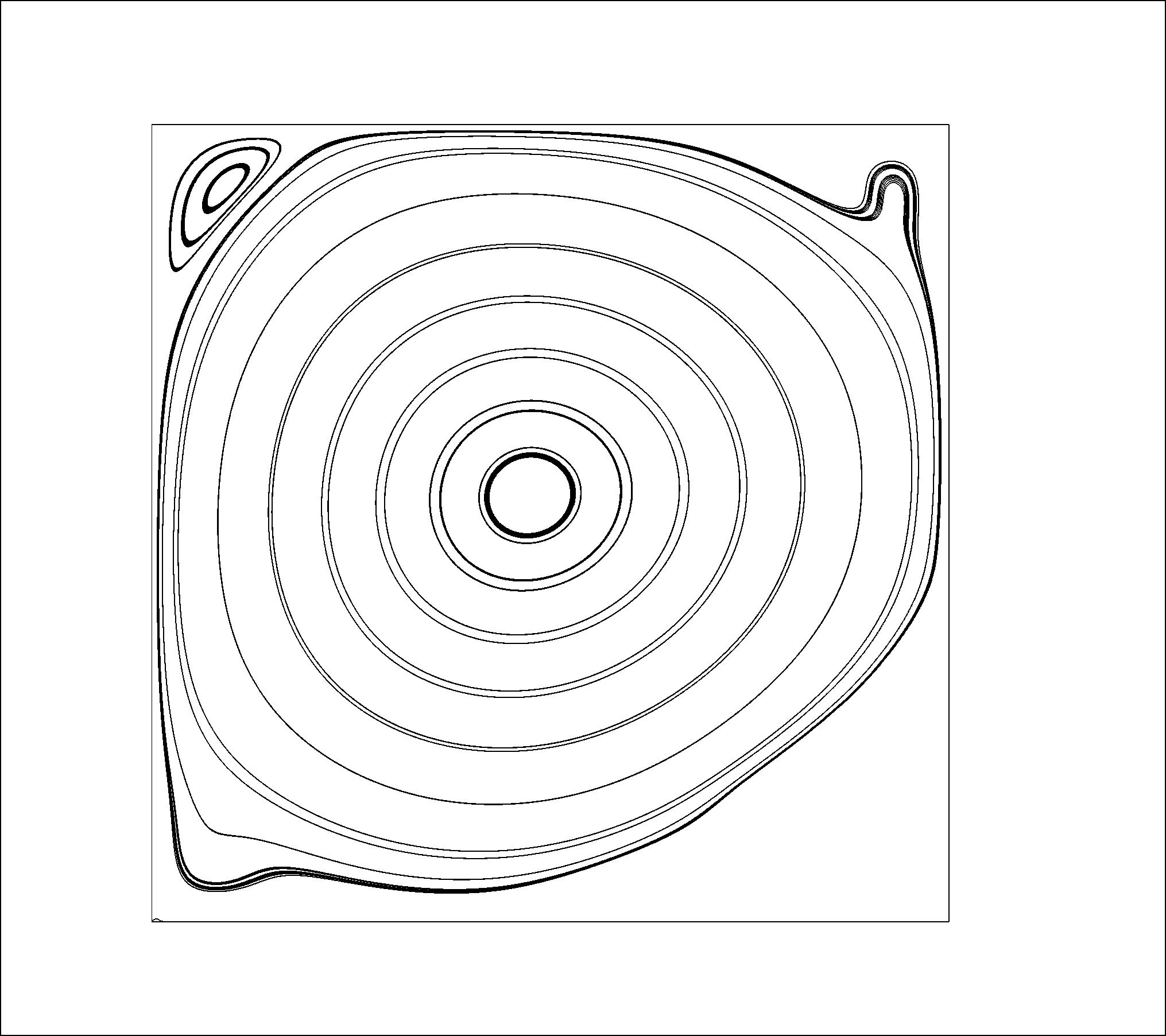}
    	\caption{$\text{Fo} = 30.0$ (Streamline)}
    	\label{fig:11-panel-12}
    \end{subfigure}

	\caption{Case1:$Pr=0.02,Ste=0.01,Ra=2.5\times 10^4$. Liquid-fraction contours, streamlines, and temperature distributions during the convection melting process for Case 1 at various Fourier numbers: (a-c) $\text{Fo} = 4.0$; (d-f) $\text{Fo} = 10.0$; (g-i) $\text{Fo} = 20.0$; (j-l) $\text{Fo} = 30.0$.}
	\label{fig:case1_contours}
\end{figure}

For Case~1, Figure~\ref{fig:case1_contours} depicts the liquid-fraction contours, temperature fields, and streamlines across several representative dimensionless times. The present predictions exhibit excellent agreement with benchmark data reported in the literature~\cite{Lu2019, Liu2022}. Under this Rayleigh number, the phase-change process is characterized by two distinct regimes: an initial conduction-dominated regime followed by a convection-dominated regime. In the early stage, the solid-liquid interface advances nearly parallel to the heated left wall. As melting proceeds, buoyant upwelling carries warm liquid toward the upper cavity, accelerating the advancement of the top portion of the front and resulting in a typical curved, top-heavy interface profile.

\begin{figure}[htbp]
	\centering
	\begin{subfigure}[b]{0.48\textwidth}
		\centering
	    \includegraphics[width=\textwidth]{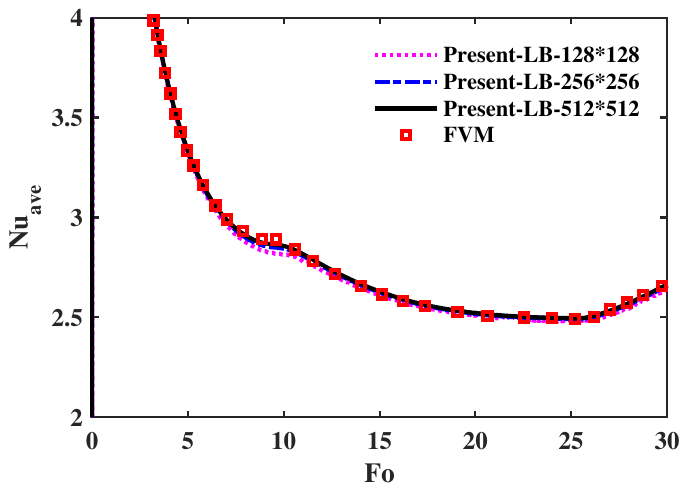}
	    \caption{Average Nusselt number}
	\end{subfigure}\hfill
\begin{subfigure}[b]{0.48\textwidth}
    	\centering
    	\includegraphics[width=\textwidth]{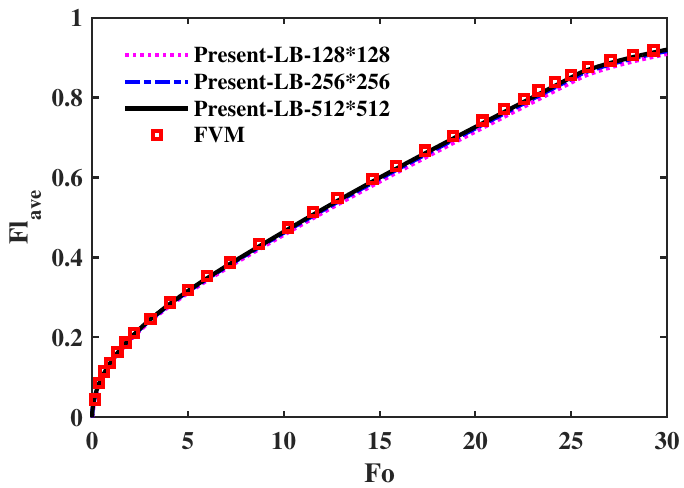}
    	\caption{Total liquid fraction}
    \end{subfigure}
	\caption{Case1:$Pr=0.02,Ste=0.01,Ra=2.5\times 10^4$. The average Nusselt number at the left wall and the total liquid fraction at different Fourier numbers: the present LB model and the FVM results~\cite{Mencinger2004}}
	\label{fig:case1_nusselt_fraction}
\end{figure}

Figure~\ref{fig:case1_nusselt_fraction} illustrates the temporal evolution of the average Nusselt number along the left (heated) wall and the total liquid fraction as functions of the Fourier number, alongside benchmark results obtained from the finite volume method (FVM). Data are sampled every 20,000 time steps over a total duration of $9{,}750{,}000$ steps, which is sufficiently dense to faithfully resolve the transient behaviors of both quantities. As the grid is refined, the predicted profiles show excellent overall agreement with the FVM benchmarks, demonstrating the grid convergence and fidelity of the present RLB model.

\begin{figure}[!t]
	\centering
	\begin{subfigure}[b]{0.31\textwidth}
		\centering
		\includegraphics[width=\linewidth]{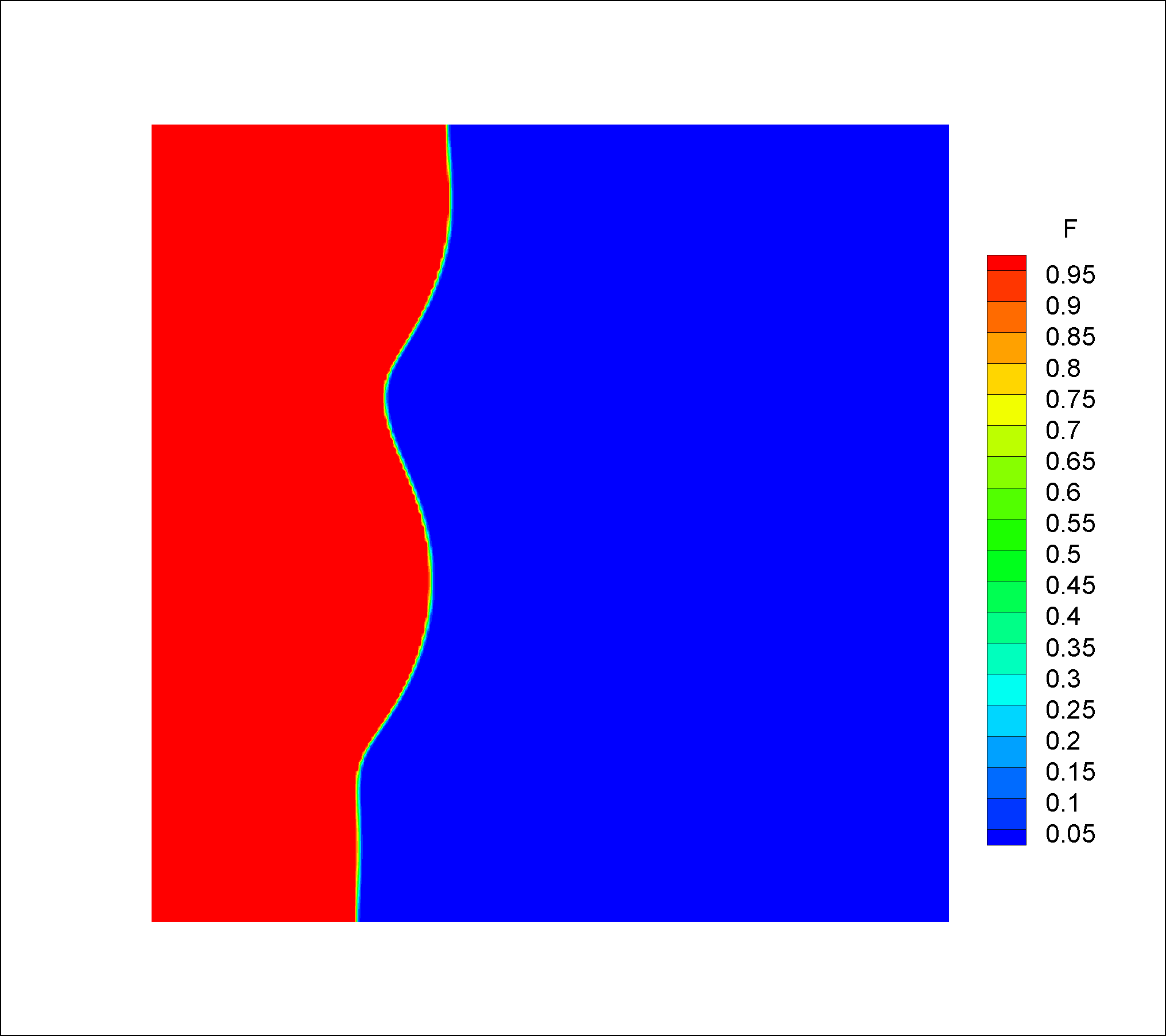}
		\caption{$\text{Fo} = 4.0$ (Liquid fraction)}
		\label{fig:13-panel-1}
	\end{subfigure}\hfill
	\begin{subfigure}[b]{0.31\textwidth}
		\centering
		\includegraphics[width=\linewidth]{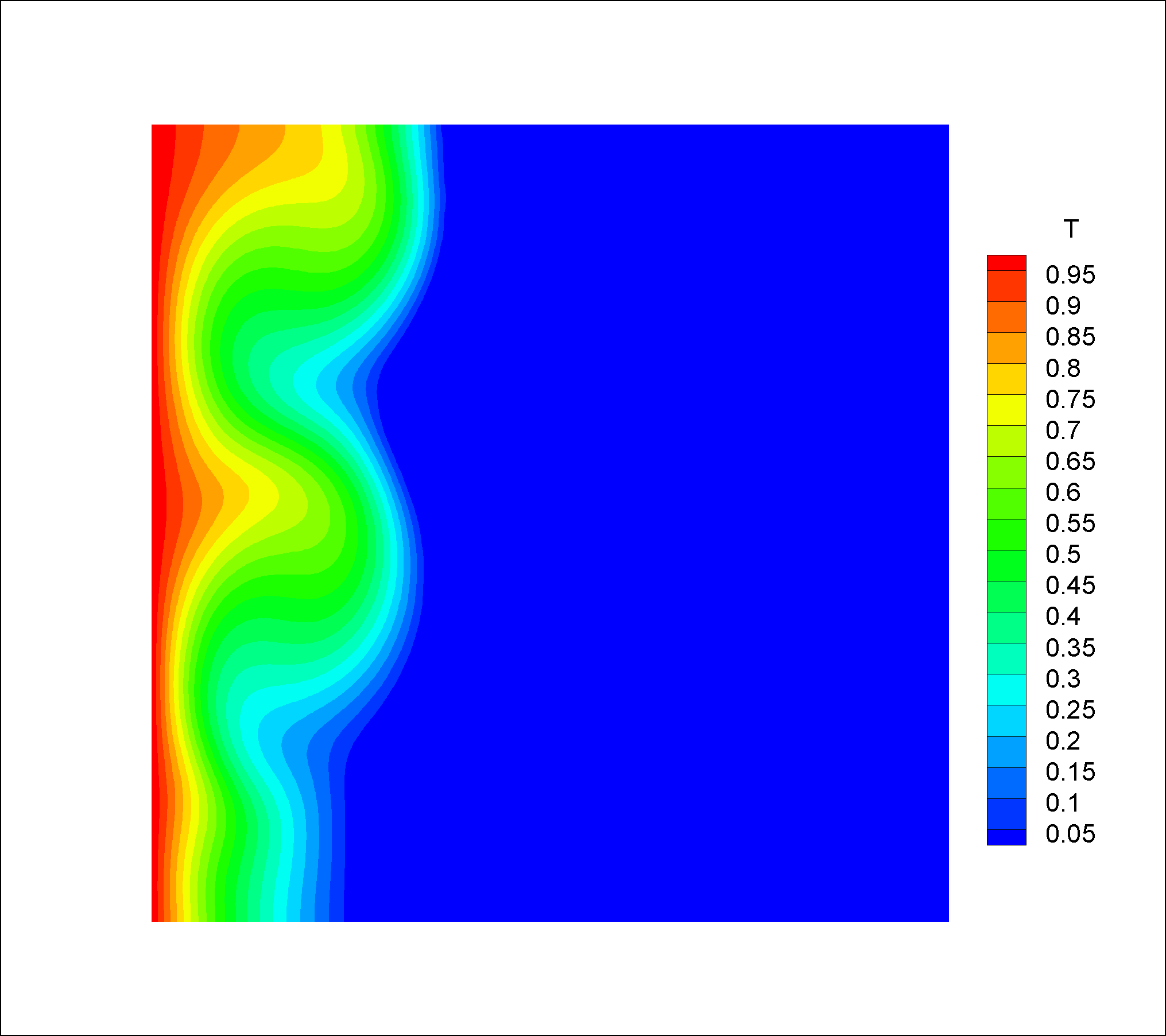}
		\caption{$\text{Fo} = 4.0$ (Temperature)}
		\label{fig:13-panel-2}
	\end{subfigure}\hfill
	\begin{subfigure}[b]{0.31\textwidth}
		\centering
		\includegraphics[width=\linewidth]{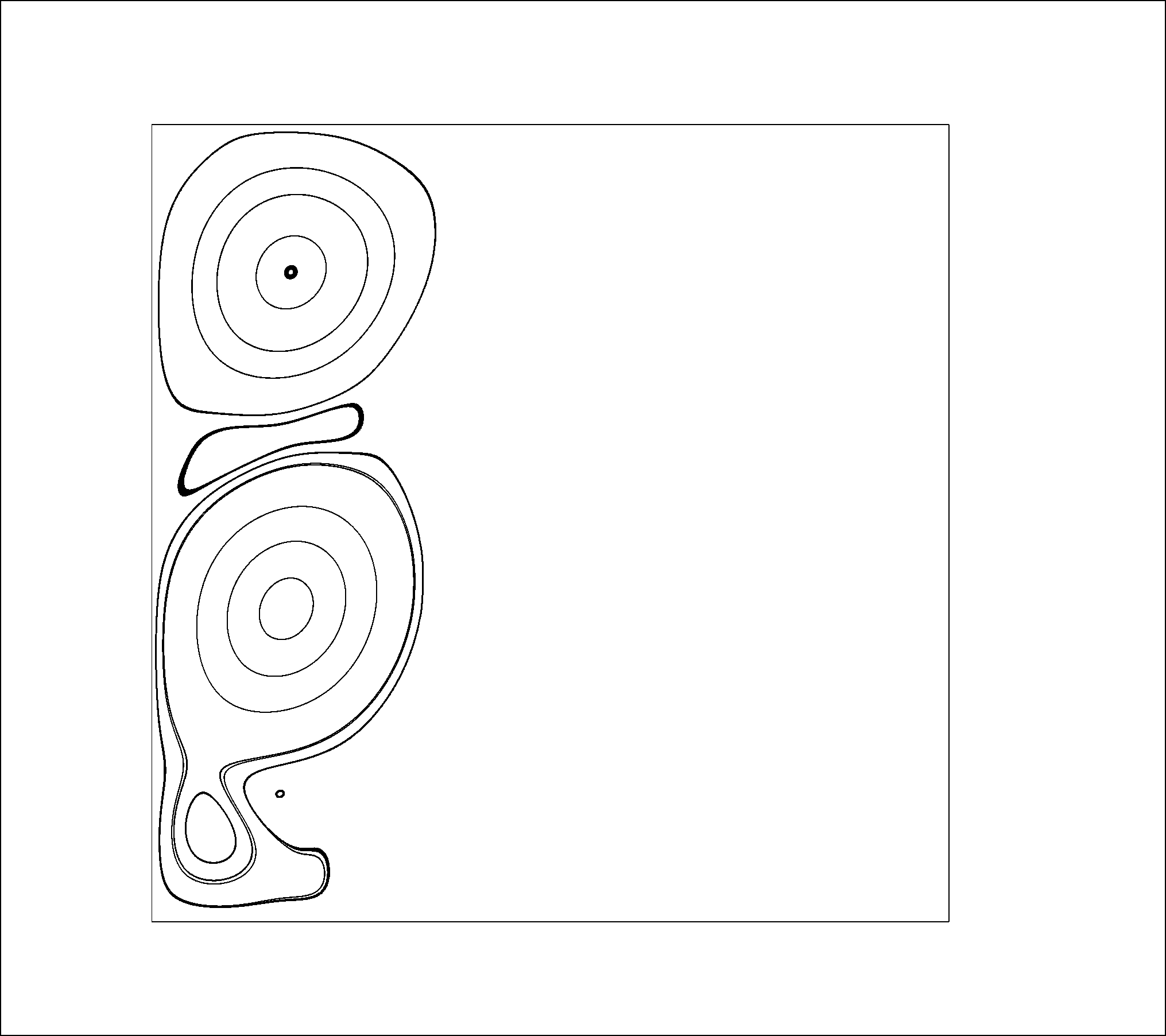}
		\caption{$\text{Fo} = 4.0$ (Streamline)}
		\label{fig:13-panel-3}
	\end{subfigure}

	\vspace{0.5em}

	\begin{subfigure}[b]{0.31\textwidth}
		\centering
		\includegraphics[width=\linewidth]{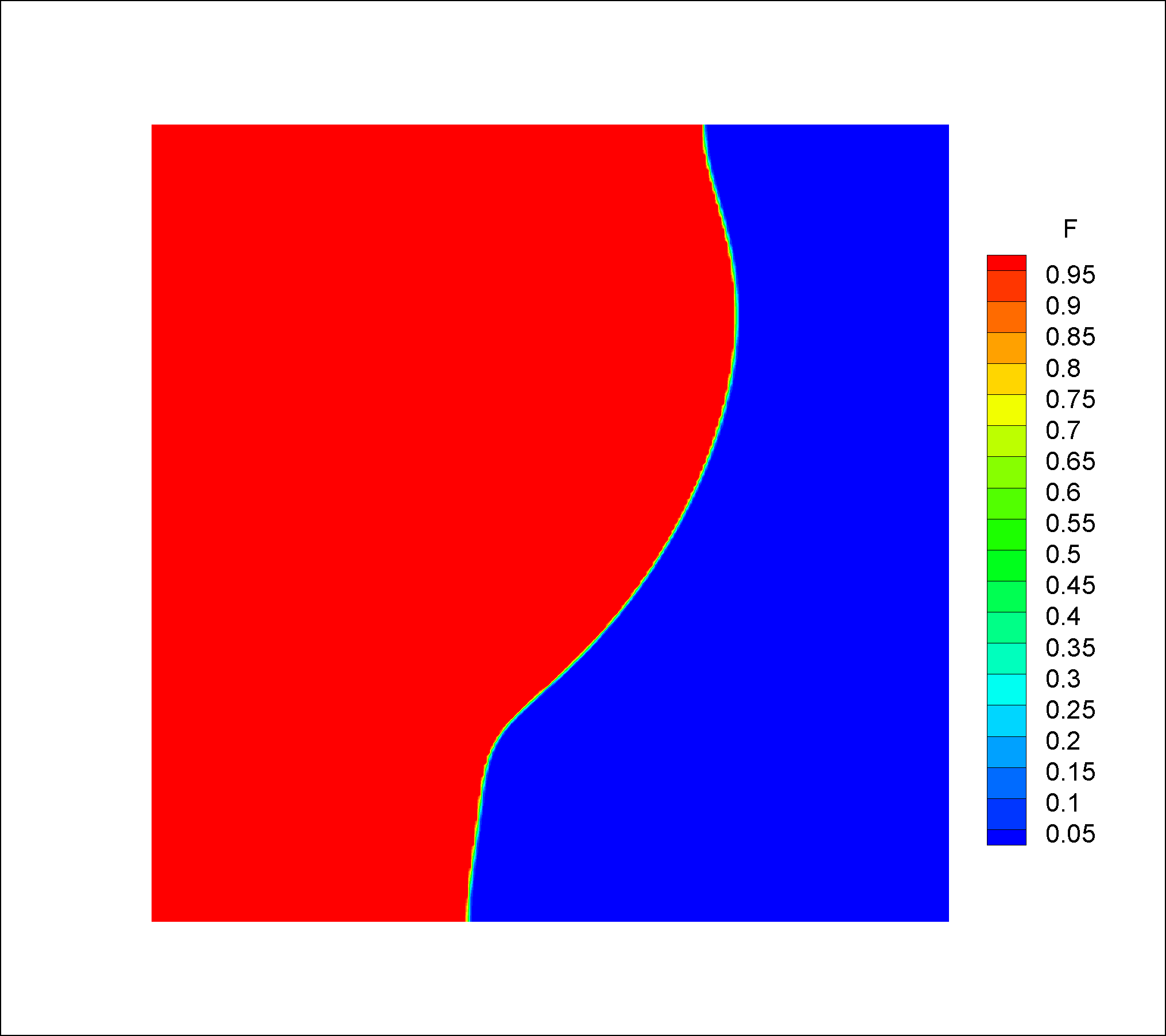}
		\caption{$\text{Fo} = 10.0$ (Liquid fraction)}
		\label{fig:13-panel-4}
	\end{subfigure}\hfill
	\begin{subfigure}[b]{0.31\textwidth}
		\centering
		\includegraphics[width=\linewidth]{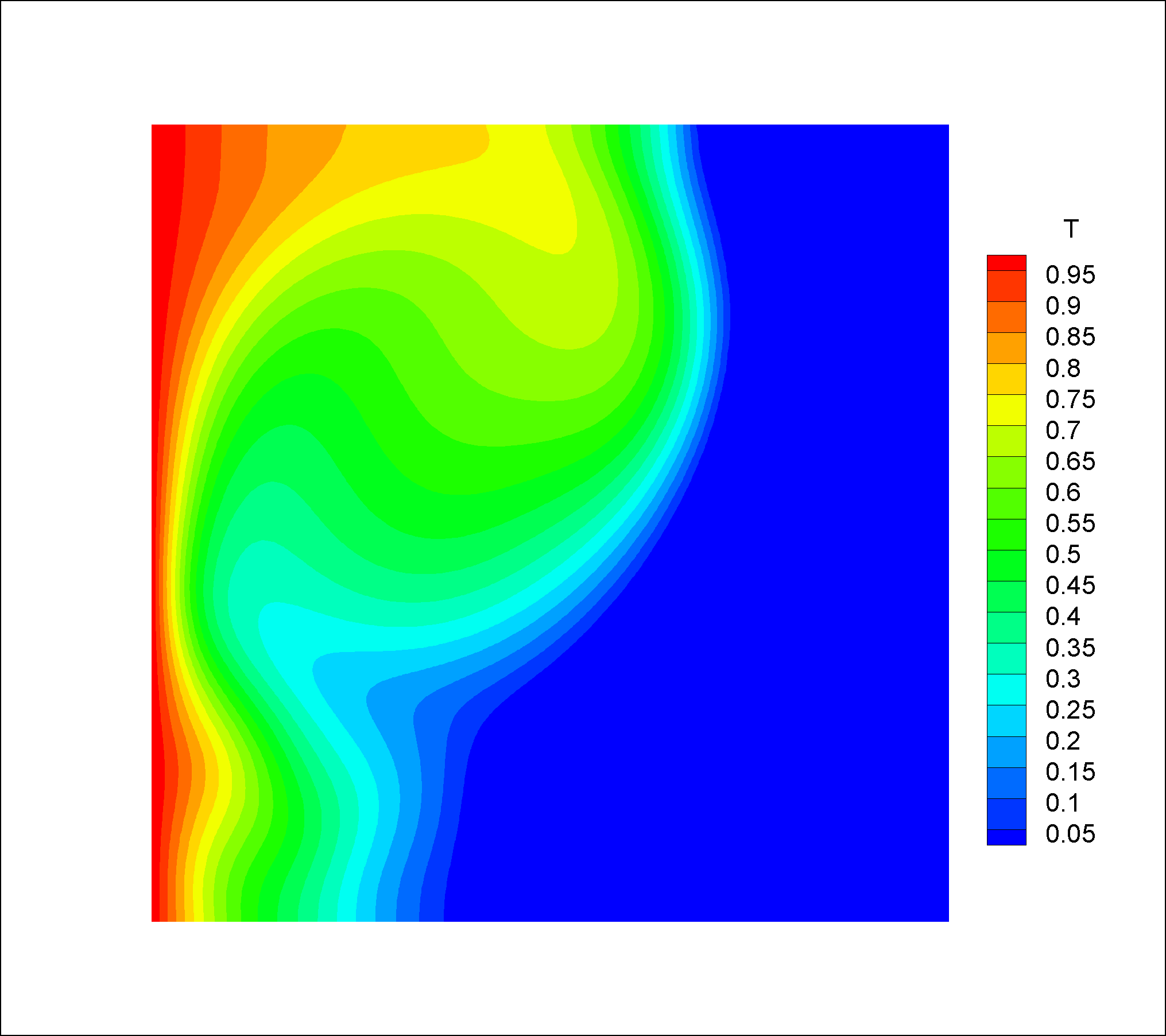}
		\caption{$\text{Fo} = 10.0$ (Temperature)}
		\label{fig:13-panel-5}
	\end{subfigure}\hfill
	\begin{subfigure}[b]{0.31\textwidth}
		\centering
		\includegraphics[width=\linewidth]{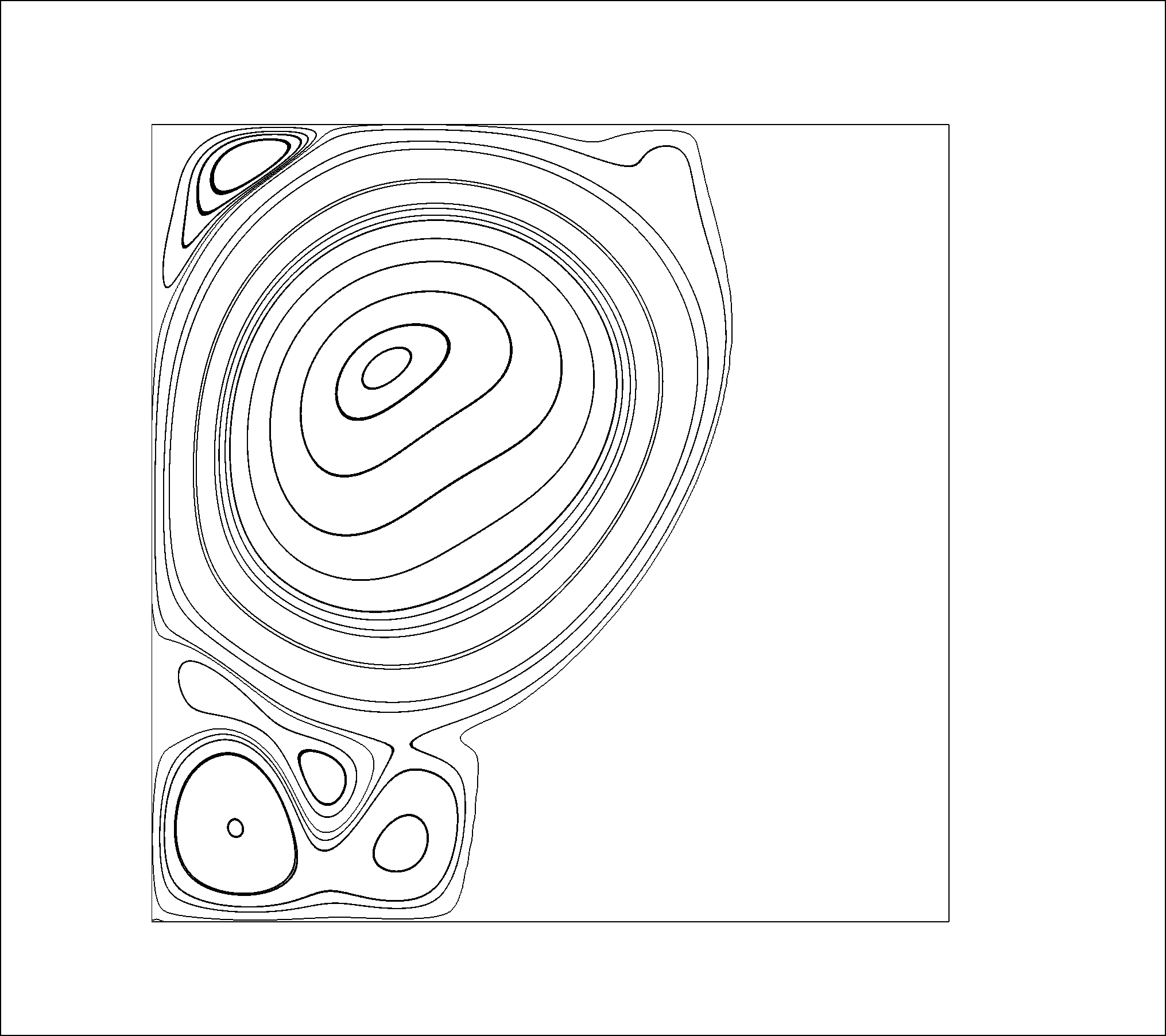}
		\caption{$\text{Fo} = 10.0$ (Streamline)}
		\label{fig:13-panel-6}
	\end{subfigure}

	\vspace{0.5em}

	\begin{subfigure}[b]{0.31\textwidth}
		\centering
		\includegraphics[width=\linewidth]{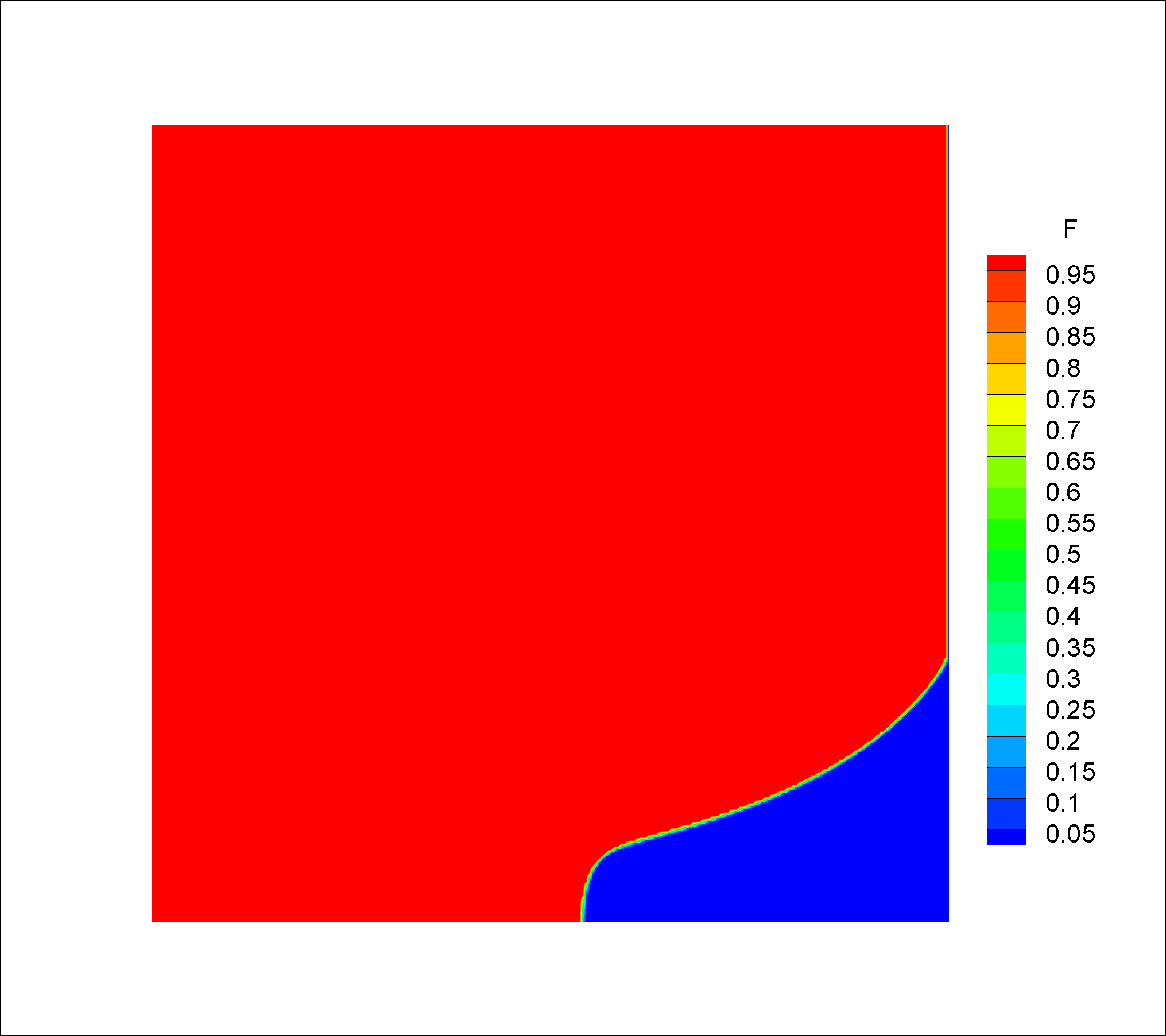}
		\caption{$\text{Fo} = 20.0$ (Liquid fraction)}
		\label{fig:13-panel-7}
	\end{subfigure}\hfill
	\begin{subfigure}[b]{0.31\textwidth}
		\centering
		\includegraphics[width=\linewidth]{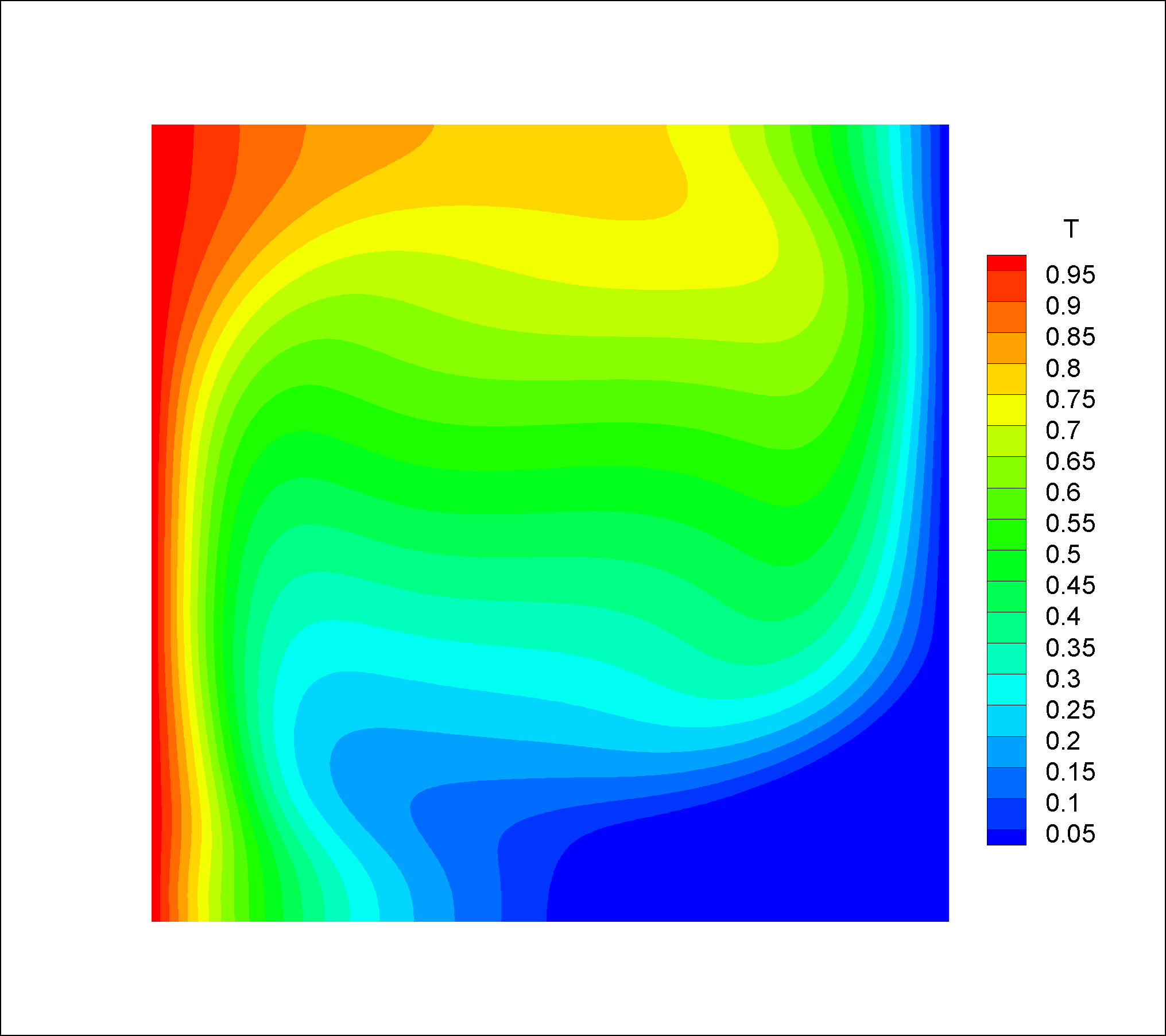}
		\caption{$\text{Fo} = 20.0$ (Temperature)}
		\label{fig:13-panel-8}
	\end{subfigure}\hfill
	\begin{subfigure}[b]{0.31\textwidth}
		\centering
		\includegraphics[width=\linewidth]{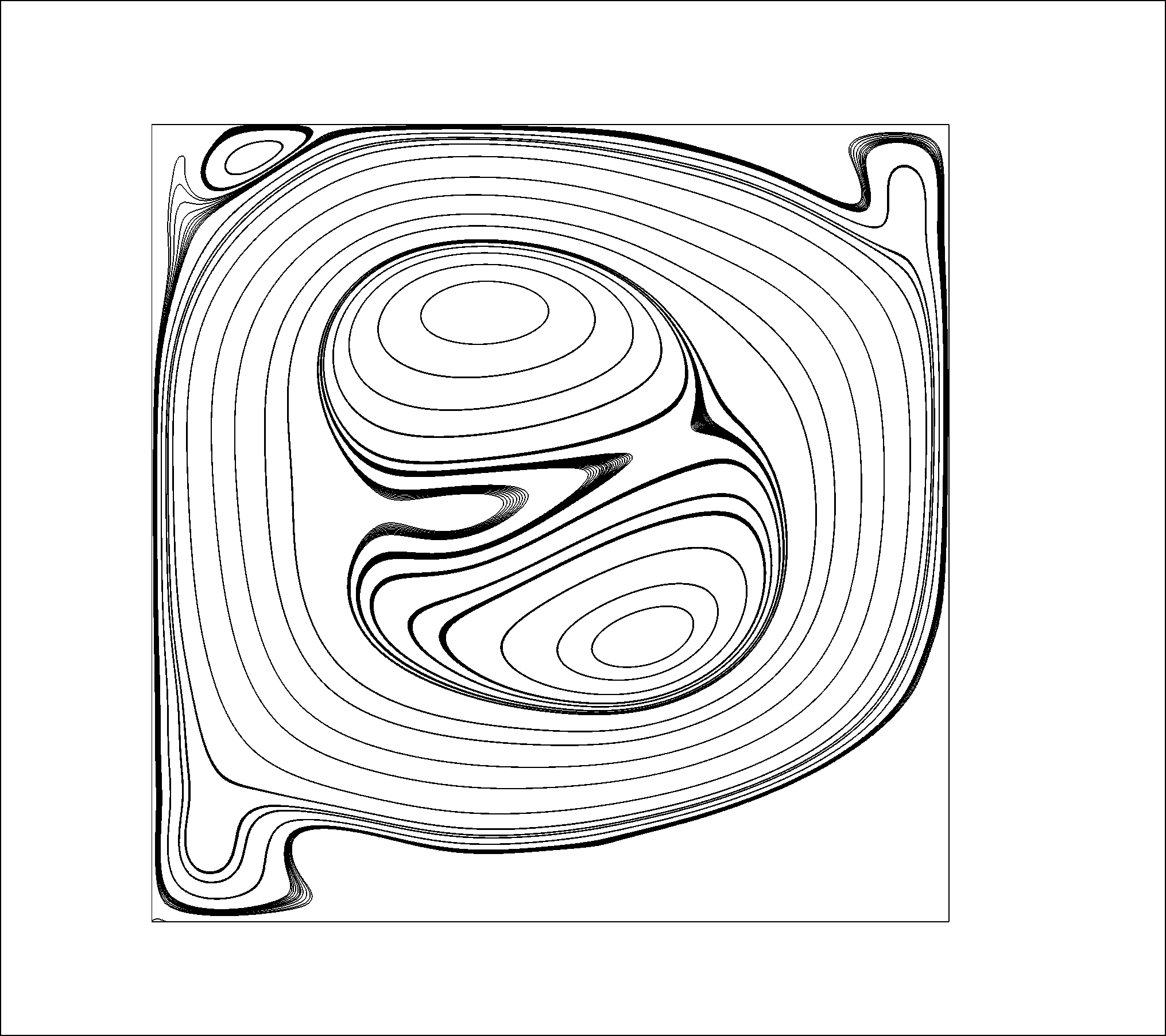}
		\caption{$\text{Fo} = 20.0$ (Streamline)}
		\label{fig:13-panel-9}
	\end{subfigure}

	\caption{Case1:$Pr=0.02,Ste=0.01,Ra=2.5\times 10^5$. Liquid-fraction contours, temperature distributions, and streamlines during the convection melting process for Case 2 at various Fourier numbers: (a) $\text{Fo} = 4.0$; (b) $\text{Fo} = 10.0$; (c) $\text{Fo} = 20.0$.}
	\label{fig:case2_contours}
\end{figure}

\begin{figure}[htbp]
	\centering
	\includegraphics[width=0.62\textwidth]{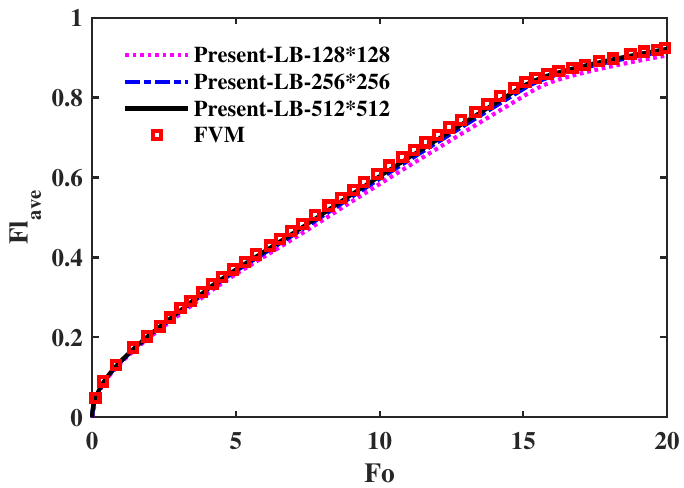}
	\caption{Case2:$Pr=0.02,Ste=0.01,Ra=2.5\times 10^5$. The total liquid fraction at different Fourier numbers: the present LB model and the FVM results~\cite{Mencinger2004} }
	\label{fig:case2_fraction}
\end{figure}

Figure~\ref{fig:case2_contours} illustrates the liquid-fraction contours, temperature fields, and streamlines for Case~2. In this regime, natural convection rapidly overtakes conduction as the governing heat-transfer mechanism, giving rise to significantly more complex recirculating flow structures. The temporal evolution of the total liquid fraction is depicted in Figure~\ref{fig:case2_fraction}. Owing to the pronounced oscillations in the benchmark Nusselt number reported by Mencinger~\cite{Mencinger2004}, the data of their result is difficult to be extracted. Thus, the validation here focuses primarily on the global liquid fraction. As the grid resolution increases, the predicted liquid fraction yields excellent agreement with the reference data. Overall, these findings substantiate the accuracy and robustness of the proposed regularized lattice Boltzmann (RLB) model in capturing convection-dominated melting dynamics.

\subsection{Convection melting in a 3D rectangular cuboid with high Rayleigh number}

\begin{figure}[htbp]
	\centering
	\includegraphics[width=0.62\textwidth]{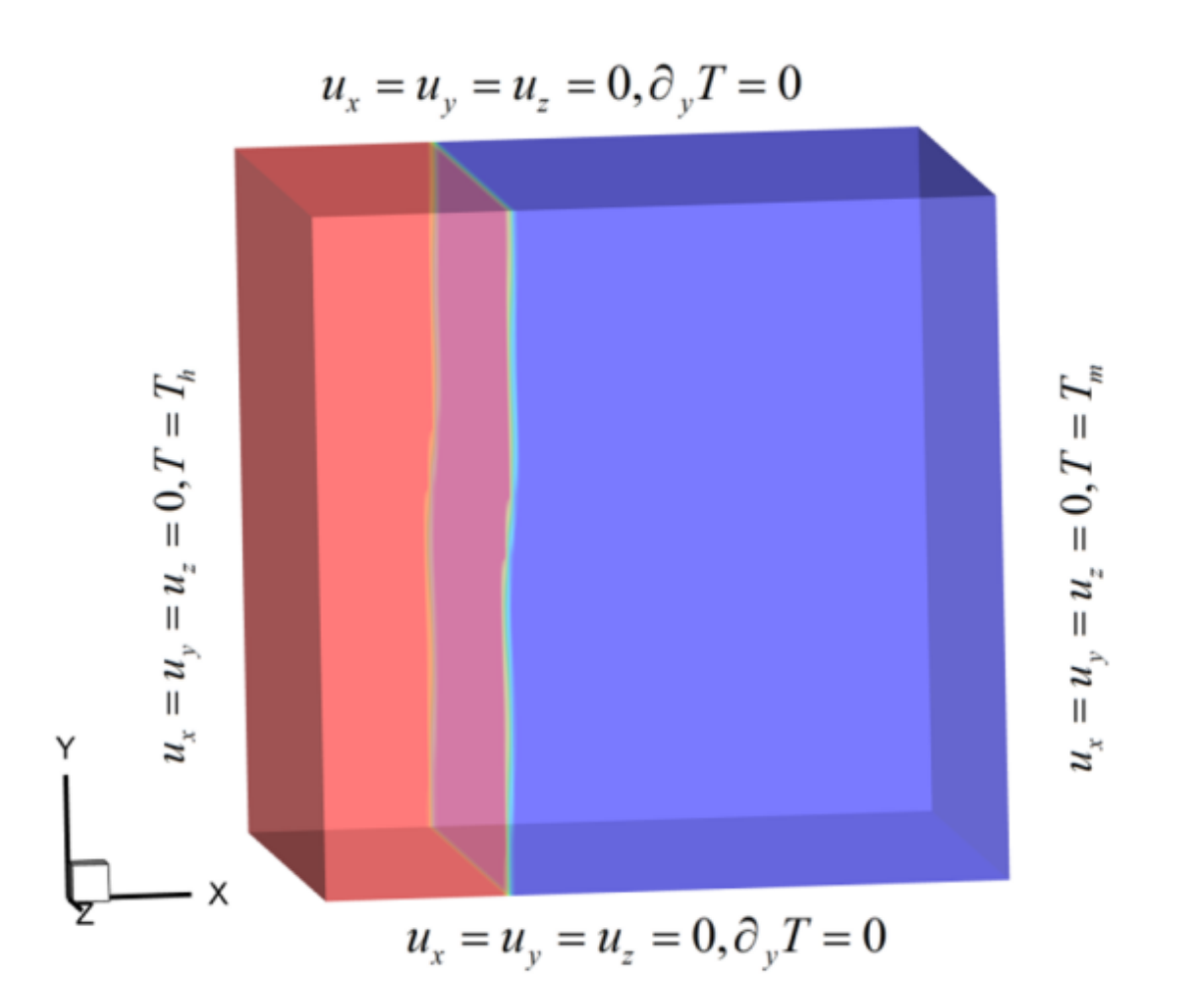}
	\caption{Schematic diagram of the 3D cavity melting:The left wall is hot wall and no-slip boundary, the right wall is cold wall and no-slip boundary, other walls are zero-normal temperature gradient and no-slip boundaries.
	}
	\label{fig:3d_cavity_schematic}
\end{figure}

In this subsection, the melting process within a three-dimensional (3D) rectangular cuboid is investigated, in which a fully three-dimensional flow with no-slip conditions for the velocity field and adiabatic conditions for the temperature field on the $z$-boundary walls. The boundary conditions in the $x$- and $y$-directions remain identical to those employed in the 2D cavity melt example. The computational configuration is schematically illustrated in Fig.~\ref{fig:3d_cavity_schematic}.

The computational performance of the present three dimensional CUDA C solver---coupling a D3Q27 lattice for the velocity field with a D3Q7 lattice for the temperature field---was evaluated on NVIDIA A800 GPU (80~GB, peak memory bandwidth: $2039.04\,\text{GB/s}$, FP64 compute: $9.7\,\text{TFLOPS}$). Resolving both distribution functions in double precision imposes severe register pressure. When enforcing \texttt{\_\_launch\_bounds\_\_(256)}, the excessive register allocation per thread restricts the active warps to only 8 per streaming multiprocessor (SM), yielding low warp occupancy that fails to effectively hide high memory latency ($\sim 500$ cycles). To address this, we employed the compiler flag \texttt{-maxrregcount=168} to constrain per-thread register usage, thereby increasing the concurrent warps to 12 per SM. Although this introduces minor register spills into local memory, the substantially improved latency-hiding capability yields a notable net gain in throughput. As detailed in Table~\ref{table:1GPU}, the performance scales monotonically with grid size, reaching up to $1833.05\,\text{MNUPS}$ and a memory bandwidth efficiency of $59.69\%$ on the $512^3$ mesh. The present performance with two distribution functions for both the flow field and temperature are comparable with those performance in many reported literatures(Table 6 in~\citep{Watanabe2026}).  Additionally, it is found that the CUDA kernels implementing Guo's extrapolation boundary condition for both the flow and temperature fields account for up to $10\%$ of the total computational time, indicating that these boundary kernels are heavily memory-bound.

\begin{table}[htbp]
	\centering
\small
	\captionsetup{justification=centering}
	\caption{The parallel performance of the three dimensional code on single A800 (Peak Bandwidth = 2039.04 GB/s)}
	\label{table:1GPU}
	\begin{tabular}{lcccc}
		\hline
		& $128 \times 128 \times 128$ & $256 \times 256 \times 256$ & $384 \times 384 \times 384$ &  $512 \times 512 \times 512$ \\
		\hline
		MNUPS & 1203.43 & 1603.01 & 1742.15 & 1833.05   \\
		Effective BW & 799.08 & 1064.39 & 1156.79 & 1217.14  \\
		Efficiency of BW & 39.19\% & 52.20\% & 56.73\% & 59.69\%  \\
		\hline
	\end{tabular}
\end{table}

For the multi-GPU implementation, the computational domain is decomposed along the $z$-direction. Each subdomain consists of an inner domain, a halo domain, and a ghost layer. The inter-subdomain communication of particle distribution functions is partially hidden by overlapping it with the computation of the inner domain. Both the weak and strong scaling of the multi-GPU code are evaluated. As listed in Table~\ref{tab:performance_weakscaling}, the weak scaling performance is measured with a fixed workload of $256\times256\times 1024$ nodes per device, where the parallel efficiency remains close to $100\%$. The strong scaling results are presented in Tables~\ref{tab:performance_strongscaling1} and~\ref{tab:performance_strongscaling2}. As the number of GPUs increases, the parallel efficiency decreases slightly owing to the growing exposed communication time with lower compuation load of the inner domain. Furthermore, comparing the results of the two grid sizes in Tables~\ref{tab:performance_strongscaling1} and~\ref{tab:performance_strongscaling2} shows that parallel efficiency improves for the larger domain, as the heavier inner-domain computation allows a greater portion of the communication latency to be concealed.

\begin{table}[htbp]
	\centering
\small
	\caption{Weak scaling efficiency up to 8 GPU cards, 256*256*1024 nodes on each device}
	\label{tab:performance_weakscaling}
	\begin{tabular}{lcccc}
		\toprule
		$256 \times 256 \times 1024$ & 2 & 4 & 6 & 8 \\
		\midrule
		MNUPS\_Total    & 3246.09  & 6480.98  & 9716.73  & 12957.85 \\
		MNUPS\_Average  & 1623.05  & 1620.25  & 1619.46  & 1619.73  \\
		BW\_Total       & 2155.41  & 4303.37  & 6451.91  & 8604.02  \\
		BW\_Average     & 1077.71  & 1075.84  & 1075.32  & 1075.50  \\
		BW Efficiency   & 52.85\%  & 52.76\%  & 52.74\%  & 52.75\%  \\
		Parallelism     & 100.0\%  & 99.8\%   & 99.8\%   & 99.8\%   \\
		\bottomrule
	\end{tabular}
\end{table}

\begin{table}[htbp]
	\centering
\small
	\caption{Strong scaling efficiency up to 8 GPU cards, 256*256*1024 nodes in total}
	\label{tab:performance_strongscaling1}
	\begin{tabular}{lccccc}
		\toprule
		$256 \times 256 \times 1024$ & 1 & 2 & 4 & 6 & 8 \\
		\midrule
		MNUPS\_Total    & 1691.78  & 3171.73  & 5982.97  & 8527.51  & 10267.73 \\
		MNUPS\_Average  & 1691.78  & 1585.87  & 1495.74  & 1421.25  & 1283.47  \\
		BW\_Total       & 1123.34  & 2106.03  & 3972.69  & 5662.27  & 6817.77  \\
		BW\_Average     & 1123.34  & 1053.02  & 993.17   & 943.71   & 852.22   \\
		BW Efficiency   & 55.09\%  & 51.64\%  & 48.71\%  & 46.28\%  & 41.80\%  \\
		Parallelism     & 100.0\%  & 93.7\%   & 88.4\%   & 84.0\%   & 75.9\%   \\
		\bottomrule
	\end{tabular}
\end{table}

\begin{table}[htbp]
	\centering
\small
	\caption{Strong scaling efficiency up to 8 GPU cards, 256*256*1792 nodes in total}
	\label{tab:performance_strongscaling2}
	\begin{tabular}{lccccc}
		\toprule
		$256 \times 256 \times 1792$ & 1 & 2 & 4 & 6 & 8 \\
		\midrule
		MNUPS\_Total    & 1692.48  & 3238.65  & 6267.56  & 9176.15  & 11755.19 \\
		MNUPS\_Average  & 1692.48  & 1619.33  & 1566.89  & 1529.36  & 1469.40  \\
		BW\_Total       & 1123.81  & 2150.46  & 4161.66  & 6092.97  & 7805.44  \\
		BW\_Average     & 1123.81  & 1075.23  & 1040.42  & 1015.50  & 975.68   \\
		BW Efficiency   & 55.11\%  & 52.73\%  & 51.02\%  & 49.80\%  & 47.85\%  \\
		Parallelism     & 100.0\%  & 95.7\%   & 92.6\%   & 90.4\%   & 86.8\%   \\
		\bottomrule
	\end{tabular}
\end{table}

\begin{figure}[htbp]
	\centering
	\begin{subfigure}[b]{0.325\textwidth}
		\centering
		\includegraphics[width=\linewidth]{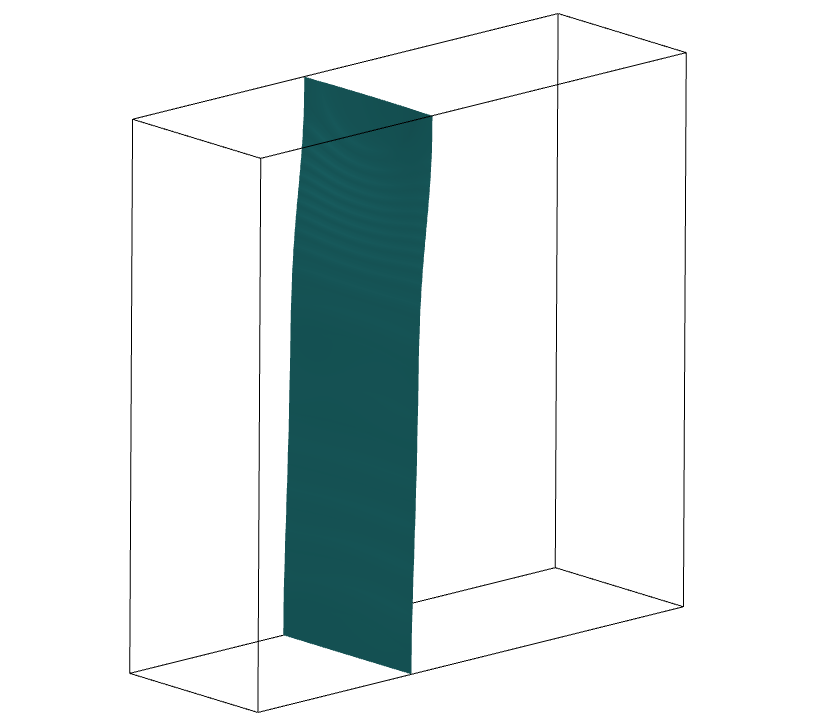}
		\caption{}
		\label{fig:16-panel-1}
	\end{subfigure}\hfill
	\begin{subfigure}[b]{0.325\textwidth}
		\centering
		\includegraphics[width=\linewidth]{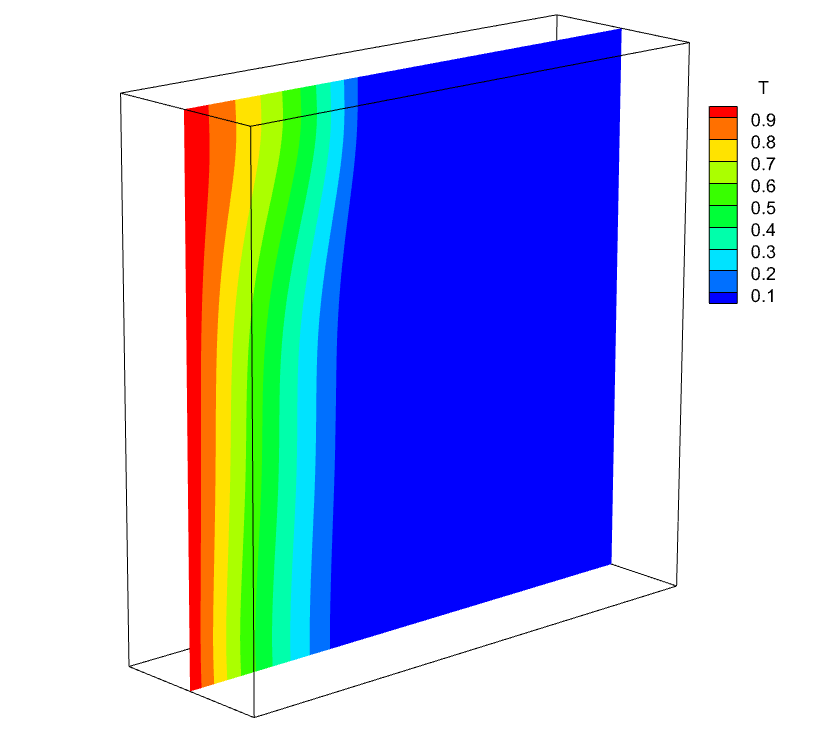}
		\caption{}
		\label{fig:16-panel-2}
	\end{subfigure}\hfill
	\begin{subfigure}[b]{0.325\textwidth}
		\centering
		\includegraphics[width=\linewidth]{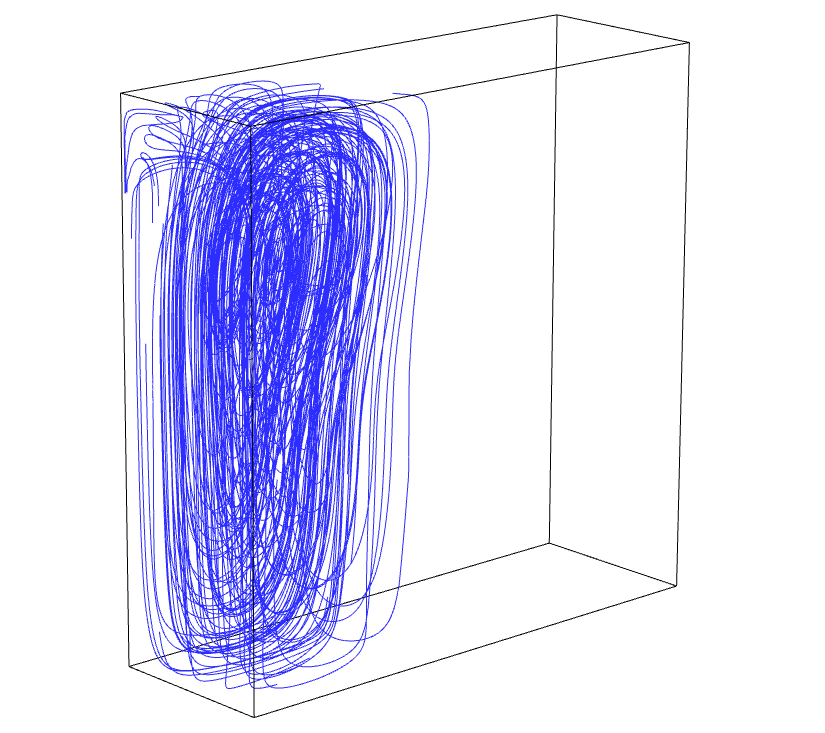}
		\caption{}
		\label{fig:16-panel-3}
	\end{subfigure}

	\vspace{0.8ex}

	\begin{subfigure}[b]{0.325\textwidth}
		\centering
		\includegraphics[width=\linewidth]{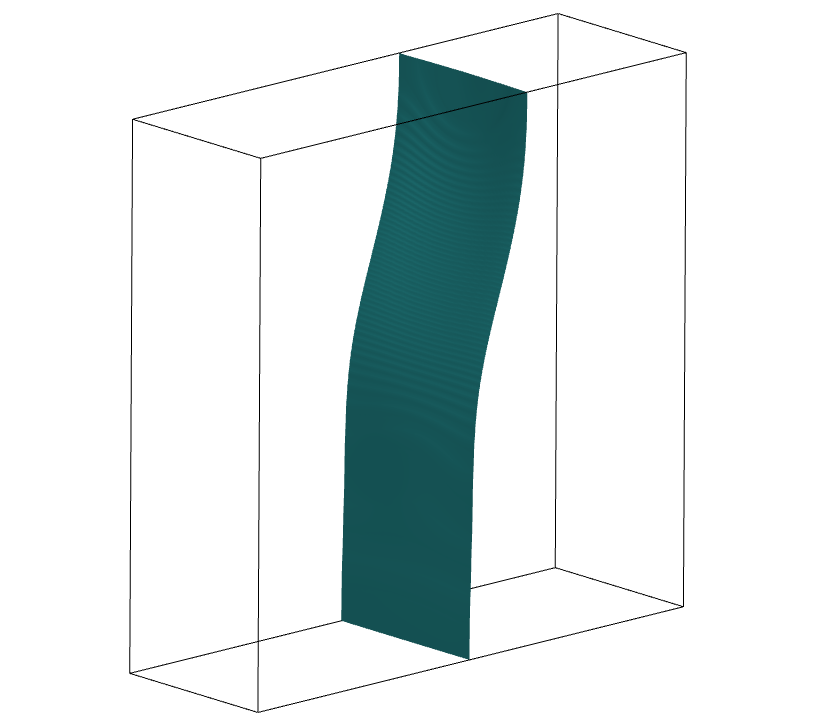}
		\caption{}
		\label{fig:16-panel-4}
	\end{subfigure}\hfill
	\begin{subfigure}[b]{0.325\textwidth}
		\centering
		\includegraphics[width=\linewidth]{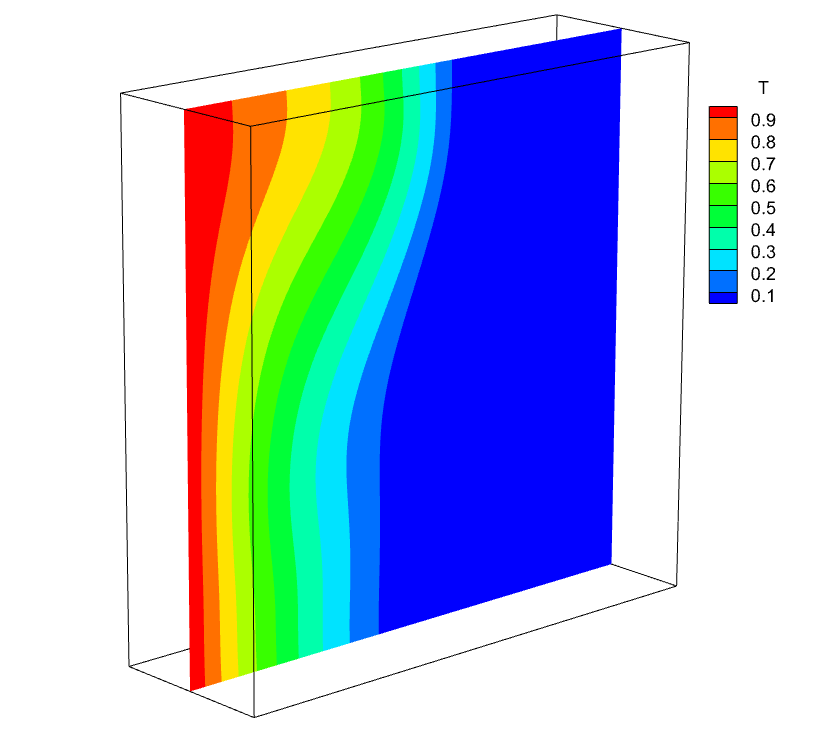}
		\caption{}
		\label{fig:16-panel-5}
	\end{subfigure}\hfill
	\begin{subfigure}[b]{0.325\textwidth}
		\centering
		\includegraphics[width=\linewidth]{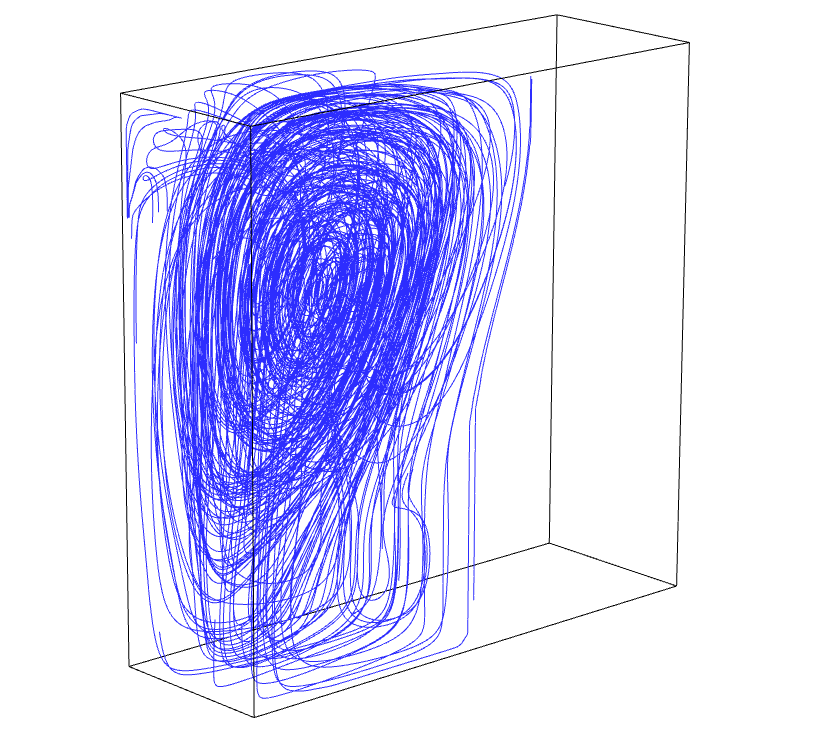}
		\caption{}
		\label{fig:16-panel-6}
	\end{subfigure}

	\vspace{0.8ex}

	\begin{subfigure}[b]{0.325\textwidth}
		\centering
		\includegraphics[width=\linewidth]{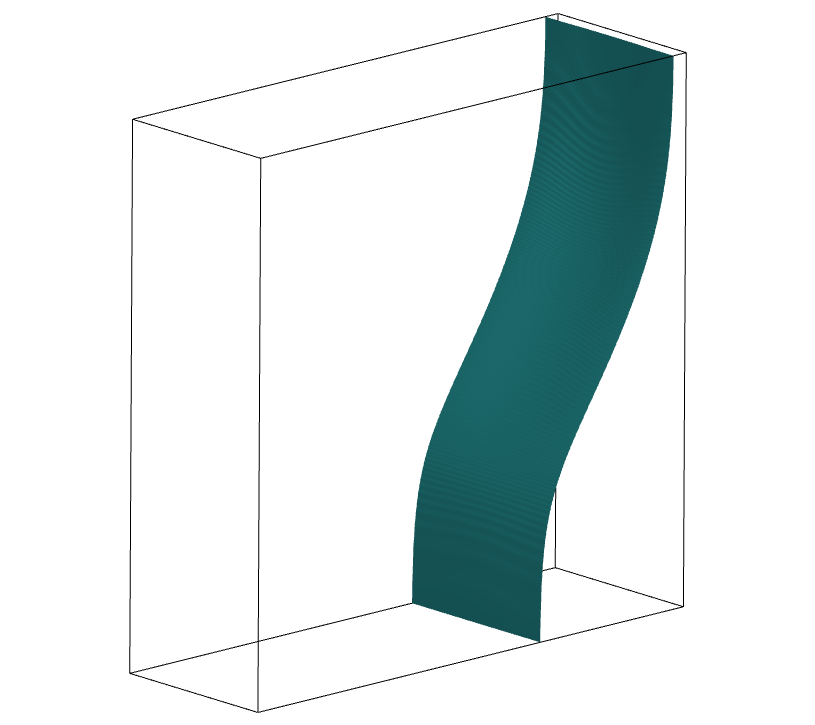}
		\caption{}
		\label{fig:16-panel-7}
	\end{subfigure}\hfill
	\begin{subfigure}[b]{0.325\textwidth}
		\centering
		\includegraphics[width=\linewidth]{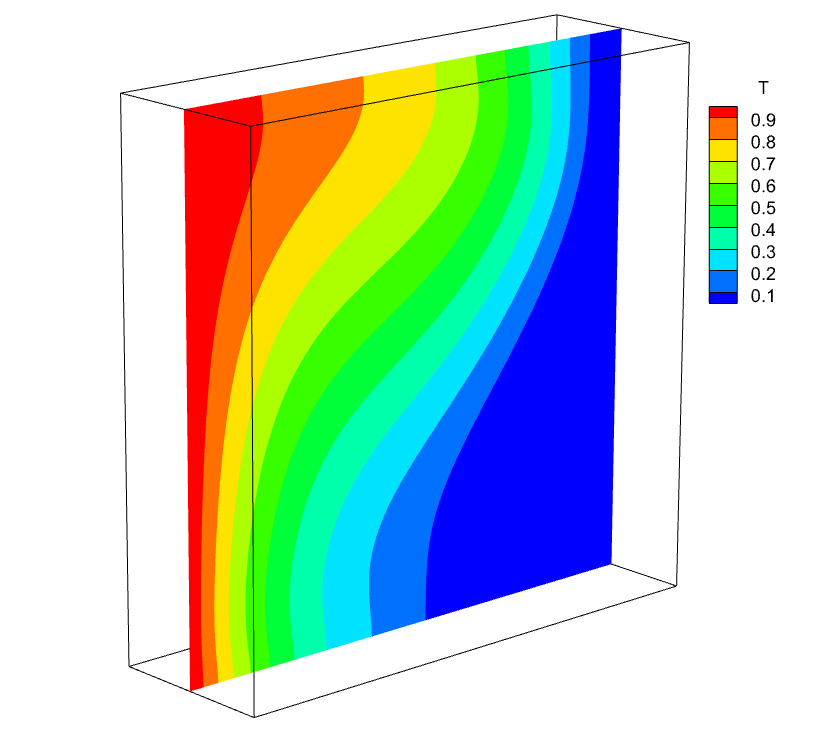}
		\caption{}
		\label{fig:16-panel-8}
	\end{subfigure}\hfill
	\begin{subfigure}[b]{0.325\textwidth}
		\centering
		\includegraphics[width=\linewidth]{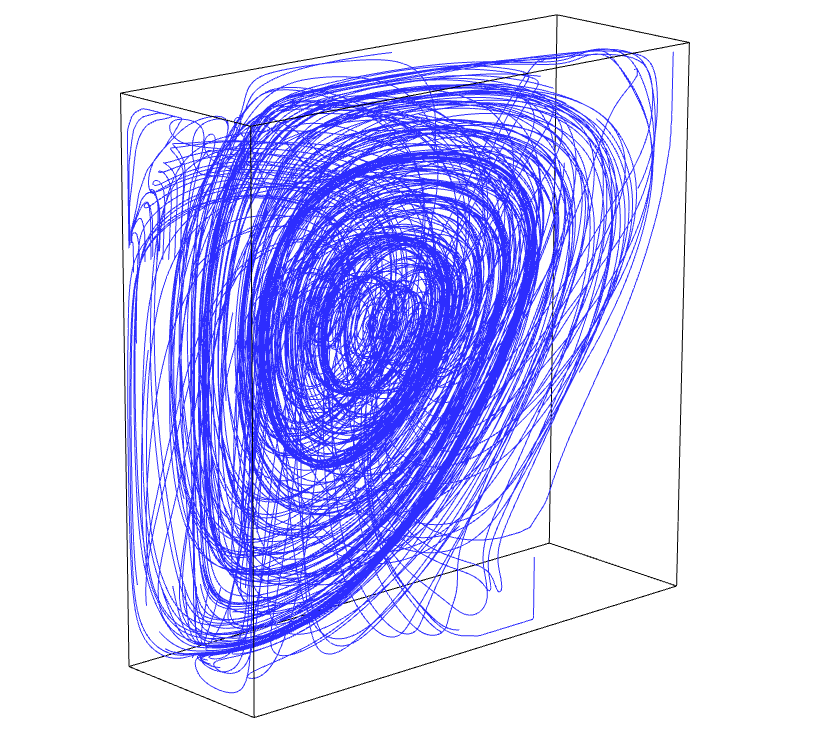}
		\caption{}
		\label{fig:16-panel-9}
	\end{subfigure}

	\caption{Case1:$Pr=0.02,Ste=0.4,Ra=2.5\times 10^4$. Phase interface positions (left column), temperature contours (middle column), and streamlines (right column) at representative dimensionless times for Case~1: (a--c)~$\mathrm{Fo} = 0.2$, (d--f)~$\mathrm{Fo} = 0.4$, and (g--i)~$\mathrm{Fo} = 0.8$.}
	\label{fig:case1_evolution}
\end{figure}

\begin{figure}[htbp]
	\centering
	\begin{subfigure}[b]{0.325\textwidth}
		\centering
		\includegraphics[width=\linewidth]{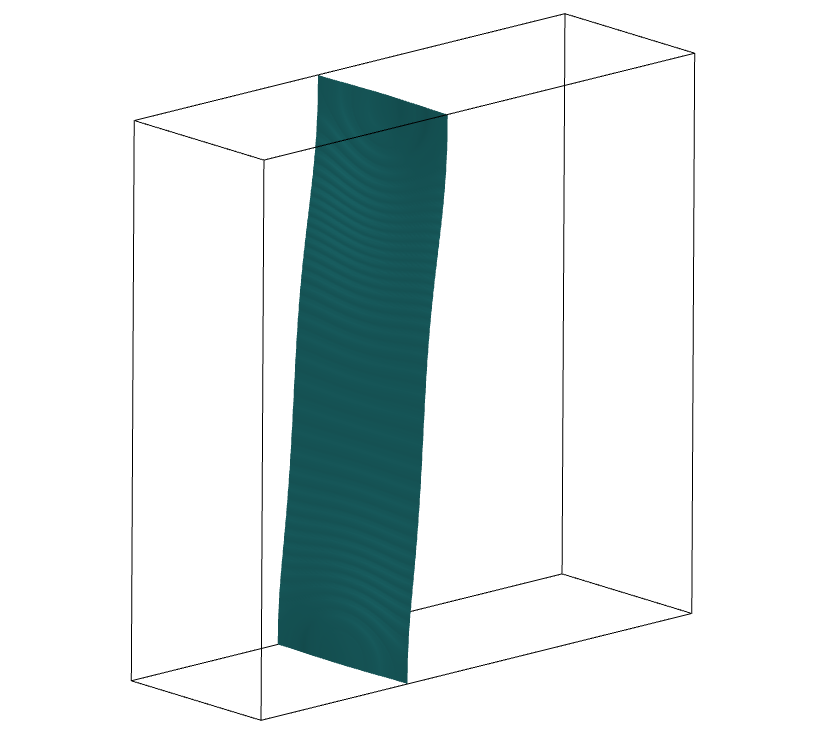}
		\caption{}
		\label{fig:17-panel-1}
	\end{subfigure}\hfill
	\begin{subfigure}[b]{0.325\textwidth}
		\centering
		\includegraphics[width=\linewidth]{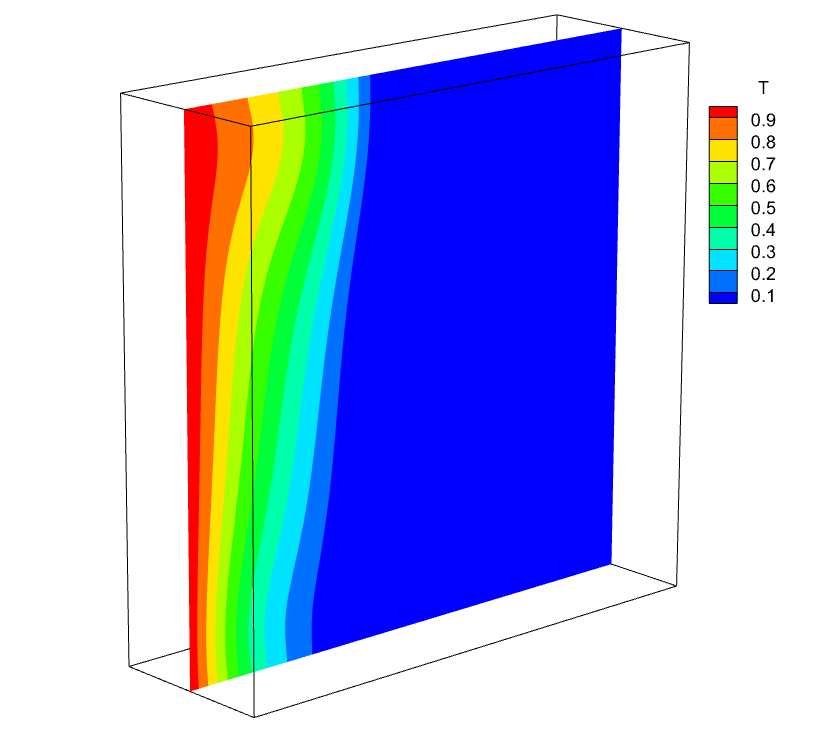}
		\caption{}
		\label{fig:17-panel-2}
	\end{subfigure}\hfill
	\begin{subfigure}[b]{0.325\textwidth}
		\centering
		\includegraphics[width=\linewidth]{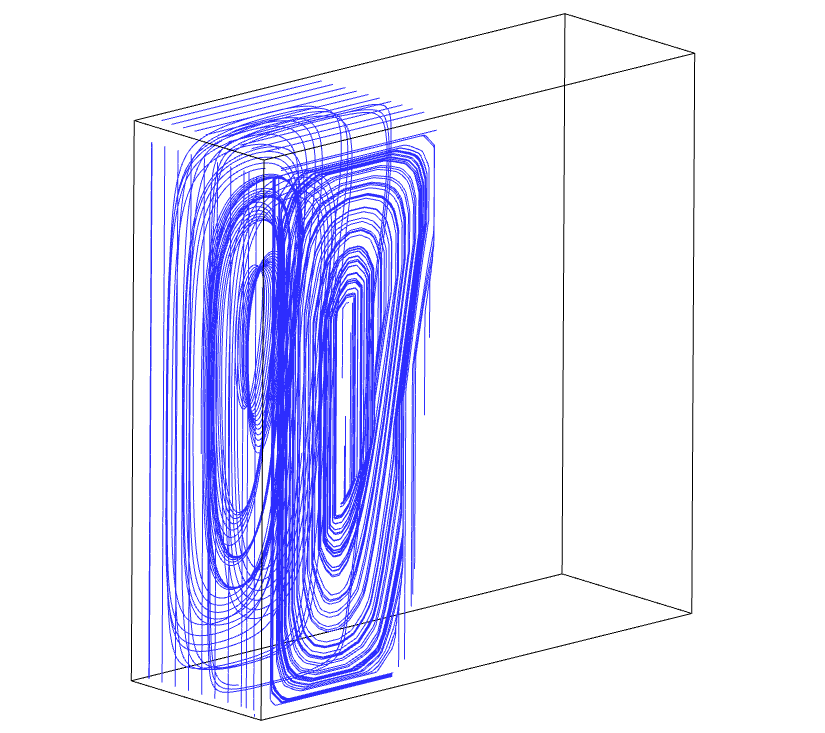}
		\caption{}
		\label{fig:17-panel-3}
	\end{subfigure}

	\vspace{0.8ex} %

	\begin{subfigure}[b]{0.325\textwidth}
		\centering
		\includegraphics[width=\linewidth]{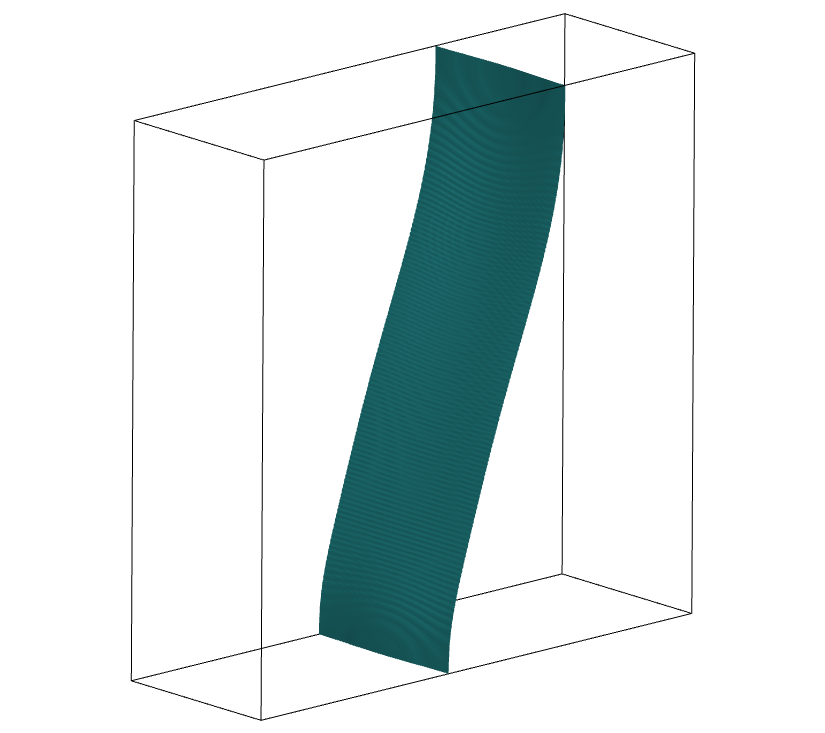}
		\caption{}
		\label{fig:17-panel-4}
	\end{subfigure}\hfill
	\begin{subfigure}[b]{0.325\textwidth}
		\centering
		\includegraphics[width=\linewidth]{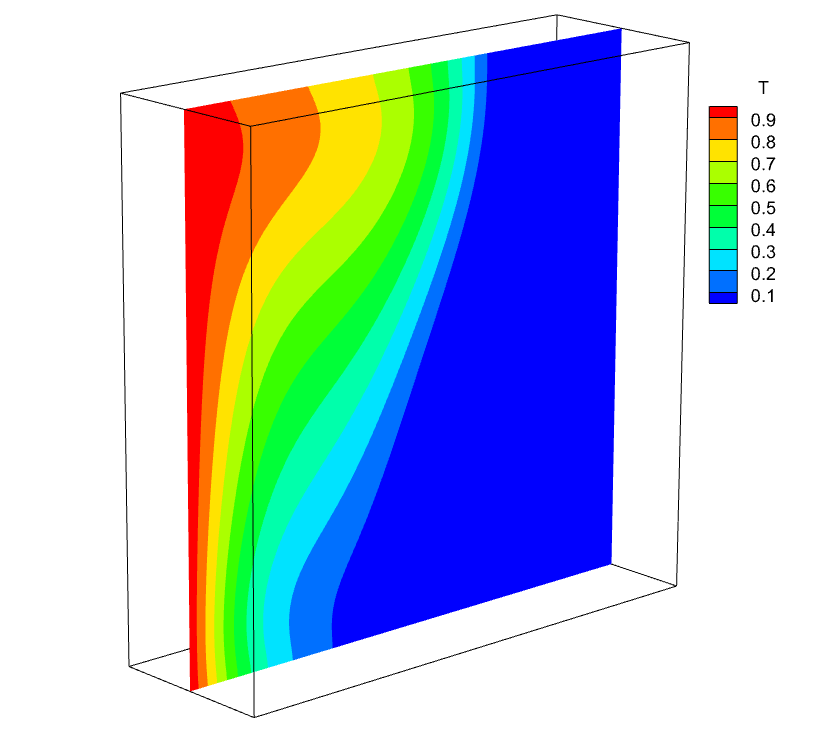}
		\caption{}
		\label{fig:17-panel-5}
	\end{subfigure}\hfill
	\begin{subfigure}[b]{0.325\textwidth}
		\centering
		\includegraphics[width=\linewidth]{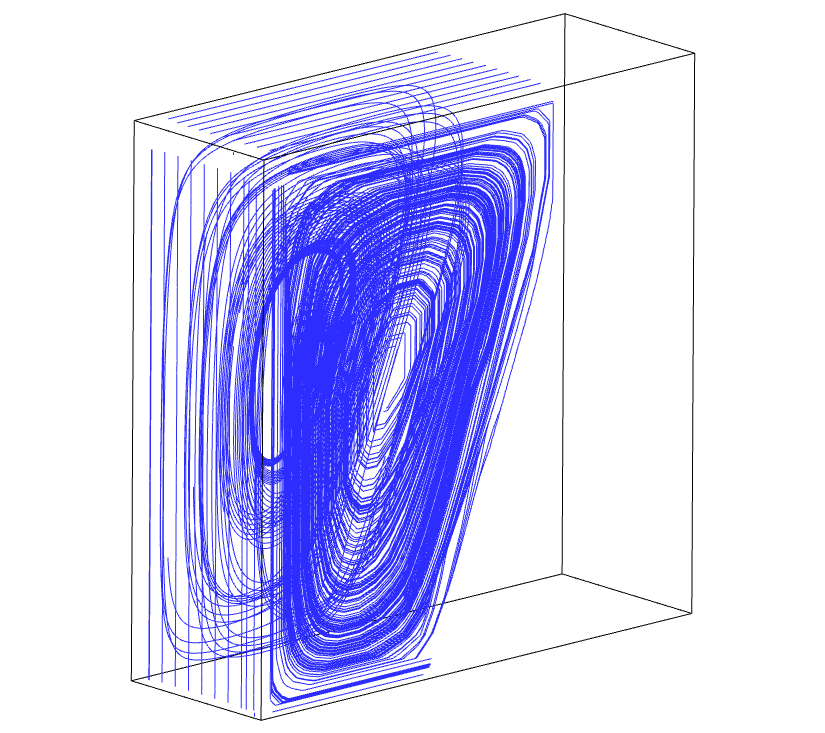}
		\caption{}
		\label{fig:17-panel-6}
	\end{subfigure}

	\vspace{0.8ex}

	\begin{subfigure}[b]{0.325\textwidth}
		\centering
		\includegraphics[width=\linewidth]{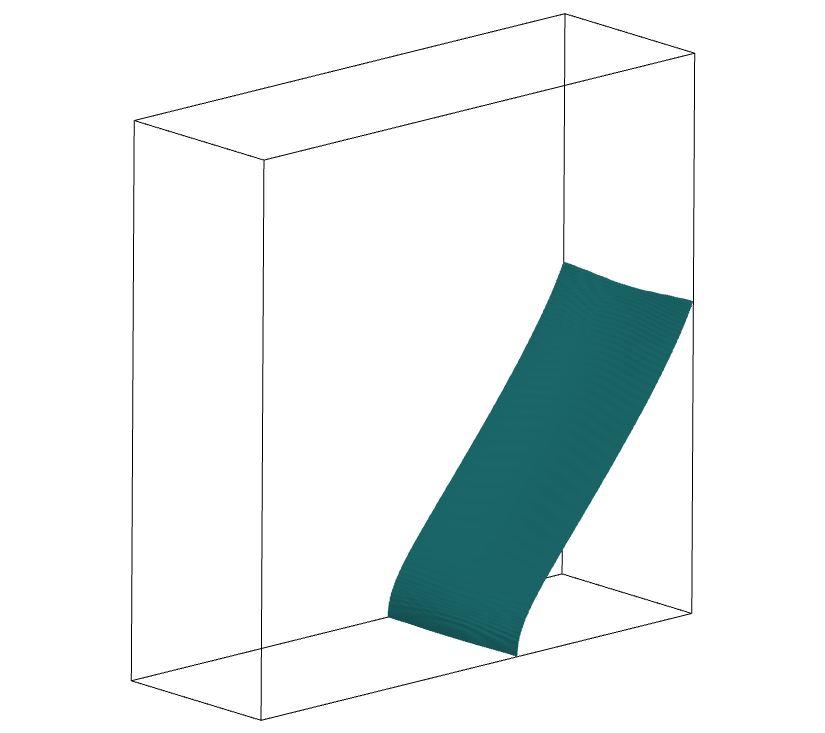}
		\caption{}
		\label{fig:17-panel-7}
	\end{subfigure}\hfill
	\begin{subfigure}[b]{0.325\textwidth}
		\centering
		\includegraphics[width=\linewidth]{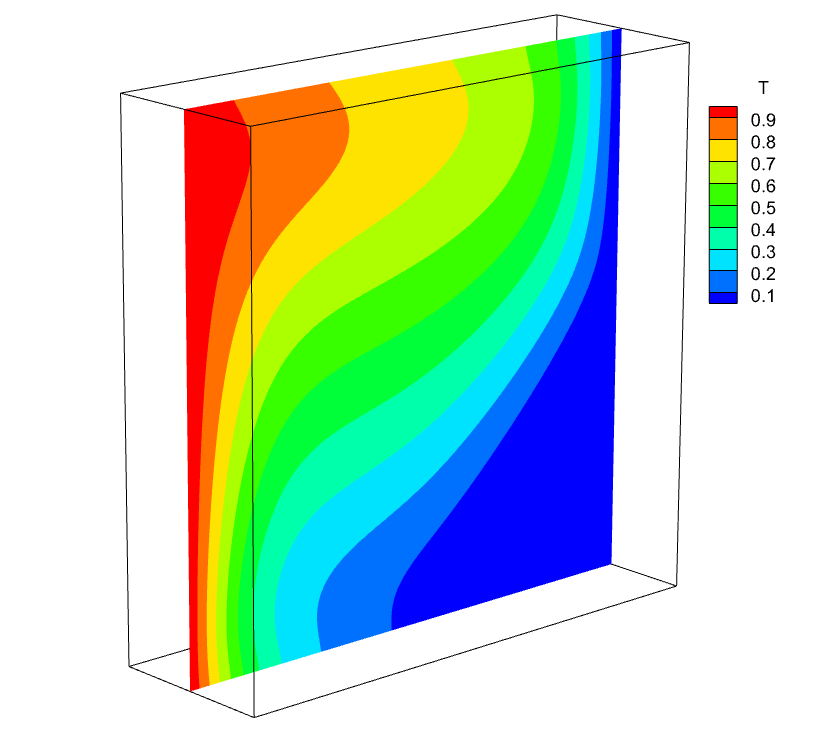}
		\caption{}
		\label{fig:17-panel-8}
	\end{subfigure}\hfill
	\begin{subfigure}[b]{0.325\textwidth}
		\centering
		\includegraphics[width=\linewidth]{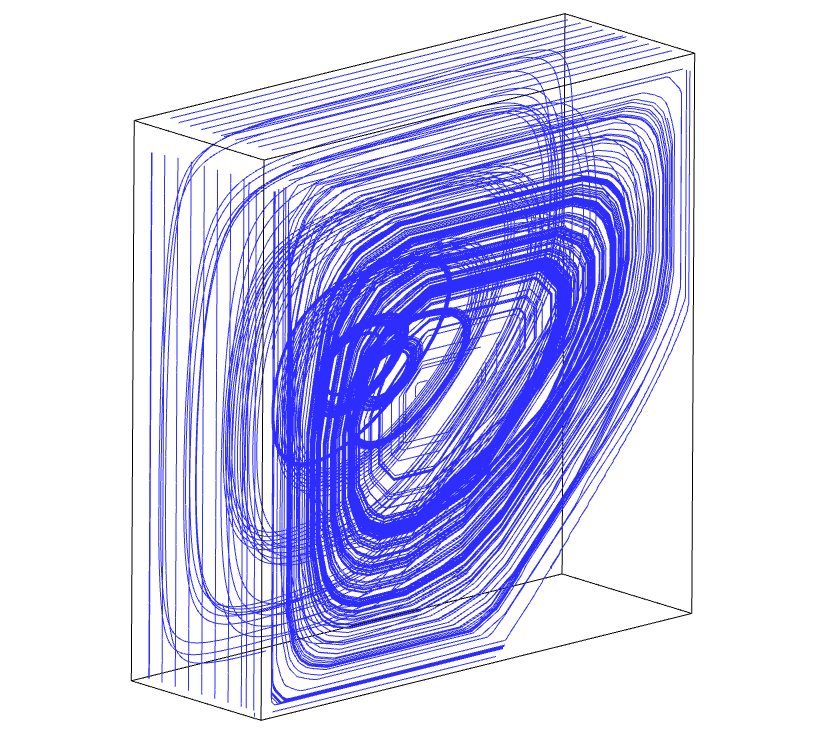}
		\caption{}
		\label{fig:17-panel-9}
	\end{subfigure}

	\caption{Case2:$Pr=10.0,Ste=0.4,Ra=2.5\times 10^4$. Phase interface positions (left column), temperature contours (middle column), and streamlines (right column) at representative dimensionless times for Case~2: (a--c)~$\mathrm{Fo} = 0.2$, (d--f)~$\mathrm{Fo} = 0.4$, and (g--i)~$\mathrm{Fo} = 0.8$.}
	\label{fig:case2_evolution}
\end{figure}

\begin{figure}[htbp]
	\centering
	\begin{subfigure}[b]{0.325\textwidth}
		\centering
		\includegraphics[width=\linewidth]{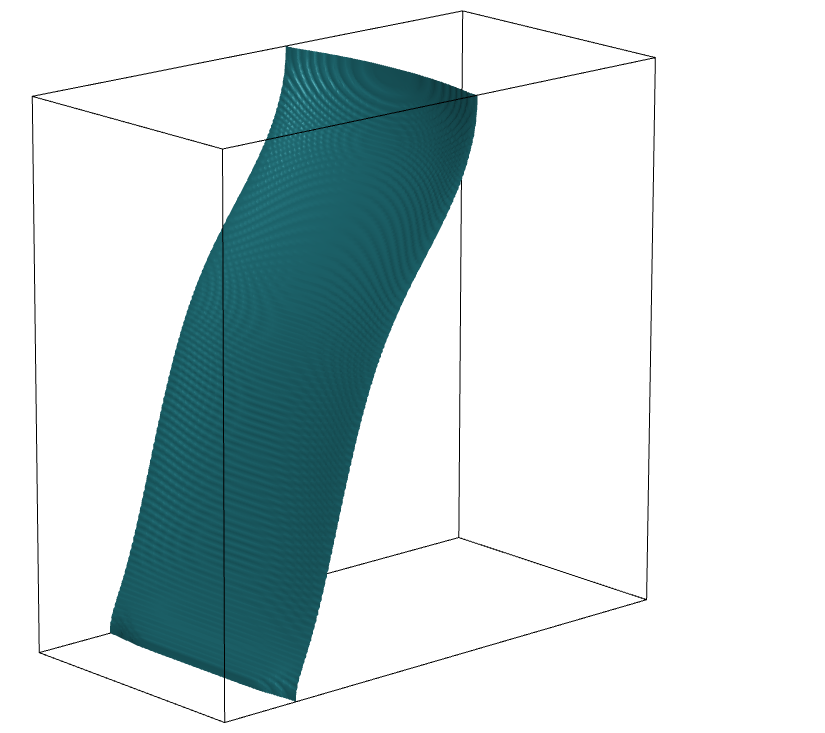}
		\caption{}
		\label{fig:18-panel-1}
	\end{subfigure}\hfill
	\begin{subfigure}[b]{0.325\textwidth}
		\centering
		\includegraphics[width=\linewidth]{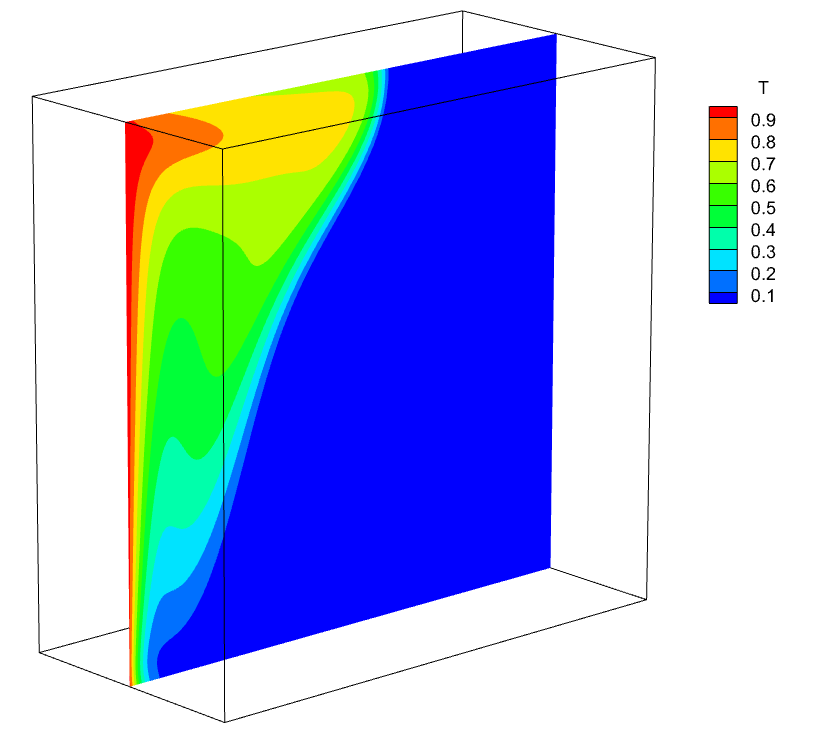}
		\caption{}
		\label{fig:18-panel-2}
	\end{subfigure}\hfill
	\begin{subfigure}[b]{0.325\textwidth}
		\centering
		\includegraphics[width=\linewidth]{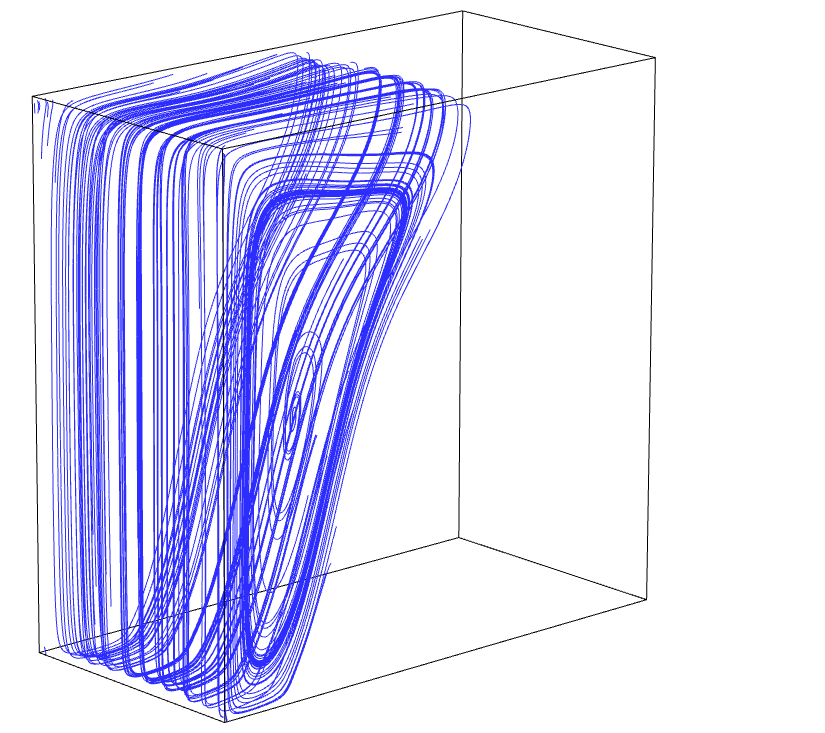}
		\caption{}
		\label{fig:18-panel-3}
	\end{subfigure}

	\vspace{0.8ex} %

	\begin{subfigure}[b]{0.325\textwidth}
		\centering
		\includegraphics[width=\linewidth]{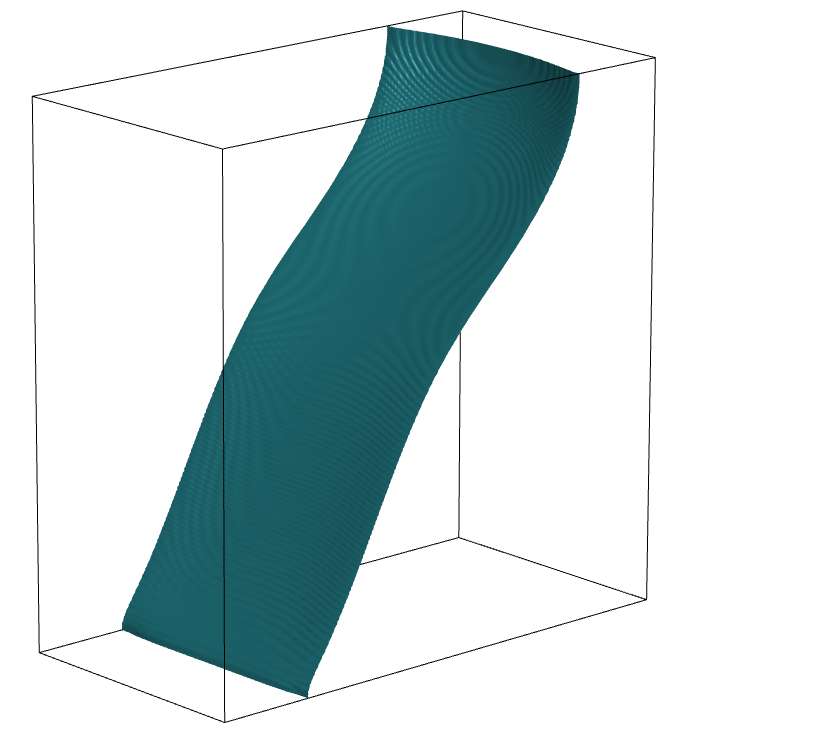}
		\caption{}
		\label{fig:18-panel-4}
	\end{subfigure}\hfill
	\begin{subfigure}[b]{0.325\textwidth}
		\centering
		\includegraphics[width=\linewidth]{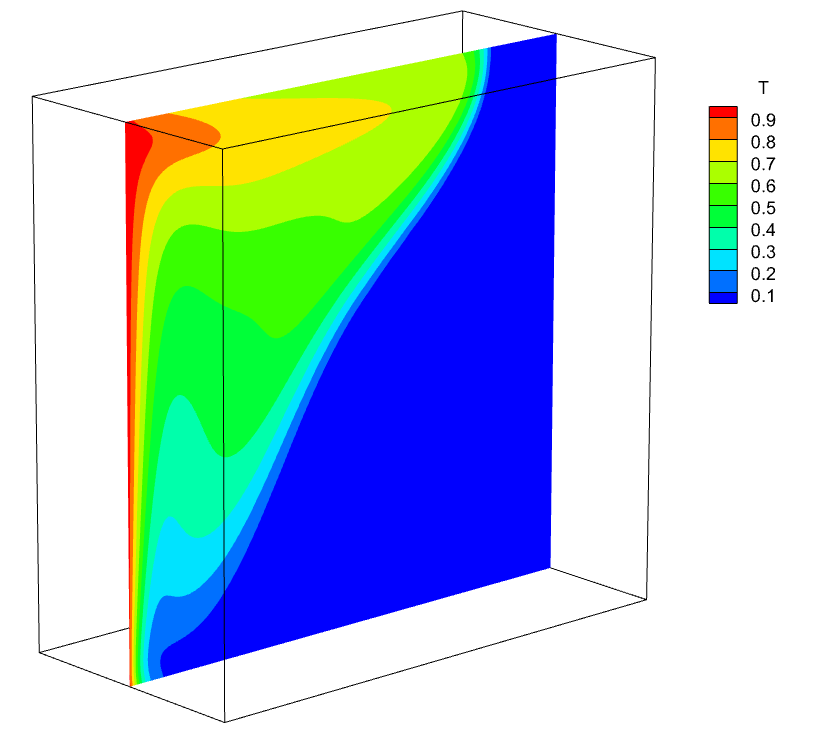}
		\caption{}
		\label{fig:18-panel-5}
	\end{subfigure}\hfill
	\begin{subfigure}[b]{0.325\textwidth}
		\centering
		\includegraphics[width=\linewidth]{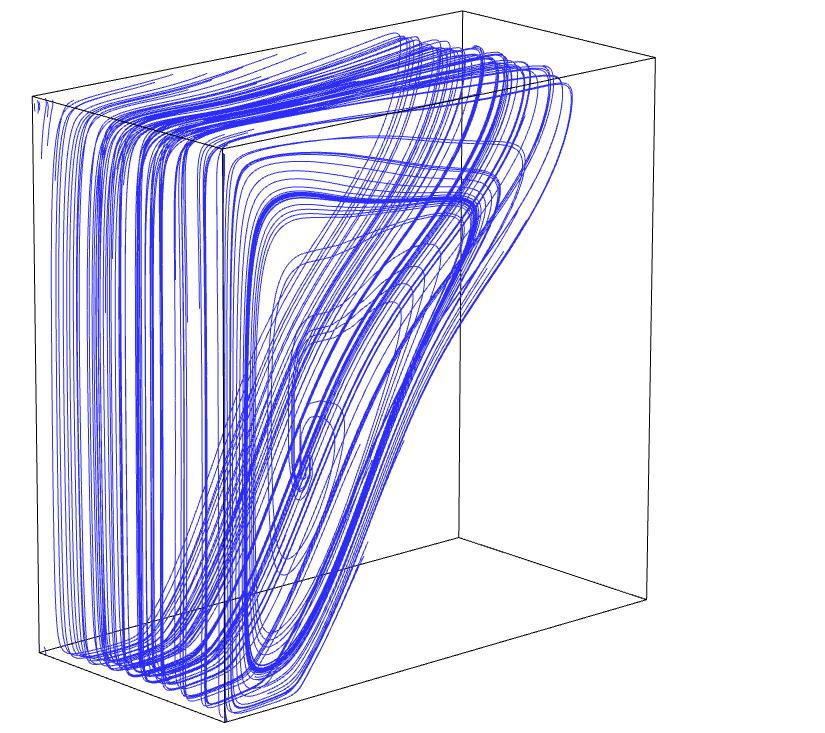}
		\caption{}
		\label{fig:18-panel-6}
	\end{subfigure}

	\vspace{0.8ex}

	\begin{subfigure}[b]{0.325\textwidth}
		\centering
		\includegraphics[width=\linewidth]{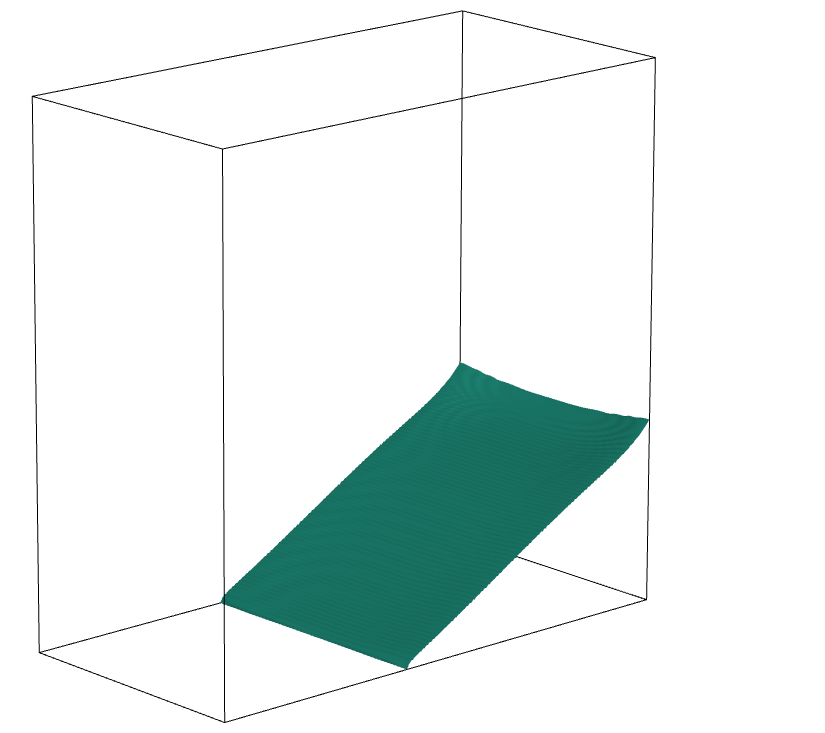}
		\caption{}
		\label{fig:18-panel-7}
	\end{subfigure}\hfill
	\begin{subfigure}[b]{0.325\textwidth}
		\centering
		\includegraphics[width=\linewidth]{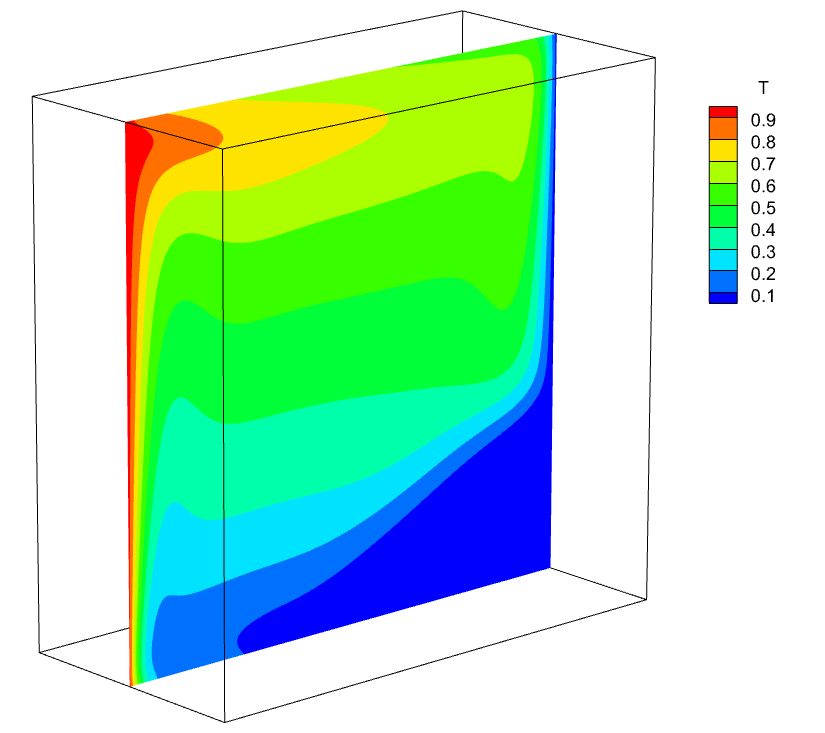}
		\caption{}
		\label{fig:18-panel-8}
	\end{subfigure}\hfill
	\begin{subfigure}[b]{0.325\textwidth}
		\centering
		\includegraphics[width=\linewidth]{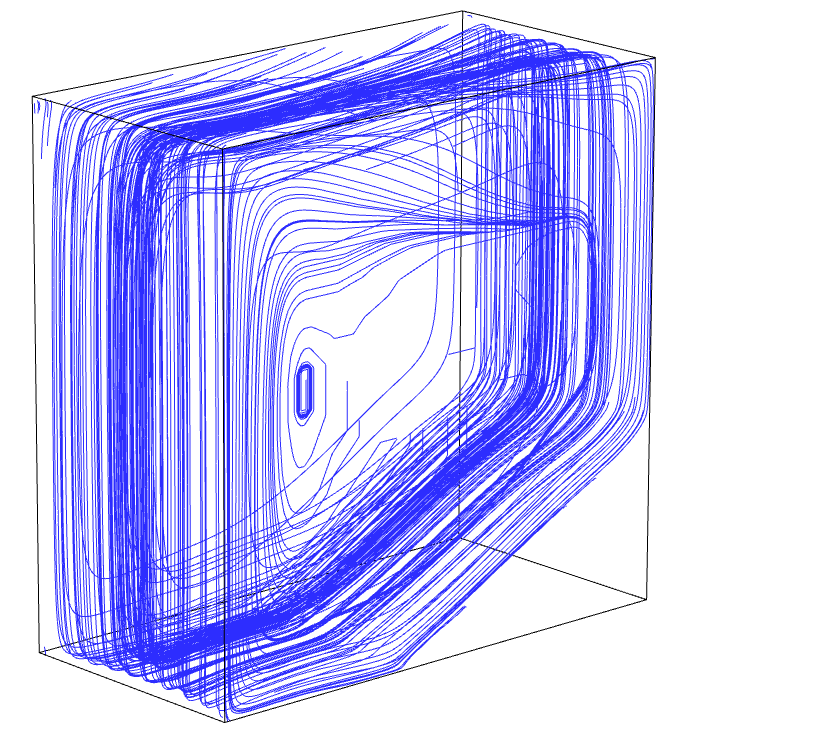}
		\caption{}
		\label{fig:18-panel-9}
	\end{subfigure}

	\caption{Case3:$Pr=10.0,Ste=0.4,Ra=1.0\times 10^6$. Phase interface positions (left column), temperature contours (middle column), and streamlines (right column) at representative dimensionless times for Case~3: (a--c)~$\mathrm{Fo} = 0.1$, (d--f)~$\mathrm{Fo} = 0.15$, and (g--i)~$\mathrm{Fo} = 0.4$.}
	\label{fig:case3_evolution}
\end{figure}

Specifically, the three-dimensional (3D) rectangular cuboid melting process is investigated under three parameter configurations, as summarized in Table~\ref{table:case_parameters}. Figures~\ref{fig:case1_evolution}, \ref{fig:case2_evolution}, and \ref{fig:case3_evolution} depict the temporal evolution of the liquid fraction distributions, temperature contours across selected $x$--$y$ plane, and 3D streamlines for Cases~1--3, respectively.

\begin{table}[htbp]
	\centering
\small
	\caption{Parameters configurations for three dimensional rectangular cuboid}
	\label{table:case_parameters}
	\begin{tabular}{ccccc}
		\hline
		& Pr & Ra & Ste \\
		\hline
		Case 1 &  0.02 & $2.5 \times 10^4$ & 0.4 \\
		Case 2 &  10 & $2.5 \times 10^4$ & 0.4 \\
		Case 3 &  10 & $1 \times 10^6$ & 0.4 \\
		\hline
	\end{tabular}
\end{table}

A comparison between Cases~1 and 2 demonstrates the pronounced influence of the Prandtl number, where $\mathrm{Ra} = 2.5 \times 10^4$ and $\mathrm{Ste} = 0.4$ remain constant while $\mathrm{Pr}$ increases from $0.02$ to $10$. In Case~1 ($\mathrm{Pr} = 0.02$), typical of low-Prandtl liquid metals, heat transfer is predominantly governed by thermal conduction; consequently, the phase interface remains relatively planar and vertical in the early stages, with isotherms stratified almost parallel to the heated wall. Although natural convection initiates within the liquid melt, the circulating flow lacks coherence and exerts minimal interfacial shear.

In sharp contrast, for Case~2 ($\mathrm{Pr} = 10$), the relative contribution of thermal diffusion diminishes, transitioning the system into a convection-dominated regime. The flow organizes into coherent, large-scale vortical structures that pump hot fluid upward along the heated wall and impinge upon the upper solid boundary. This convective thermal advection deflects the isotherms upward and concentrates heat flux at the top of the interface, resulting in a distinctly curved, top-heavy melting front.
As a result, melting becomes more non-uniform, with faster melting in the upper region and a more asymmetric phase front.

A comparison between Case~2 and Case~3 shows that while the Prandtl and Stefan numbers are kept fixed at $\mathrm{Pr} = 10$ and $\mathrm{Ste} = 0.4$, the Rayleigh number increases substantially from $\mathrm{Ra} = 2.5 \times 10^4$ to $\mathrm{Ra} = 1.0 \times 10^6$. In Case~2, a large-scale circulation is already established, causing the melting front to curve gradually under the effect of convective heat transfer. In Case~3, however, the much higher Rayleigh number triggers substantially stronger natural convection, as evidenced by more vigorous and intricate streamline patterns. The resulting temperature contours in Case~3 exhibit severe distortion, with the high-temperature zone penetrating deeper into the upper portion of the cavity, which indicates that thermal transport is heavily dominated by convection rather than conduction. Consequently, the solid--liquid interface undergoes pronounced deformation, and the overall melting process is markedly accelerated. Furthermore, Case~3 attains an advanced melting stage at much smaller Fourier numbers ($\mathrm{Fo} = 0.1, 0.15, 0.4$) than Cases~1 and 2 ($\mathrm{Fo} = 0.2, 0.4, 0.8$). This observation corroborates that elevating $\mathrm{Ra}$ intensifies natural convection, enhances heat transfer, and ultimately drives a faster and more non-uniform phase-change process within the three-dimensional rectangular cuboid.

\begin{figure}[htbp]
	\centering
	\begin{subfigure}[b]{0.48\textwidth}
		\includegraphics[width=\textwidth]{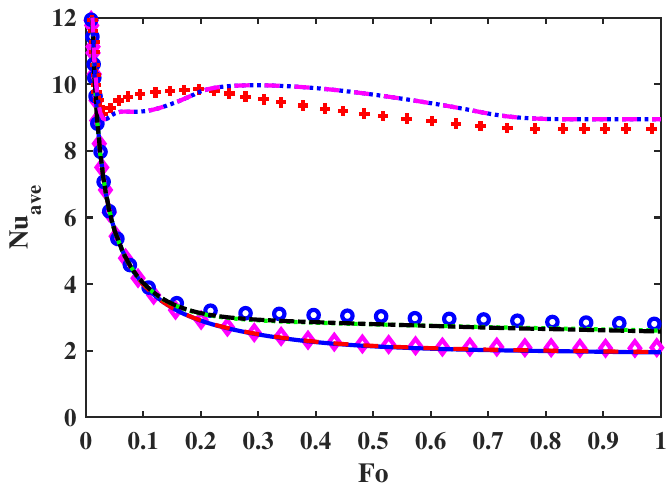}
		\caption{Average nusselt number along the hot wall}
	\end{subfigure}\hfill
\begin{subfigure}[b]{0.48\textwidth}
		\includegraphics[width=\textwidth]{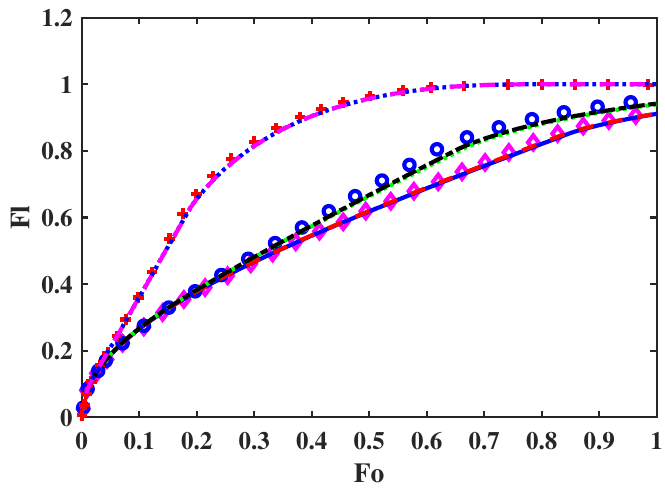}
		\caption{Total fraction number of the domain}
	\end{subfigure}
	\caption{Comparison of (a) averaged Nusselt numbers and (b) total fraction number in contrast with the reference results of Li D. et al.~\citep{Li2017}. The present numerical results are resolved with two resolutions for each case. The denotations of the lines and markers in the figures are as followings: 1) Reference data: Case1:magenta diamond,Case2:blue circular,Case3:red cross; 2) Present results: Case1-mesh1($192\times192\times64$):blue solid line, Case1-mesh2($384\times384\times128$):red dash line;
	Case2-mesh1($192\times192\times64$):green dotted line,
	Case2-mesh2($384\times384\times128$):black dash-dot line;
	Case3-mesh1($256\times256\times128$):blue dotted line;
	Case3-mesh2($512\times512\times256$):magenta dashed line.}
	\label{fig:3D_cavitymelt_allwalls}
\end{figure}

Figure~\ref{fig:3D_cavitymelt_allwalls} compares the average Nusselt numbers and total liquid fractions of the three cases simulated by the present RLB model with the benchmark results from the original MRT model by \cite{Li2017}. It should be addressed that their results were not compared with any other literatures. A grid independence test was conducted using two mesh configurations, and the results demonstrate sufficient convergence without the need for further grid refinement. Overall, the present numerical predictions show good quantitative agreement with the benchmark data for Case~1 and Case~2. During the inital stage of Case~3, the Nusselt number of the present RLB model also agree with those of reference results very well, while there exist some discrepancy during the late stage that might be attributed to the violent, convection-dominated flow at the high Rayleigh number ($Ra$). It should be addressed that the results in the reference paper are reported alone without any comparison.

\section{Conclusion}\label{Conclusion}
In this work, an advanced regularized lattice Boltzmann model with D2Q5 and D3Q7 lattices is designed for the enthalpy-based solid-liquid phase-change simulations. The numerical results have demonstrated that the desinged regularized collision operator can effectively suppress the spurious numerical diffusion across the diffuse solid-liquid interface. Meanwhile, the advanced MRT-RLB collison model is firstly developed to resolve the flow filed with the Boussinesq-type body force term to couple the temperature field. A standard $\text{D3Q27}$ lattice velocity set is adopted in the present MRT-RLB model, wherein the equilibrium distribution functions are recursively expanded up to the sixth order and the non-equilibrium distribution functions are expanded up to fourth order in Hermite polynomials. The main contribution of the present work is that both the total enthalpy equation and the flow field equation are resolved by the designed regularized collision models. The numerical simulations demonstrated that the present MRT-RLB model is capable of the coupled flow field and temperature field simulations with large parameter space, including the Prandtl, Stefan and Rayleigh numbers. Furthermore, the multi-GPU implementation of the present RLB model significantly accelerates the simulation of such computationally intensive solid-liquid phase-change processes with great weak and strong scalability. Generally, an alternative accurate, robust and efficient solid-liquid phase-change model is established in the present work. In the future work, more comprehensive flow and heat-transfer mechanisms will be investigated with the present model.

\section{Acknowledgement}
This work was supported by the National Natural Science Foundation of China No.12472288; Foreign Expert Project of the Ministry of Human Resources and Social Security (MOHRSS) No. 017630130553424. The author X. H. Li would like to extend sincere gratitude and appreciation to Dr. Liu Qing and Dr. Liu Fei.

\FloatBarrier
\bibliographystyle{plainnat}

\bibliography{cas-refs}

\end{document}